%% file: main.tex
\documentclass[acmsmall]{acmart}

\usepackage{amsmath}
\usepackage{algorithmic}
\usepackage{enumitem}
\usepackage{pgfplots}
\usepackage{prettyref}
\usepackage{tikz}
\usepackage{xcolor,colortbl}
\usepackage{multirow}
\usepackage[ruled, linesnumbered, vlined, commentsnumbered]{algorithm2e}
\usepackage[section]{placeins}
\usepackage{booktabs}
\usepackage{threeparttable}
\usepackage{makecell}
\usepackage{graphicx}
\usepackage[listings,skins,breakable]{tcolorbox}
\usepackage{csquotes}
\usepackage{subcaption}
\usepackage{standalone}
\usepackage[skip=0.5\baselineskip]{caption}
\usepackage{wrapfig}
\usepackage{rotating}
\usepackage{arydshln}
\usepackage{siunitx}
\usepackage{hhline}
\usepackage{hyperref}
\usepackage{balance}
\usepackage{xfp}
\usetikzlibrary{calc,positioning,tikzmark,patterns}
\newtcolorbox{quotebox}{colback=steel!10,boxrule=0.4pt,colframe=black,fonttitle=\bfseries,top=2pt,bottom=2pt}

\newtcolorbox{implbox}{%
  sidebyside,sidebyside align=top,lower separated=true,lefthand width=0.5em,
  arc=0pt,
  boxsep=0pt,
  left=2pt,right=2pt,top=2pt,bottom=2pt,colframe=white,
  skin=bicolor,
  colback=black!50, 
  colupper=white,
  colbacklower=white,boxrule=0pt,colframe=white,
  sidebyside gap=5pt,
}

\mathchardef\mhyphen="2D
\newcommand{\vect}[1]{\boldsymbol{#1}}
\DeclareMathAlphabet\mathbfcal{OMS}{cmsy}{b}{n}

\newbox\aMark
\setbox\aMark\hbox{\begin{pgfpicture}\textcolor{blue}{\pgfuseplotmark{o}}\end{pgfpicture}}

\newbox\bMark
\setbox\bMark\hbox{\begin{pgfpicture}\textcolor{red}{\pgfuseplotmark{x}}\end{pgfpicture}}

\newrefformat{fig}{Fig.~\ref{#1}}
\newrefformat{tab}{Table~\ref{#1}}
\newrefformat{sec}{Section~\ref{#1}}
\newrefformat{app}{Appendix~\ref{#1}}
\newrefformat{alg}{Algorithm~\ref{#1}}
\newrefformat{property}{Property~\ref{#1}}
\newrefformat{theorem}{Theorem~\ref{#1}}
\newrefformat{lemma}{Lemma~\ref{#1}}
\newrefformat{corollary}{Corollary~\ref{#1}}
\newrefformat{proposition}{Proposition~\ref{#1}}
\newrefformat{def}{Definition~\ref{#1}}
\newrefformat{eq}{equation~(\ref{#1})}

\definecolor{steel}{rgb}{0, 0.2, 0.9}

\SetCommentSty{mycommfont}

\newcommand{\rev}[1]{\textcolor{black}{#1}}

\definecolor{mycolor}{rgb}{0.122, 0.435, 0.698}

\newtcbox{\mytag}{nobeforeafter,colframe=mycolor,colback=mycolor!30!white,boxrule=0.7pt,arc=0pt,
  boxsep=-3pt,left=6pt,right=6pt,top=4pt,bottom=5pt,tcbox raise base}

\NewDocumentCommand{\gtikzmarkin}{m D(){0.725,-0.10} D(){-0.155,0.27}}{%
      \tikz[remember picture,overlay]
      \draw[line width=0.8pt,red,rectangle,dotted]
      (pic cs:#1) ++(#2) rectangle (#3)
      ;}

\NewDocumentCommand{\ctikzmarkin}{m D(){0.725,-0.10} D(){-0.155,0.27}}{%
      \tikz[remember picture,overlay]
      \draw[line width=0.5pt,black,rectangle]
      (pic cs:#1) ++(#2) rectangle (#3)
      ;}

  \newcommand{\quart}[4]{\begin{adjustbox}{max width=.1\textwidth}\begin{picture}(100,5)%1
    {\color{black}\put(#1,5){\line(1,0){#2}}\color{black}\put(#1,2){\line(0,1){6}}\color{black}\put(\fpeval{#1+#2},2){\line(0,1){6}}\color{steel}\put(#3,5){\circle*{10}}\color{black}\put(#3,5){\circle{10}}}\end{picture}\end{adjustbox}}
    
      \newcommand{\bquart}[4]{\begin{adjustbox}{max width=.1\textwidth}\begin{picture}(100,5)%1
    {\color{black}\put(#1,5){\line(1,0){#2}}\color{black}\put(#1,2){\line(0,1){6}}\color{black}\put(\fpeval{#1+#2},2){\line(0,1){6}}\color{red}\put(#3,5){\circle*{10}}\color{black}\put(#3,5){\circle{10}}}\end{picture}\end{adjustbox}}
    
         \newcommand{\tquart}[4]{\begin{adjustbox}{max width=.1\textwidth}\begin{picture}(100,5)%1
    {\color{black}\put(#1,5){\line(1,0){#2}}\color{black}\put(#1,2){\line(0,1){6}}\color{black}\put(\fpeval{#1+#2},2){\line(0,1){6}}\color{white}\put(#3,5){\circle*{10}}\color{black}\put(#3,5){\circle{10}}}\end{picture}\end{adjustbox}}

      \newcommand{\quartexp}[4]{\begin{adjustbox}{max width=.1\textwidth}\begin{picture}(20,5)%1
    {\color{black}\put(#1,3){\line(1,0){#2}}\color{black}\put(#1,0){\line(0,1){6}}\color{black}\put(\fpeval{#1+#2},0){\line(0,1){6}}\color{steel}\put(#3,3){\circle*{4}}\color{black}\put(#3,3){\circle{4}}}\end{picture}\end{adjustbox}}
    
          \newcommand{\bquartexp}[4]{\begin{adjustbox}{max width=.1\textwidth}\begin{picture}(20,5)%1
    {\color{black}\put(#1,3){\line(1,0){#2}}\color{black}\put(#1,0){\line(0,1){6}}\color{black}\put(\fpeval{#1+#2},0){\line(0,1){6}}\color{red}\put(#3,3){\circle*{4}}\color{black}\put(#3,3){\circle{4}}}\end{picture}\end{adjustbox}}

              \newcommand{\tquartexp}[4]{\begin{adjustbox}{max width=.1\textwidth}\begin{picture}(20,5)%1
    {\color{black}\put(#1,3){\line(1,0){#2}}\color{black}\put(#1,0){\line(0,1){6}}\color{black}\put(\fpeval{#1+#2},0){\line(0,1){6}}\color{white}\put(#3,3){\circle*{4}}\color{black}\put(#3,3){\circle{4}}}\end{picture}\end{adjustbox}}

\setcopyright{acmlicensed}
\acmJournal{TOSEM}
\acmYear{2023} \acmVolume{1} \acmNumber{1} \acmArticle{1} \acmMonth{1} \acmPrice{15.00}\acmDOI{10.1145/3571853}

\begin{document}
	
\title[Do Performance Aspirations Matter for Guiding Software Configuration Tuning?]{Do Performance Aspirations Matter for Guiding Software Configuration Tuning?}
%An Empirical Investigation under Dual Performance Objectives

%%
%% The "author" command and its associated commands are used to define
%% the authors and their affiliations.
%% Of note is the shared affiliation of the first two authors, and the
%% "authornote" and "authornotemark" commands
%% used to denote shared contribution to the research.
\author{Tao Chen}
\email{t.t.chen@lboro.ac.uk}
\affiliation{%
  \institution{Loughborough University}
  \city{Loughborough}
  \country{United Kingdom}
}

\author{Miqing Li}
\affiliation{%
  \institution{University of Birmingham}
  \city{Birmingham}
  \country{United Kingdom}}
\email{m.li.8@bham.ac.uk}

%%
%% By default, the full list of authors will be used in the page
%% headers. Often, this list is too long, and will overlap
%% other information printed in the page headers. This command allows
%% the author to define a more concise list
%% of authors' names for this purpose.
\renewcommand{\shortauthors}{Chen and Li.}


% note the % following the last \IEEEmembership and also \thanks - 
% these prevent an unwanted space from occurring between the last author name
% and the end of the author line. i.e., if you had this:
% 
% \author{....lastname \thanks{...} \thanks{...} }
%                     ^------------^------------^----Do not want these spaces!
%
% a space would be appended to the last name and could cause every name on that
% line to be shifted left slightly. This is one of those "LaTeX things". For
% instance, "\textbf{A} \textbf{B}" will typeset as "A B" not "AB". To get
% "AB" then you have to do: "\textbf{A}\textbf{B}"
% \thanks is no different in this regard, so shield the last } of each \thanks
% that ends a line with a % and do not let a space in before the next \thanks.
% Spaces after \IEEEmembership other than the last one are OK (and needed) as
% you are supposed to have spaces between the names. For what it is worth,
% this is a minor point as most people would not even notice if the said evil
% space somehow managed to creep in.

%As the complexity of performance requirements increases, the configurations of software systems need to be tuned such that the aspirations of not one, but two performance attributes can be satisfied, e.g., ``\texttt{the latency is less than 10s}'' while ``\texttt{the memory usage is no more than 1GB}''. 

\begin{abstract}

%This problem, namely bi-objective software configuration tuning with aspirations, has been widely studied by tailoring Pareto optimizers from the Search-Based Software Engineering paradigm. 

Configurable software systems can be tuned for better performance. Leveraging on some Pareto optimizers, recent work has shifted from tuning for a single, time-related performance objective to two intrinsically different objectives that assess distinct performance aspects of the system, each with varying aspirations to be satisfied, 
e.g., ``\texttt{the latency is less than 10s}'' while ``\texttt{the memory usage is no more than 1GB}''. Before we design better optimizers, a crucial engineering decision to make therein is how to handle the performance requirements with clear aspirations in the tuning process. 
For this, 
the community takes two alternative optimization models: 
either quantifying and incorporating the aspirations into the search objectives that guide the tuning, 
or not considering the aspirations during the search but purely using them in the later decision-making process only.  
However, despite being a crucial decision that determines how an optimizer can be designed and tailored, there is a rather limited understanding of which optimization model should be chosen under what particular circumstance, and why.

In this paper, 
we seek to close this gap. Firstly,
we do that through a review of over 426 papers in the literature and 14 real-world requirements datasets, from which we summarize four performance requirement patterns that quantify the aspirations in the configuration tuning. Drawing on these, we then conduct a comprehensive empirical study that covers 15 combinations of the state-of-the-art performance requirement patterns, 
four types of aspiration space, 
three Pareto optimizers, 
and eight real-world systems/environments, 
leading to 1,296 cases of investigation. 
Our findings reveal that (1) the realism of aspirations is the key factor that determines whether they should be used to guide the tuning; (2) the given patterns and the position of the realistic aspirations in the objective landscape are less important for the choice, but they do matter to the extents of improvement; (3) the available tuning budget can also influence the choice for unrealistic aspirations but it is insignificant under realistic ones. To promote open science practice, we make our code and dataset publicly available at: \href{https://github.com/ideas-labo/aspiration-study}{\texttt{\textcolor{blue}{https://github.com/ideas-labo/aspiration-study}}}.

%\footnote{All data and code can be found at our repository: \href{https://tinyurl.com/y8jczpeo}{\texttt{\textcolor{blue}{https://tinyurl.com/y8jczpeo}}}.}.

%(1) the aspirations,  if realistic,  can help to find considerably better or similar results on 84\% of the cases (including 23\% tie) compared with when tuning without.  (2) When the aspirations are unrealistic  (i.e., no configuration can reach the aspirations of both objectives simultaneously),  they can be fairly harmful to up to 64\% of the cases. (3) Surprisingly,  the best state-of-the-art requirement patterns and combinations in guiding the tuning may not be the one that reflects the given performance requirements. 
\end{abstract}

%Configurable software systems can be tuned for better performance. Recent work has shifted from tuning for a single performance objective to two intrinsically different objectives, each with varying aspirations to be satisfied, e.g., ``latency is less than 10s'' while ``memory usage is no more than 1GB''. A crucial engineering decision to make therein is how to handle the performance requirements with clear aspirations in the tuning. For this, two optimization models are used: either quantifying the aspirations into the search objectives that guide the tuning, or not considering them during the search at all. However, there is very little understanding of which optimization model should be chosen under what circumstance, and why. In this paper, we seek to close this gap through a comprehensive empirical study that covers 1,296 cases. We found that (1) the realism of aspirations is the key factor that determines whether they should guide the tuning; (2) the given patterns and the position of the realistic aspirations in the objective landscape are less important for the choice, but they do matter to the extents of improvement; (3) the tuning budget can also influence the choice for unrealistic aspirations but it is insignificant under realistic ones.


% make the title area
\maketitle

\input{introduction}

\input{problem}

\input{requirements} % + problem formulation?

\input{study}

\input{results}

\input{lessons}

\input{threats}

\input{related}

\input{conclusion}

\bibliographystyle{ACM-Reference-Format}
\bibliography{references}

\end{document}

%% file: introduction.tex
\section{Introduction}
\label{sec:intr}

Many software systems are highly configurable, such that there is a daunting number of configuration options (e.g., the \texttt{max\_spout} in \textsc{Apache Storm}), which the software engineers can tune to meet the requirements of some performance objectives, e.g., improving latency, throughput, and resource consumption~\cite{DBLP:conf/sigsoft/XuJFZPT15,DBLP:journals/computer/ChenB15,DBLP:conf/msr/GongC22,DBLP:journals/csur/ChenBY18,DBLP:conf/kbse/LiXCT20}. Configuration tuning for software systems plays an integral role in Software Engineering as a recent interview reveals that industrial practitioners have recognized it as a key to the success of software products~\cite{DBLP:journals/tse/SayaghKAP20}. Indeed, it has been reported that globally 59\% of the software performance issues---wherein the performance requirements were severely violated---are related to ill-suited configuration rather than code~\cite{DBLP:conf/esem/HanY16}, leading to serious consequences. For example, in 2017-2018, configuration-related performance issues cost at least 400,000 USD per hour for more than 50\% of the software companies worldwide\footnote{https://www.evolven.com/blog/downtime-outages-and-failures-understanding-their-true-costs.html}.

Finding good configurations (i.e., the possible combinational settings of the configuration options) is challenging, because:

\begin{itemize}
    \item The default configuration is often far from ideal. Jamshidi and Casale~\cite{DBLP:conf/mascots/JamshidiC16} show that the defaults for \textsc{Apache Storm} can lead to 480 times worse performance than some others.
    \item The configuration space can be large and the measurement is often expensive~\cite{nair2018finding}, rendering greedy search unrealistic. 
    \item While traditionally software configuration tuning has been focusing on a single performance objective~\cite{DBLP:conf/sigmetrics/YeK03,DBLP:conf/sigsoft/OhBMS17,DBLP:conf/www/XiLRXZ04,DBLP:conf/hpdc/LiZMTZBF14,DBLP:conf/sc/BehzadLHBPAKS13,DBLP:conf/icac/RamirezKCM09,DBLP:conf/ssbse/SinhaCC20,guo2010evaluating,DBLP:conf/wcre/Chen22}, recent work raises the necessity of simultaneously tuning for multiple performance objectives. Our review (see Section~\ref{sec:req}) found that considering two performance objectives is the most common case~\cite{DBLP:journals/ase/GerasimouCT18,Chen2018FEMOSAA,nair2018finding,9397392}. For example, naturally, improving the image quality while reducing the energy consumption are both critical for video encoders like \textsc{x264}; higher accuracy with shorter training time are two inherent performance objectives for deep learning models, e.g., the deep neural network supported by frameworks such as \textsc{Keras}. This further complicates the tuning process as the performance objectives may be conflicting and the extents to such a conflict are often unknown a priori~\cite{Chen2018FEMOSAA,nair2018finding}.
\end{itemize}

%\footnote{Indeed, more than two performance attributes are possible, but considering two is the most common case from the literature when more than one performance aspect is of concern~\cite{DBLP:journals/ase/GerasimouCT18,Chen2018FEMOSAA,nair2018finding,9397392}, which is the focus of this paper.}

%There have been different approaches proposed for multi-objective software configuration,

To automatically tune software configuration for better performance, different approaches have been proposed, such as rule-based~\cite{DBLP:conf/icdcs/GiasCW19,garlan2004rainbow}, learning-based~\cite{DBLP:conf/kbse/BaoLWF19,DBLP:conf/sigsoft/JamshidiVKS18}, and search-based~\cite{Chen2018FEMOSAA,nair2018finding,DBLP:conf/wosp/SinghBSH16,Calinescu2017Designing,DBLP:journals/jss/CalinescuCGKP18,DBLP:conf/icse/Kumar0BB20}. Among these, search-based approach, primarily relying on the Pareto optimizers widely used in the Search-Based Software Engineering (SBSE) paradigm~\cite{Harman2012}, has been a promising way to handle all the aforementioned challenges in software configuration tuning, especially in the presence of more than one performance objective~\cite{DBLP:conf/wosp/SinghBSH16,Calinescu2017Designing,DBLP:journals/jss/CalinescuCGKP18,Chen2018FEMOSAA}. In a nutshell, a Pareto optimizer most commonly maintains a population (or at least an archive) of configurations, which can be repeatedly reproduced and evaluated by directly profiling the software, aiming to find the Pareto optimal ones. The output is a set of configurations that are nondominated to each other, which approximates the Pareto front of the software system.

\subsection{The Problem and Significance}

\rev{An important factor in software configuration tuning is the possible requirements with clear aspirations for the performance objectives~\cite{DBLP:conf/models/RamirezC11,Calinescu2017Designing,DBLP:conf/sigsoft/EsfahaniKM11,DBLP:journals/jss/CalinescuCGKP18}, for which we distinguish two important notions in this work:}

\begin{itemize}
    \item \rev{\textbf{Aspiration:} The information that allows us to quantify the extent to which the performance is considered satisfactory (or unsatisfactory).} 
    \item \rev{\textbf{Performance requirement:} The context under which the preference of the performance is defined.}
\end{itemize}

\rev{For example, according to the research in the Requirement Engineering community~\cite{DBLP:conf/re/WhittleSBCB09,DBLP:conf/re/BaresiPS10}, it is not uncommon to have performance requirements from the requirement documents, such as ``\texttt{the latency shall be less than $x$}'' while ``\texttt{the memory usage shall be no more than $y$}'', where ``\texttt{less than $x$}'' and ``\texttt{no more than $y$}'' are the clear aspirations therein. It is worth noting that not all performance requirements would contain aspiration, e.g., ``\texttt{the latency shall be low}'' is a requirement with no aspiration since nothing can be quantified with respect to the level of satisfaction.} Indeed, given a scenario with clear aspirations in the performance requirements, it has been well-acknowledged that the information provided serves as useful metrics for the software engineers to conduct a posterior cherry-picking after the tuning completes, extracting the satisficing configuration(s) from the set produced by a Pareto optimizer~\cite{9252185}. The natural motivations behind this are:

%the complex configuration space and
\begin{itemize}
    \item Given a fixed tuning budget, finding the optimal performance is not always feasible or even desirable to the stakeholders.
    \item The clear aspiration levels allow an implicit trade-off/preferences between the conflicting performance objectives according to the stakeholders.
\end{itemize}

Regardless of the Pareto optimizer used, in the tuning process, existing work takes one of two intrinsically different optimization models to handle aspirations when tuning for two performance objectives, namely: \textbf{Pareto search with aspirations} (denoted as \texttt{PS-w})~\cite{Calinescu2017Designing,DBLP:conf/wosp/MartensKBR10,Gerasimou2016Search,DBLP:journals/jss/CalinescuCGKP18} and \textbf{Pareto search without aspirations} (denoted as \texttt{PS-w/o})~\cite{Chen2018FEMOSAA,DBLP:conf/wosp/SinghBSH16,nair2018finding,DBLP:conf/qosa/KoziolekKR11}. In \texttt{PS-w}, the performance requirements with aspirations are quantified in certain forms (we will elaborate on this in Section~\ref{sec:req}), which then serve as new search objectives in the tuning. The motivation is simple: since the aspirations provide information on the degree of satisficing, one can exploit this advantage to guide the tuning process. \texttt{PS-w/o}, in contrast, is more classic and simply ignores the aspirations in the tuning. \rev{The assumption here is that, since the search in whatever a Pareto optimizer is essentially an optimization process that seeks to find the Pareto optimal configurations, the tuning always aims to achieve the best possible performance, which preserves the tendency towards satisficing whatever aspirations\footnote{This assumes the most common case that the best possible performance is at least equally preferred than some other values.}}. For example, finding the Pareto optimal configuration \texttt{latency=10s} and \texttt{memory usage=1GB} will certainly meet the requirement and aspiration of ``\texttt{latency shall be less than 20s}'' while ``\texttt{memory usage shall be no more than 2GB}''. This matches with Odhnoff's argument that ``\textit{optimizing}'' and ``\textit{satisficing}'' are merely stylistically different but fundamentally the same~\cite{odhnoff1965techniques}.

%there has been no systematic and quantitative study to justify the choices. 

Despite either of the two optimization models being respectively used by their corresponding research groups, the choice was mostly ad-hoc and there is often an implied belief that \textit{``they do not differ much hence can be used arbitrarily.''} As such, there remains a rather limited understanding of which optimization model should be chosen under what particular circumstance, and why. This has been well-echoed by some researchers. Ghanbari \textit{et al.}~\cite{DBLP:journals/fgcs/GhanbariSLI12} have stated that it is important to consider the choice, as the shape of the function that guides the tuning, especially after passing the aspirations, may impact the behavior of the optimizer; but they did not discuss what implication that would be. Yet another example from a recent work by Fekry \textit{et al.}~\cite{DBLP:conf/icdcs/FekryCPRH19} recommends that studying whether to leverage aspirations for guiding the optimizers and measuring its effectiveness is an important future challenge for software configuration tuning. Indeed, \textbf{understanding in this regard is non-trivial as it will help practitioners to make more informed-decision,} especially when given the expensive measurements of configurable software systems, it is unrealistic to always empirically compare the two models in a case-by-case manner. Furthermore, \textbf{the insights can hint at future research directions for software configuration tuning:} if the \texttt{PS-w/o} is more promising, then we can largely simplify the research to the design of an effective optimizer without considering the given requirements since the human inputs (i.e., the requirements/aspirations) are less important in the overall tuning process. On the other hand, if \texttt{PS-w} is overall more effective, then the problem can become more complicated but also provide more opportunities, e.g., future research can largely focus on how to better quantify those performance requirements and aspirations, together with how to better embed them into more specialized optimizers.

To understand this, 
we have also tuned into the literature on general multi-objective optimization, 
with a particular focus on preference-driven multi-objective optimization~\cite{wang2017mini,DBLP:journals/ac/BechikhKSG15,9066927,DBLP:conf/cec/YuJO19}. However, we did not find answers that are directly relevant to our case, 
due to two reasons: 
(1) the representation of the preferences (e.g., weights and ranks) in preference-driven multi-objective optimization is different from the requirement patterns we summarized from the work for software configuration tuning; 
(2) they mainly develop algorithms/optimizers that are tailored to a specific preference representation 
while software configuration tuning often relies on a vanilla optimizer~\cite{Calinescu2017Designing,DBLP:journals/jss/CalinescuCGKP18}.

Our work is, therefore, motivated by the desirability of the community to understand the following:

	\begin{displayquote} 
		\textit{Should we incorporate requirements and aspirations to guide the software configuration tuning process? If so, in what context and why?}
	\end{displayquote}

\subsection{Research Questions}

In this paper, we seek to fill the above gap via an empirical study that systematically compares \texttt{PS-w} and \texttt{PS-w/o} for tuning software configuration under two performance objectives. \rev{Suppose that there are some \textit{realistic} aspirations (i.e., all the aspirations are achievable by tuning the configuration of the software system),} the first research question (RQ) we wish to answer is:

% To the best of our knowledge, this is the first work that studies the effectiveness between \texttt{PS-w} and \texttt{PS-w/o} for bi-objective software configuration tuning, aiming to build ``\textit{a tale of two cities}".

\begin{quotebox}
\noindent
\textit{\textbf{RQ1:} Given performance requirements with realistic aspirations, of \texttt{PS-w} and \texttt{PS-w/o}, which can find a better set of configurations?}
\end{quotebox}

%In particular, we are specifically interested in two sub-research questions:

%\begin{itemize}
%    \item \textbf{RQ1.1:} What is the result under reasonably sufficient convergence\footnote{By reasonably sufficient convergence, we mean that increases the tuning budget is unlikely to significantly change the result of both \texttt{PS-w} and  \texttt{PS-w/o}.}?
%    \item \textbf{RQ1.2:} What if the tuning budget is limited, i.e., no sign of clear convergence? 
%\end{itemize}

\textbf{RQ1} seeks to provide a global picture of the comparison between the two optimization models. However, the diverse possible requirement scenarios imply that the specific aspirations can be radically different in the objective landscape. For example, one may have higher expectations on latency while lower needs on throughput, or vice versus. Therefore, what we would like to understand in more detail is: 

\begin{quotebox}
\noindent
\textit{\textbf{RQ2:} How do different realistic aspirations influence the result?}
\end{quotebox}

\textbf{RQ1} and \textbf{RQ2} investigate under the normal context where the given aspirations are reasonable and achievable. \rev{However, since the actual aspirations are negotiated by the software engineers and stakeholders a priori, they could turn out to be \textit{unrealistic} and may require attention beyond configurations, i.e., no configuration in the search landscape can reach the required aspiration levels for all performance objectives simultaneously, despite that may be possible for a single objective.} This brings our next RQ, in which we ask:

\begin{quotebox}
\noindent
\textit{\textbf{RQ3:} What if the given aspirations are unrealistic?}
\end{quotebox}

While we are interested in cases where the tuning budget is reasonably sufficient to achieve a good convergence, it is possible that, in real-world scenarios, there is a limited resource for tuning software configuration due to, e.g., pressure for quick release or task prioritization. Therefore, our last question aims to explore:

\begin{quotebox}
\noindent
\textit{\textbf{RQ4:} Is the given tuning resource (tuning budget) important to the choice between \texttt{PS-w} and \texttt{PS-w/o}?}
\end{quotebox}

%Since each performance objective may come with a different pattern of performance requirement, these patterns can be arbitrarily combined in a bi-objective software configuration tuning scenario. This is interesting for  \texttt{PS-w}, as the patterns provided in a requirement scenario do not necessarily be identical to the ones used to guide the tuning, therefore our final RQ seeks to understand:

%\begin{quotebox}
%\noindent
%\textit{\textbf{RQ3:} Given the required combination of state-of-the-art patterns $\mathbfcal{P}$ in a requirement scenario, will \texttt{PS-w} achieve the best result only when it is guided by the same $\mathbfcal{P}$, among the other combinations?}
%\textit{\textbf{RQ3:} For \texttt{PS-w}, what is the best combination of the state-of-the-art patterns for guiding the tuning under a requirement scenario?}
%\end{quotebox}

%Clearly, we would only be interested in \textbf{RQ3} if \texttt{PS-w} can indeed demonstrate some benefits over its \texttt{PS-w/o} counterpart.

%\begin{itemize}
   % \item \textbf{RQ3.1:} What if the given requirements are guaranteed to be realistic?
    %\item \textbf{RQ3.2:} What if the given requirements are unrealistic? 
%\end{itemize}

%\textit{\textbf{RQ3:} Will the combination of state-of-the-art patterns used in \texttt{PS-w}, which is consistent with what is required in the given requirement scenario, always lead to the best result?}

\subsection{Contributions}

%Addressing these RQs requires us to at first understand what kind of common requirements with aspirations exist in the industry, and how they have been quantified for a performance objective in the literature (see Section~\ref{sec:req}). During the process, we have identified three state-of-the-art performance requirement patterns that can quantify the aspirations (such as whether a certain violation is permitted or whether outperforming the aspiration is more preferred), together with one classic pattern for the cases where no aspiration exists. 

To address these RQs , we conducted an extensive empirical study on 15 combinations of patterns to quantify aspirations, four types of aspiration space in the objective landscape, three Pareto optimizers, and eight real-world systems/environments with diverse performance objectives, leading to 1,296 cases of investigations. Briefly, the \textit{\underline{first contribution}} in this paper is a set of performance requirement patterns (for individual performance objectives) summarized from 426 papers in the literature \rev{from the Software Engineering community} and 14 widely-used real-world requirements datasets. \rev{These patterns are}:

\begin{enumerate}

   \item \rev{No aspiration is given but assuming that the optimal possible performance is preferred, e.g., ``the lower latency is preferred'', meaning that one prefers the best possible latency.}

       \item \rev{The performance in the aspiration space is equally good or otherwise there is a certain degree of tolerance, e.g., ``the minimum latency shall ideally be {500ms}'', implying that anything better than \textit{500ms} is equally good while a performance worse than that is acceptable but not ideal.}
    
    \item \rev{The performance in the aspiration space is equally good while anything outside the space is unacceptable, e.g., ``the latency shall be {500ms}''. This suggests that a latency better than \textit{500ms} is equally good and no tolerance is allowed for performance worse than that.}

    \item \rev{Preferring the optimal performance while anything outside the aspiration space is unacceptable, e.g., ``the latency shall be at most {500ms}'', reflecting that no tolerance is allowed for worse than \textit{500ms} while the lower the latency, the better.}
\end{enumerate}

Our \textit{\underline{second contribution}} is the pragmatic findings that answer the aforementioned RQs over the 1,296 cases as follows:

\begin{itemize}
    \item \textbf{To RQ1:} \texttt{PS-w} performs considerably better or similar to \texttt{PS-w/o} on 84\% of the cases, out of which over 60\% show statistically significant improvement. 
    \item \textbf{To RQ2:} The improvement of \texttt{PS-w} over \texttt{PS-w/o} is often largely biased to a certain position of the aspiration space in the objective landscape, e.g., centered or left-shifted.
    \item \textbf{To RQ3:} \texttt{PS-w/o} is no worse than \texttt{PS-w} for 70\% cases, wherein the difference is considerable with statistical significance for more than 85\%. 
    \item \textbf{To RQ4:} Under realistic aspirations, \texttt{PS-w} obtains consistently better outcomes than \texttt{PS-w/o} throughout the trajectory and with a speedup up to $10\times$. When the aspirations are unrealistic, in contrast, the two optimization models are competitive in the early stage of tuning but soon \texttt{PS-w/o} would lead to better results with considerably high speedup.
    %The conclusion still hold even under limited tuning budget.
    
    %\item \textbf{To RQ4:} To our surprise, guiding the tuning by \texttt{PS-w} with the combination of patterns that reflects exactly the given scenario rarely leads to the sole best results therein.
\end{itemize}

Hence, we conjecture that \textbf{the performance aspirations do matter for guiding bi-objective software configuration tuning in general. Yet, depending on the context, it can either be helpful or harmful.} We provide, as part of the \textit{\underline{third contribution}}, some in-depth analysis and discussions on the reasons behind the above observations. More importantly, these findings allow us to derive our  \textit{\underline{fourth contribution}}: the key lessons learned on the choice between \texttt{PS-w} and \texttt{PS-w/o} for bi-objective software configuration tuning, which are:

%our empirical study provides new insights that help to further advance our understanding on multi-objective software configuration tuning:

\begin{itemize}
    \item {\textbf{Lesson 1:} The choice on whether to exploit aspirations for guiding the tuning is primarily dependent on their realism.}
    \item  {\textbf{Lesson 2:} It is unlikely that the combinations of patterns can change the decision on whether to incorporate aspiration in the tuning, but it can influence the benefit/detriment of aspiration-guided tuning.}
    \item   {\textbf{Lesson 3:} The positions of realistic aspiration space in the objective space can largely affect the benefits brought by considering aspirations within tuning, but it is less likely to influence the choice.}
    \item  {\textbf{Lesson 4:} The given tuning budget has a marginal impact on the choice when the aspirations are realistic. However, it can be an important factor to consider under unrealistic aspirations.}
    %\item The comparison results between \texttt{PS-w} and \texttt{PS-w/o} are, in general, less significant and divergent in cases for Machine Learning (ML) software than in those for other domains. This is because the effect of configurations on performance attributes in ML software often has lower sparsity.
   % \item The findings for \textbf{RQ3} imply that, surprisingly, most state-of-the-art patterns/combinations, which reflect the given requirement scenario, can be substituted by the others with no negative impact on guiding the tuning towards the desired configurations, the details of which are vital for future research on their provable relationships.

    %certain state-of-the-art patterns/combinations are exchangeable to guide the tuning with no negative impact on satisficing most of the given performance requirements and their aspirations, 
    
    %\item The findings for \textbf{RQ3} imply that some state-of-the-art patterns (and their combinations) can, in fact, create similar impacts to the search, the details of which are vital for future research on their provable ``relaxation'' relationships.
\end{itemize}

Drawing on those lessons, our  \textit{\underline{fifth contribution}} outlines three future opportunities for this field of research, namely:

\begin{itemize}

\item Landscape Analysis for Configurable Software Systems.
\item Requirement-Robust Optimizer for Configuration Tuning.
\item Study on the Relative Impact between Requirement Patterns to the Tuning.

\end{itemize}

To promote open science practice, all the code, dataset, and necessary supplementary documents for this work can be publicly accessed at: \href{https://github.com/ideas-labo/aspiration-study}{\texttt{\textcolor{blue}{https://github.com/ideas-labo/aspiration-study}}}.

The rest of this paper is organized as follows: Section~\ref{sec:prob} formalizes the problem and presents the motivating example. Section~\ref{sec:req} discusses the patterns that quantify performance requirements with aspirations and how they were identified. Section~\ref{sec:emp} elaborates the design of our empirical study. Section~\ref{sec:result} presents and analyzes the experiment results. Thereafter, Section~\ref{sec:lessons} discusses the lessons learned and future opportunities, followed by threats to validity in Section~\ref{sec:threats}. Finally, Sections~\ref{sec:related} and~\ref{sec:con} review the related work and conclude the paper, respectively.  

%The related work is reviewed in Section~\ref{sec:related}. Finally, Section~\ref{sec:con} concludes the paper with pointers for future work.

% SCT
% SBSE for SCT
% (requirement) two types of Pareto search
% limitation
% finding 
% insight

%% file: problem.tex
%\section{Problem in software configuration tuning}

\section{Theory}
\label{sec:prob}

In this section, we present the theoretical knowledge for understanding the purpose of this work.

\subsection{Formal Definition}

\subsubsection{Background and Problem Formalization}

In the DevOps era, software configuration tuning involves two fundamental roles that interact frequently~\cite{DBLP:journals/tse/SayaghKAP20} ---
the stakeholders (whose benefit is directly affected by the software performance) negotiate their performance requirements with the software engineers, who then act as the operators to tune the configurations for satisfying these requirements. Beyond a single performance concern, recently there has been an increasing demand for considering multiple performance objectives~\cite{9397392,DBLP:journals/corr/abs-2001-08236}. Among those, our literature review from Section~\ref{sec:req} shows that 90\% of the recent work has considered two performance objectives~\cite{DBLP:journals/ase/GerasimouCT18,Chen2018FEMOSAA,nair2018finding}, such as the latency versus throughput for \textsc{Storm}; image quality versus energy usage for \textsc{x264}. This makes software configuration tuning with requirements in mind even more complex. 

Without loss of generality, we assume that a configurable software comes with a set of configuration options, whereby the $i$th option is denoted as $c_i$, which can be a binary, integer, or enumerate variable. A particular configuration is denoted as $\vect{\overline{c}}$. The search space, $\mathbfcal{C}$, is the Cartesian product of the possible values for all the $c_i$. Formally, given a scenario of requirements with clear aspirations for two performance objectives, the goal of \texttt{PS-w} for software configuration tuning is to find the configuration(s) that achieve:
\begin{align}
	maximize~p_x(f_1(\vect{\overline{c}})), p_y(f_2(\vect{\overline{c}})),~~\vect{\overline{c}} \in \mathbfcal{C}
	\label{Eq:SOP}
\end{align}
whereby $f$ is the raw measurement of the performance value achieved by $\vect{\overline{c}}$; $p$ is the corresponding requirement pattern, which quantifies the degree of satisficing given $f(\vect{\overline{c}})$ (see Section~\ref{sec:req}). In this work, we consider cases where at least one $p$ contains a clear aspiration level\footnote{We use $p_x$ and $p_y$ to distinguish two performance requirement patterns.}.

In contrast, the goal of \texttt{PS-w/o} is to:

\begin{align}
	minimize~f_1(\vect{\overline{c}}), f_2(\vect{\overline{c}}),~~\vect{\overline{c}} \in \mathbfcal{C}
	\label{Eq:SOP-wo}
\end{align}

As can be seen, \texttt{PS-w} explicitly leverages information about the given requirements with clear aspirations to guide the search and tuning while \texttt{PS-w/o} assumes the basic Pareto optimality\footnote{We assume that all performance objectives are to be minimized; maximizing ones can be easily converted.}. 

% \begin{align}
% 	\argmax~g_1(f_1(\vect{\overline{c}})), g_2(f_2(\vect{\overline{c}})), ..., g_n(f_n(\vect{\overline{c}})),~~\vect{\overline{c}} \in \mathbfcal{C}
% 	\label{Eq:SOP}
% \end{align}

\begin{figure}[t!]
\centering
\includegraphics[width=\textwidth]{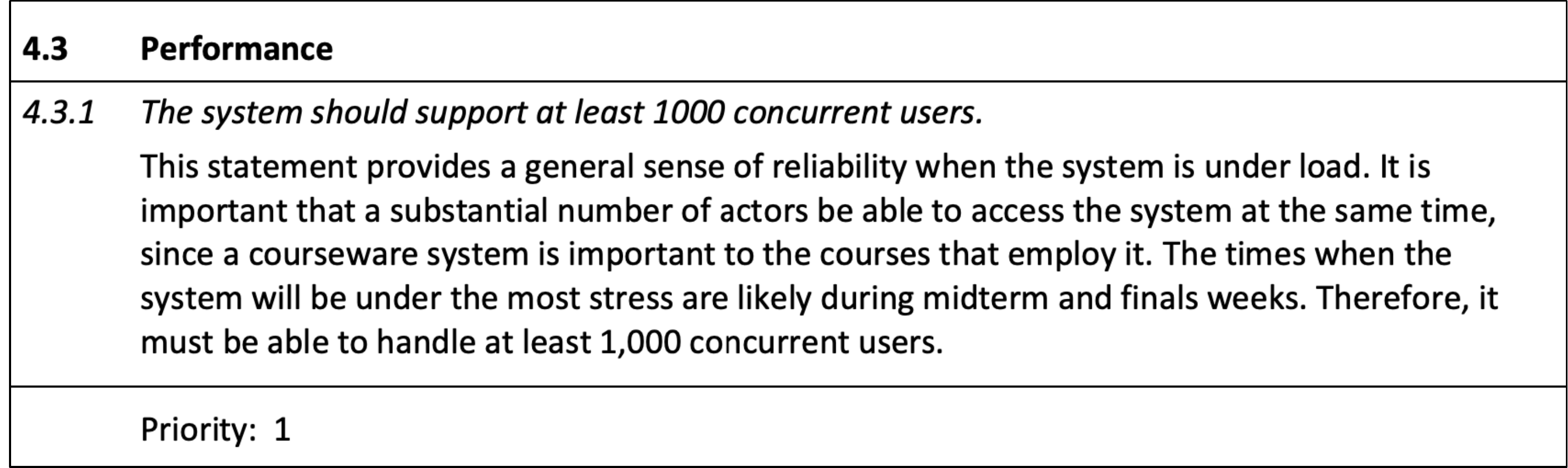}
\caption{A performance requirement snippet from the requirement document of a real-world project in the PURE dataset~\cite{DBLP:conf/re/FerrariSG17}.}
\label{fig:perf-req}
\end{figure}

\begin{figure}[t!]
\centering
\includegraphics[width=0.5\textwidth]{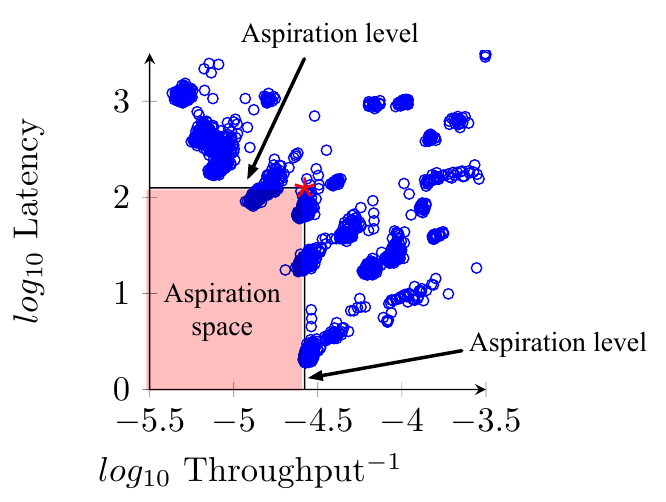}
\caption{The aspiration space (highlighted by color) and aspiration levels within the bi-objective space (latency and throughput) for \textsc{Storm} under the \textsc{Rolling Sort} benchmark.}
\label{fig:aspir-exp}
\end{figure}

\subsubsection{Aspiration Space}

Following the normal software engineering practice of requirement negotiation, it is likely that a single performance requirement can come with a \textbf{clear aspiration}, in which case we define \textbf{aspiration space} as the portion of performance points that are not inferior to the given aspiration level. A real-world example has been shown in Figure~\ref{fig:perf-req}. Here, ``\texttt{the system shall support at least 1,000 concurrent users}'' contains a clear \textbf{aspiration level} of \textit{1,000 users}, meaning that the aspiration space covers throughput between 1,000 (inclusive) and the true optimum (which is case-dependent). Beyond such a one-dimensional case, it is easy to know that the aspiration space can be generalized to a two-dimensional case when the aspiration levels of two performance objectives are involved. For example, Figure~\ref{fig:aspir-exp} shows the aspiration space for the requirements ``\texttt{the system shall perform with 39800 users at a time}'' while ``\texttt{the latency shall be no worse than 160 seconds}'' for \textsc{Storm} (with log-transformed values ($\log_{10}$) and all performance objectives are to be minimized as we consider the reciprocal of Throughput). This forms the foundation of our analysis in what follows.

\begin{algorithm}[t]
    \DontPrintSemicolon
    \footnotesize
    %\scriptsize
    \caption{Unified code for \texttt{PS-w} and \texttt{PS-w/o} with NSGA-II.}
    \label{alg:nsgaii}
\KwIn{Configuration space $\mathcal{V}$; the system $\mathcal{F}$; a matrix of fitness quantified by the requirements $\Gamma$}
    %\KwIn{Portfolio of transfer learners $\mathcal{T}$ and classifier $\mathcal{C}$}
    \KwOut{A set of nondominated configurations $\mathcal{S'}$}
    Randomly initialize a population of $n$ configurations $\mathcal{P}$\\
     \tcc{measuring on the actual configurable system}
    \textsc{measure($\mathcal{P},\mathcal{F}$)}\\
    
    \tcc{\textcolor{blue}{for \texttt{PS-w}, the fitness that guides the search is computed according to Equation (\ref{Eq:SOP})}}
      \lIf{\texttt{PS-w}}{
        $\Gamma\leftarrow$\textsc{getFitnessBasedonRequirements($\mathcal{P}$)}
        }
    
    %\tcc*[r]{$t\in\mathcal{T}$ and $c\in\mathcal{C}$ are a candidate transfer learner and classifier}
    \While{The search budget is not exhausted}
    {  
      $\mathcal{P'}=\emptyset$\\
      \While{$\mathcal{P'}<n$}
      { 
         
           \tcc{for \texttt{PS-w}, selecting parents with respect to their compliance to the requirements}
          \lIf{\texttt{PS-w}}{
      $\{s_x,s_y\}\leftarrow$\textsc{mating($\mathcal{P}, \Gamma$)}
        } \lElse {
         $\{s_x,s_y\}\leftarrow$\textsc{mating($\mathcal{P}$)}
        }
        $\{o_x,o_y\}\leftarrow$\textsc{doCrossoverAndMutation($\mathcal{V}, s_x,s_y$)}\\
        \textsc{measure($o_x, o_y, \mathcal{{F}}$)}\\

      \lIf{\texttt{PS-w}}{
        $\Gamma\leftarrow$\textsc{getFitnessBasedonRequirements($o_x, o_y$)}
        }
        $\mathcal{P'}\leftarrow\mathcal{P'}\bigcup\{o_x,o_y\}$\\
      }
      
          \tcc{\textcolor{blue}{for \texttt{PS-w}, the configurations are preserved according to the fitness computed with respect to the requirements}}   
         \lIf{\texttt{PS-w}}{
        $\mathcal{U}\leftarrow$\textsc{nondominatedSorting($\mathcal{P}\bigcup\mathcal{P'},\Gamma$)}
        } \lElse {
             $\mathcal{U}\leftarrow$\textsc{nondominatedSorting($\mathcal{P}\bigcup\mathcal{P'}$)}
        }

      $\mathcal{P}\leftarrow$top $n$ configurations from $\mathcal{U}$\\  

    }

   \lIf{\texttt{PS-w}}{
       \Return $\mathcal{S'}\leftarrow$\textsc{nondominatedConfigurations($\mathcal{P},\Gamma$)}
        } \lElse {
           \Return $\mathcal{S'}\leftarrow$\textsc{nondominatedConfigurations($\mathcal{P}$)}
        }

\end{algorithm}

\subsubsection{Pareto search with and without Aspirations for Tuning Software}

To illustrate the difference between \texttt{PS-w} and \texttt{PS-w/o}, a pseudo-code using NSGA-II as the underlying optimizer has been shown in Algorithm~\ref{alg:nsgaii}. \rev{As can be seen, \texttt{PS-w} and \texttt{PS-w/o} mainly differ in the fact that the former is guided by the information extracted from the given requirements and aspirations (denoted as $\Gamma$) while the latter runs without, i.e., it uses the raw values of the measured performance objectives. This means that all the fitness of configurations evaluated in the \texttt{PS-w} makes use of the $\Gamma$ while that of the configuration in \texttt{PS-w/o} does not. For example, under the raw performance, a latency of \textit{500ms} is certainly more preferred than the case of \textit{700ms}. However, under the requirement and aspiration that any latency less than \textit{900ms} is equally preferred, 
they are actually equivalent therein and hence \texttt{PS-w} reflects precisely that.}

\rev{As a result, the above generates two differences between \texttt{PS-w} and \texttt{PS-w/o}. Firstly, the process of deciding on which two configurations to be selected as parents for generating new configurations is guided differently (i.e., lines 10--11). Secondly, the environmental selection that determines what configurations to be preserved in the next iteration is also guided by different fitness (i.e., lines 17--18).}

\rev{As we will show, even with such a simple deviation the leading results can be radically different depending on the circumstances.}

%As can be seen, \texttt{PS-w} mainly differs in the fact that the tuning is guided by the given requirements, which also determine which configuration to be preserved and the Pareto dominance relation between the configurations. However, as we will show, even with such a simple deviation the leading results can be radically different depending on the circumstances.

\subsection{Motivating Scenario}
\label{sec:motivation}

%Depending on the actual requirements, 

Taking \textsc{x264} --- a configurable video encoder --- as a concrete example, a possible \textbf{requirement scenario} could involve \textbf{performance requirements} (denoted as $\mathbfcal{P}_1$) \texttt{``the PNSR\footnote{PNSR stands for Peak signal-to-noise ratio, which measures the reconstruction quality for images; the larger the PNSR, the better.} shall be at least 40dB''} and \texttt{``the energy usage shall be at most 80 watts''}. Here, there is a clear \textbf{aspiration level} \textit{40dB} and \textit{80 watts} for the performance attribute PNSR and energy usage, respectively. Indeed, depending on the requirement scenario, the preference for performance deviating from the aspiration level could vary even with a clear aspiration level (as we will discuss in Section~\ref{sec:req}). For instance, the above example may imply that \textit{one would not accept any performance worse than 40dB or 80 watts but prefers any configurations with better PNSR and energy usage.} This means that, suppose there are three configurations $\vect{A} = \{65dB,30 watts\}$, $\vect{B} = \{80dB,25 watts\}$, and $\vect{C} = \{35dB,10 watts\}$, the $\vect{C}$, although it has the best energy usage, would be ruled out as it fails to meet aspiration for PNSR; $\vect{B}$ would certainly be more ideal under such a requirement scenario since it has better results on both performance objectives than $\vect{A}$. In a different requirement scenario, the requirements (denoted as $\mathbfcal{P}_2$) may become \texttt{``the PNSR shall be no worse than 40dB''} while \texttt{``the energy usage shall be no worse than 80 watts''}, which implies that  \textit{one would not accept any performance worse than 40dB or 80 watts, but equally prefer anything that goes beyond 40dB and 80 watts.} Here, $\vect{C}$ is ruled out again but $\vect{A}$ and $\vect{B}$ would become equally preferred as their PNSR and energy usages are better than \texttt{40dB} and \texttt{80 watts}, respectively. Of course, the given \textit{40dB} and/or \textit{80 watts} may well be unrealistic aspirations, i.e., none of the configurations would reach them (or at least no one can be found under the possible tuning budget).

\begin{figure}[t!]
\centering
\includegraphics[width=\textwidth]{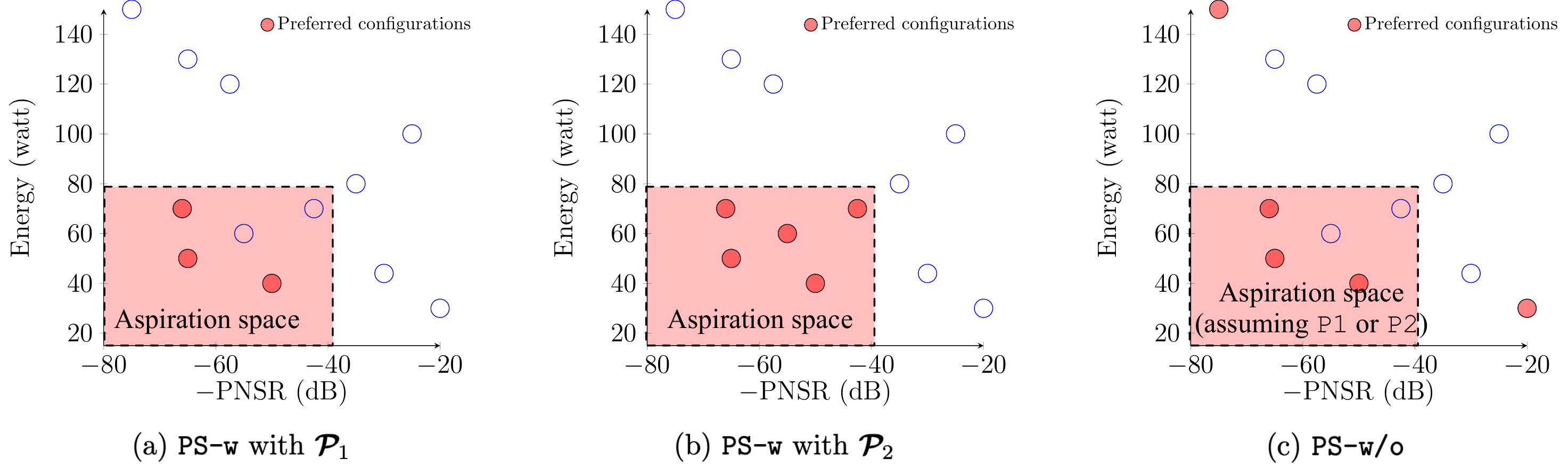}
\caption{The preferred configurations for \textsc{x264} by \texttt{PS-w/o} and \texttt{PS-w} given different requirement scenarios.}
\label{fig:exp-search}
\end{figure}

To make the meaning of the above clear for \texttt{PS-w} and \texttt{PS-w/o}, Figure~\ref{fig:exp-search} illustrates what configurations are preferred when using \texttt{PS-w} and \texttt{PS-w/o} in the tuning under $\mathbfcal{P}_1$ or $\mathbfcal{P}_2$. Here, the quality of the configurations produced would need to be evaluated with respect to the requirements and \texttt{PS-w} prefers precisely what is needed therein. \texttt{PS-w/o}, in contrast, naturally prefers all configurations on the Pareto front. Intuitively, we note that \texttt{PS-w/o} would also prefer some configurations that are preferred by its \texttt{PS-w} counterpart. For example, when comparing Figure~\ref{fig:exp-search}a and~\ref{fig:exp-search}c, all the points preferred by \texttt{PS-w} are also preferred by \texttt{PS-w/o} (but not vice versus), hence they should converge to the same satisfiability under $\mathbfcal{P}_1$. In Figure~\ref{fig:exp-search}b and~\ref{fig:exp-search}c, although \texttt{PS-w/o} prefers different points to that of \texttt{PS-w} in the aspiration space, they should be able to reach the same degree of satisfaction with respect to the requirements because all configurations within the aspiration space are deemed equivalent when being evaluated by $\mathbfcal{P}_2$. Indeed, if both \texttt{PS-w} and \texttt{PS-w/o} can find all their preferred points in the space, then the engineers can simply cherry-pick the fully satisfied ones according to the given performance requirements from the final set of configurations returned. Yet, the unanswered question would be: is the above assumption true and hence there would be no difference regarding whether \texttt{PS-w} or \texttt{PS-w/o} is chosen?

% Clearly, when determining which configurations to be preserved into the next iteration/generation during the search and tuning, each of them has a different set of preferred points but some of which are overlapping. Therefore, given the requirements (says $\mathbfcal{P}_1$), an unanswered question is: can \texttt{PS-w} leverage the aspiration information to better locate those satisfied configurations than its \texttt{PS-w/o} counterpart? 

%But the question is, given expensive measurement and limited search budget, will that really happen?

The rest of this paper provides an empirical understanding of the above confusion.

%the configurations $\vect{A}$ and $\vect{B}$ from above become nondominated, i.e., a trade-off is required.

%The same applied to the cases where clear aspiration levels exist for both performance attributes.

%Our goal in this paper is to understand whether taking the above requirement qualification to guide the search, when used in the bi-objective software configuration tuning, would have any implication for the results with respect to the given requirement scenario. 

%% file: requirements.tex
%\section{Quantifying Requirement Patterns}
\section{How Requirements are Handled}
\label{sec:req}

Here we describe the process of mining, classifying, and analyzing the real-world performance requirements with aspirations. We use Cohen's Kappa coefficient ($\kappa$)~\cite{mchugh2012interrater} to mitigate bias between authors --- the classification is often regarded as unbiased and sustainable when $\kappa > 0.7$ . \rev{In a nutshell, Cohen’s Kappa coefficient is generally thought to be a more robust measure than a simple percent agreement calculation between the raters, as it takes into account the possibility of the agreement occurring by chance. In this work, we use the coefficient in two aspects:}

\begin{itemize}

\item \rev{Measure the agreement on which implication category a requirement belongs to (we have $\kappa=0.85$).}

\item \rev{Measure the agreement on which patterns that a paper assumes (we have $\kappa=0.76$).}

\end{itemize}

% find dataset (requirements with aspiration, which needs to be satisfied)
% - on performance or non-functional requirements
% - performance attribute the can be applicable to many software (non feature specific)
% - has one quantifiable/numeric aspiration level (or can be converted or combined)
% - share the same optimum when no asipratiion level is avaiable

% put into the 3 categories, the one can be p1 or p2; the ones can be p2 or p3; the ones can be p1, p2 or p3; p2 can be further splited into p4

% then summarie these four categories and find papers that do their quantificaition on configrable software.

% find papers
% fuzzy qunatification in a unified way
\subsection{Real-world Requirements with Aspirations}
\label{sec:real-req}

\begin{figure}[t!]
\centering
\includegraphics[width=0.7\columnwidth]{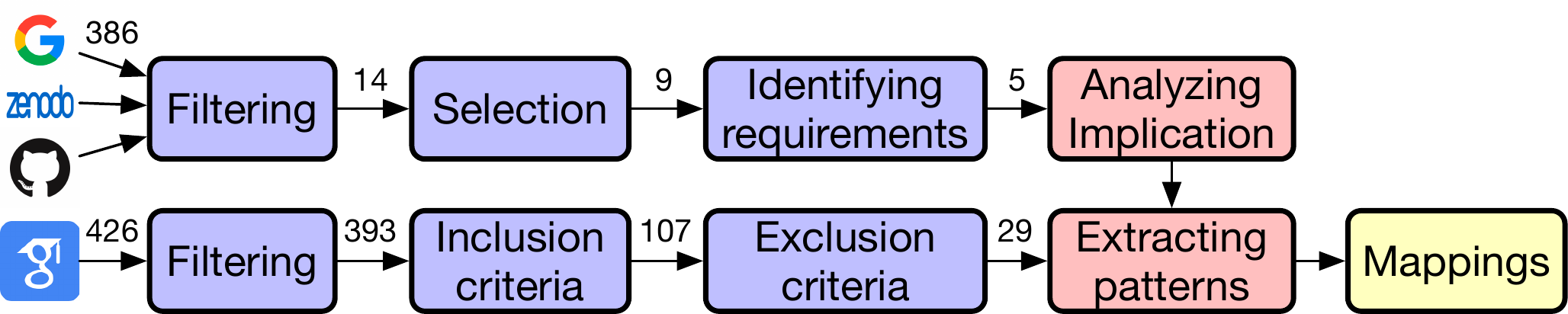}
\caption{Overview of dataset analysis and literature review.}
\label{fig:req}
\end{figure}

To understand what are the common real-world performance requirements with aspirations and their implications in the industry, in Jan 2021, we mined the publicly available requirement dataset from Zenodo (under the \textit{Empirical Software Engineering} label), GitHub, and the Google Dataset Search, using a keyword ``requirement dataset'', as shown in Figure~\ref{fig:req}. The results led to 386 items, including duplication and many irrelevant ones which can be easily identified from their titles. As such, we filtered the candidates down to 14, within which we followed the criteria below to extract the most relevant ones for this study:

%However, this includes duplication and it is easy to know from the titles that most of them are irrelevant to our purpose. 

\begin{itemize}
    \item The dataset has clearly documented requirement statements for the software systems to be built. 
    \item The dataset contains labeled requirements for performance objectives or there is readily available code to do so.
    %\item The dataset contains requirements for performance attributes.
    %\item To meet our goal of multi-objective software configuration tuning, the dataset should contain at least two different performance attributes.
    \item To ensure external validity, the dataset contains performance requirements for systems from different domains.
\end{itemize}

The process has resulted in nine shortlisted datasets, based on which we attempted to identify the statements of performance requirements according to the following rules:

\begin{itemize}
    \item The performance requirement should contain a quantifiable aspiration level, such as ``\texttt{the system shall perform with 1500 users at a time}''. In contrast, ``\texttt{the system shall be fast}'' is too vague to be quantified.
    
    %\item Long and complex requirements with more than one aspiration level are broken down, such that we ensure each decomposed requirement has one related aspiration level. 
    
    %For example, ``\texttt{ReqView Desktop shall start in less than 10s, but the very first time it shall start in less than 30s.}'' would be considered as ``\texttt{ReqView shall start in less than 10s}'' and ``\texttt{the very first time ReqView shall start in less than 30s}''. This is because one aspiration level is the simplest (and most common) form as we found from the literature review and we leave the investigation of multiple levels to future work.
  
    \item \rev{To ensure fairness when comparing with the \texttt{PS-w/o}, we eliminate the performance requirements that do not prefer one extreme of the objective, such as ``\texttt{the display shall be refreshed every 60 seconds}''. This is because such requirements prefer the performance to reach a clear aspiration (e.g., 60 seconds) instead of a maximum/minimum of the performance objective. Therefore, in such a case, \texttt{PS-w} should always be preferred, since there is no point to use \texttt{PS-w/o} which naturally maximizes/minimizes the objectives while does not take aspiration into account\footnote{\rev{Note that, indeed, in some cases, the preference of this kind of requirement can be derived by inferring from the context. Using the same example, if the display could not be refreshed because some long-running analyses could not be terminated within 60 seconds, then the preference would be to guarantee the ability to refresh every 60-sec or less. However, in our cases, 
most of those requirements come from the PROMISE dataset, 
which has no extra information other than some sentences describing the requirement. 
This makes it difficult for us to correctly infer the preferences implied. 
Hence, in the above example, 
we stick with the literal meaning that one would prefer and only prefer a refresh rate of 60 seconds; 
no more and no less.}}.}
    
    %completely contradict the intention of approximating Pareto front in Pareto search, despite having a clear aspiration level. 
\end{itemize}

The above has led us to rule out four datasets that contain no appropriate requirements. Table~\ref{tb:reqs} shows details of the final five datasets used in our study (removing duplication). 

%Note that the identified requirements are generic to the software systems, which can be assured by improving various aspects, including the software configuration. Therefore, with the extracted implications in mind, we also conducted a systematic literature search, aiming to verify whether and how the implications of the requirements with aspirations have been quantified when tuning software configuration in academia.

\input{tables/req}

\subsection{Literature Search of Patterns}

As from Figure~\ref{fig:req}, we also conducted a literature search according to the best practice of a systematic literature review in software engineering~\cite{DBLP:journals/infsof/KitchenhamBBTBL09}, containing search protocol, inclusion, and exclusion criteria. Our goal is to understand a single question: how are the implications of real-world performance requirements with aspirations, which are generic to the software systems as identified from Section~\ref{sec:real-req}, have been specifically quantified in current bi-objective software configuration tuning work? Note that we do not intend to be comprehensive, but rather to gather representatives.

\rev{In Feb 2021, we conducted a full-text search over Google Scholar for papers published since 2010 from the software engineering community (we exclude the system-related papers for better representation in the community), using a focused search string below:}

%\begin{tcolorbox}[left=2pt,right=2pt,top=1pt,bottom=1pt]
\begin{displayquote} 
\textit{``requirement'' AND (``multi objective'' OR ``multi goal'' OR ``multi criteria'') AND (``performance'' OR ``non-functional'') AND (``configurable software'' OR ``adaptive software'') AND (``tuning'' OR ``optimization'')}
\end{displayquote}
%\end{tcolorbox}

This gives us 426 papers. We then filtered patents, inaccessible papers, and any non-English documents, leading to 393 papers. Next, we further extracted the papers by using the following inclusion criteria on the title and abstract:

	\begin{itemize}
	  \item The paper is relevant to tuning the configuration of the software system.
	  \item The paper seeks to improve or evaluate the performance objectives of the software system.
	  \item The paper considers performance requirements.
	  \item The paper is peer-reviewed and is not a survey or tutorial.
      %\item The paper is published in a peer-reviewed public venue.
	\end{itemize}

A paper was ruled out if it does not meet all the above criteria, which resulted in 107 papers. Then, we removed papers based on the following exclusion criteria by reviewing the content:

	\begin{itemize}
	  \item The considered performance requirements do not have a clear aspiration level.
	  \item The paper tackle only a single performance objective.
	  \item The paper does not have quantitatively experimental results with clear instructions on how the results were obtained.
	\end{itemize}

A paper was ruled out if it met any of the above criteria. Finally, we obtained 29 papers, as shown in Table~\ref{tb:papers}.

%Our next stage is to extract the data about the patterns of requirement quantification according to the implications identified in Section, which we will elaborate in the next section.

\input{tables/papers}

\subsection{Results Analysis}

\subsubsection{Number of Performance Objectives}

From the review, we found that 26 out of 29 (90\%) of the papers considered two performance objectives in their tuning process. The remaining three papers take into account three or more. This is a clear sign that two performance objectives remain a state-of-the-art setting for tuning software configuration, which is consistent with the finding from the recent survey for a related field~\cite{DBLP:journals/corr/abs-2001-08236}. Therefore, in this work, we focus on bi-objective software configuration tuning.

\subsubsection{Implications}

We analyzed all 151 performance requirements with aspirations from Section~\ref{sec:real-req}, and found three possible implications on the aspiration space for a given performance objective:

\begin{itemize}
%\textbf{Anything superior to the aspiration level is equally preferred.}
    \item $\mathbf{\mathcal{I}_1}$: \textbf{Anything in the aspiration space is equally preferred.} This gives a clear upper aspiration bound without other information, e.g., ``\texttt{the server will support a maximum of 1,000 simultaneous users}''; or there is a lower aspiration bound but clear information has been given for the cases when the performance reaches the aspiration space, e.g., ``\texttt{results shall be returned in under 15 seconds}''.
    
    %\textbf{Anything inferior to the aspiration level is equally non-preferred.}
    
    \item $\mathbf{\mathcal{I}_2}$: \textbf{Anything not in the aspiration space is equally non-preferred.} For example, ``\texttt{the system shall allow for a minimum of 6 users at the same time}'', in which case there is only information for a clear lower aspiration bound.
    
    \item $\mathbf{\mathcal{I}_3}$: \textbf{No information is available with respect to the aspiration space.} This often refers to the requirements where there is a clear aspiration level, but no indication about whether it is an upper or lower aspiration bound while any other information is unavailable. For example ``\texttt{the system shall cater to 10 simultaneous users}''.
\end{itemize}

\begin{figure}[!t]
  \centering
  \hspace{-0.5cm}
   \begin{subfigure}[t]{0.31\columnwidth}
        \centering
%\includestandalone[width=\columnwidth]{figures/imp-bar}
\includegraphics[width=0.8\columnwidth]{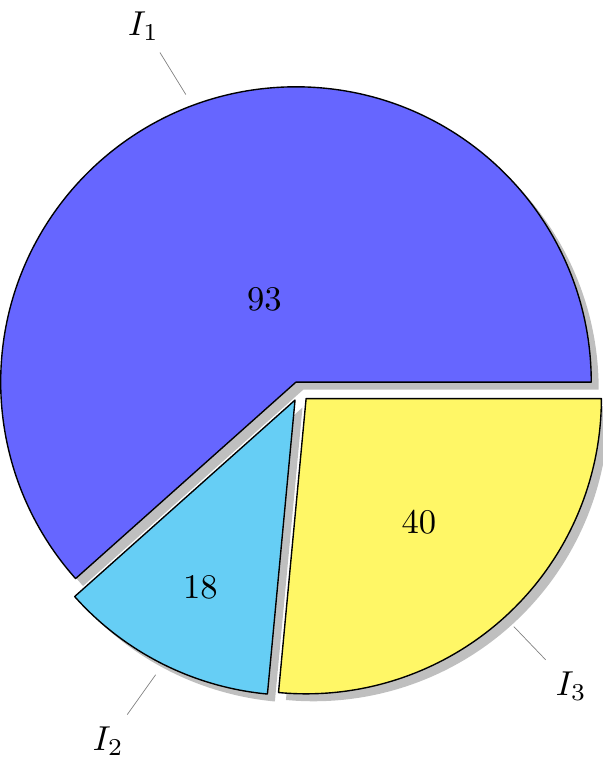}
 %\vspace{-0.1cm}
    \subcaption{$\#$ requirements per implication}
 \label{fig:imp}
   \end{subfigure}
  ~\hspace{0.0cm}
     \begin{subfigure}[t]{0.335\columnwidth}
     \vspace{-4.4cm}
     \centering
%\includestandalone[width=\columnwidth]{figures/pattern-bar}
\includegraphics[width=0.8\columnwidth]{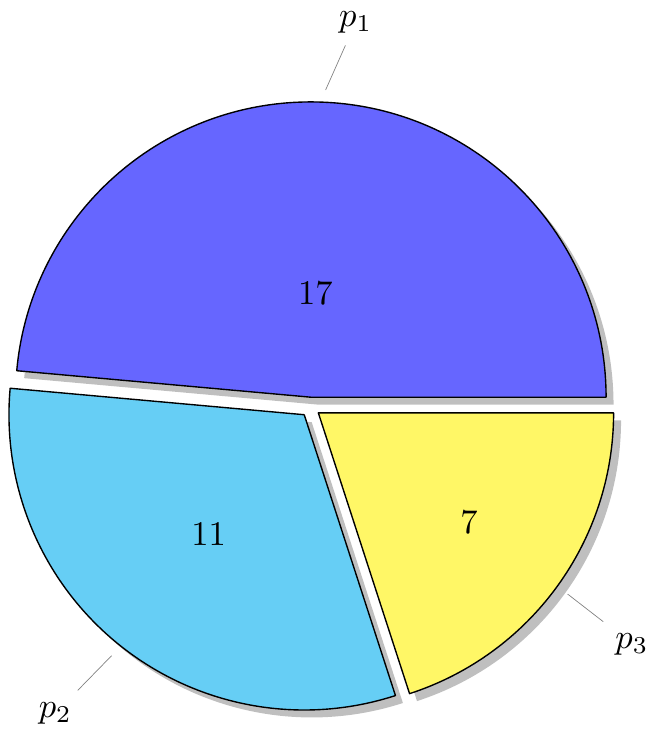}
 \vspace{0.25cm}
    \subcaption{$\#$ papers per pattern}
 \label{fig:pat}
   \end{subfigure}
  ~\hspace{0.0cm}
  \begin{subfigure}[t]{0.3\columnwidth}
  \vspace{-3.1cm}

\input{tables/req-mapping}
 \vspace{1.3cm}
\subcaption{Mappings}
 \label{fig:mappings}
    \end{subfigure}
    \caption{Distribution of implications, patterns and their mappings (six papers consider more than one pattern).}
     
  \end{figure}

The distribution of the implications can be found in Figure~\ref{fig:imp} and we achieve a Kappa coefficient $\kappa=0.85$ for this.

\begin{figure}[t!]
%\vspace{0.2cm}
\centering
\includegraphics[width=\columnwidth]{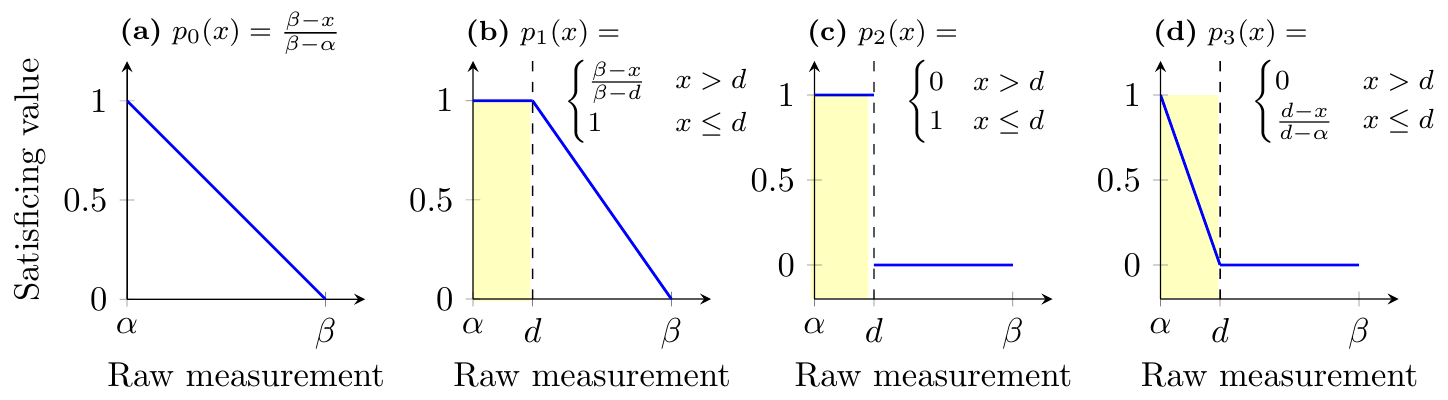}
\caption{Requirement patterns with (and without) aspiration from the literature. $\alpha$ and $\beta$ denote the lower and upper bound of the performance objective, respectively. $d$ is the aspiration level and the aspiration space has been shaded.}
\label{fig:fun}
\end{figure}

\subsubsection{Patterns}

Next, with the above implications in mind, we seek to understand how they are quantified within the 29 papers identified. This led to three state-of-the-art patterns on the functions to quantify requirements with aspiration level (assuming the lower bound is optimum). Suppose that $\alpha$ and $\beta$ denote the lower and upper bound of the performance objective, respectively; $d$ is the aspiration level, the patterns, and their quantification have been shown in Figure~\ref{fig:fun} and are explained below: 

\begin{itemize}
    \item $\vect{p_1}$: The performance in the aspiration space is equally good or otherwise there is a certain degree of tolerance (Figure~\ref{fig:fun}b). The function can be formulated as:
    \begin{align}
    \vect{p_1}(x) =
    \begin{cases}
        {{\beta - x} \over {\beta - d}} &x> d\\
        1 &x \leq d
    \end{cases}
\end{align}
    
    \item $\vect{p_2}$: The performance in the aspiration space is equally good while anything outside the space is unacceptable (Figure~\ref{fig:fun}c), such that:
       \begin{align}
    \vect{p_2}(x) =
   \begin{cases}
        0 &x> d\\
        1 &x \leq d
        \end{cases}
\end{align}

    \item $\vect{p_3}$: Preferring the optimal performance while anything outside the aspiration space is unacceptable (Figure~\ref{fig:fun}d), which is defined as:
       \begin{align}
    \vect{p_3}(x) =
   \begin{cases}
        0 &x> d\\
        {{d - x} \over {d - \alpha}} &x \leq d
        \end{cases}
\end{align}
\end{itemize}

Similarly, we can also formalize the requirement with no clear aspiration level involved (e.g., ``\texttt{the latency shall be small}''), denoted as $\vect{p_0}$, which is illustrated in Figure~\ref{fig:fun}a and can be formulated as follow: 
   \begin{align}
    \vect{p_0}(x) = {{\beta - x} \over {\beta - \alpha}}
\end{align}
The distribution of the patterns has been shown in Figure~\ref{fig:pat} where we have $\kappa = 0.76$, which is sustainable~\cite{mchugh2012interrater}.

Through normalization in those patterns, the raw measurement of a performance objective is transformed into the satisficing degree with respect to a given aspiration space (if any), ranging between 0 and 1 where the latter means fully satisfied. As such, the transformation depends on the assumption of satisficing over measurements included or excluded by the aspiration space, which distinguishes the patterns. At this point, we can immediately see the mappings between the patterns and the extracted implications from the real-world dataset. Such mappings have been illustrated in Figure~\ref{fig:mappings}, from which we see that each pattern, except $\vect{p_0}$, can fit with at least two implications from the real-world requirements. For example, $\vect{p_1}$ can fit with $\mathcal{I}_1$ and $\mathcal{I}_3$, because the former prefers anything within the aspiration space and specifies nothing on the other extreme, while the latter has no information at all and thus one needs to rely on an assumption when quantifying $\mathcal{I}_3$ to guide the search, meaning that it has the possibility to fit with all the three patterns.

%could be valid under different assumptions. Since no information is given for $\mathcal{I}_3$, one needs to make different assumptions when using it to guide the search, and thereby it fits with all three patterns.

%due to their imprecision nature

%Although the patterns aim for a single performance attribute, they can be arbitrarily combined (together with $\vect{p_0}$) under multiple attributes. 

%In fact, it is straightforward to apply these patterns as objectives to steer the Pareto search for multi-objective software configuration tuning, as we will elaborate in Section.

%Clearly, the simplest way for a Pareto search to consider requirements with aspiration is to apply them as the objectives during the search,

From the above, it is confirmed that there exist patterns from current work which can reflect the implication of real-world performance requirements and their aspirations. We, therefore, will seek to examine all of them in our empirical study.

\begin{figure}[t!]
%\vspace{0.2cm}
\centering
\includegraphics[width=\columnwidth]{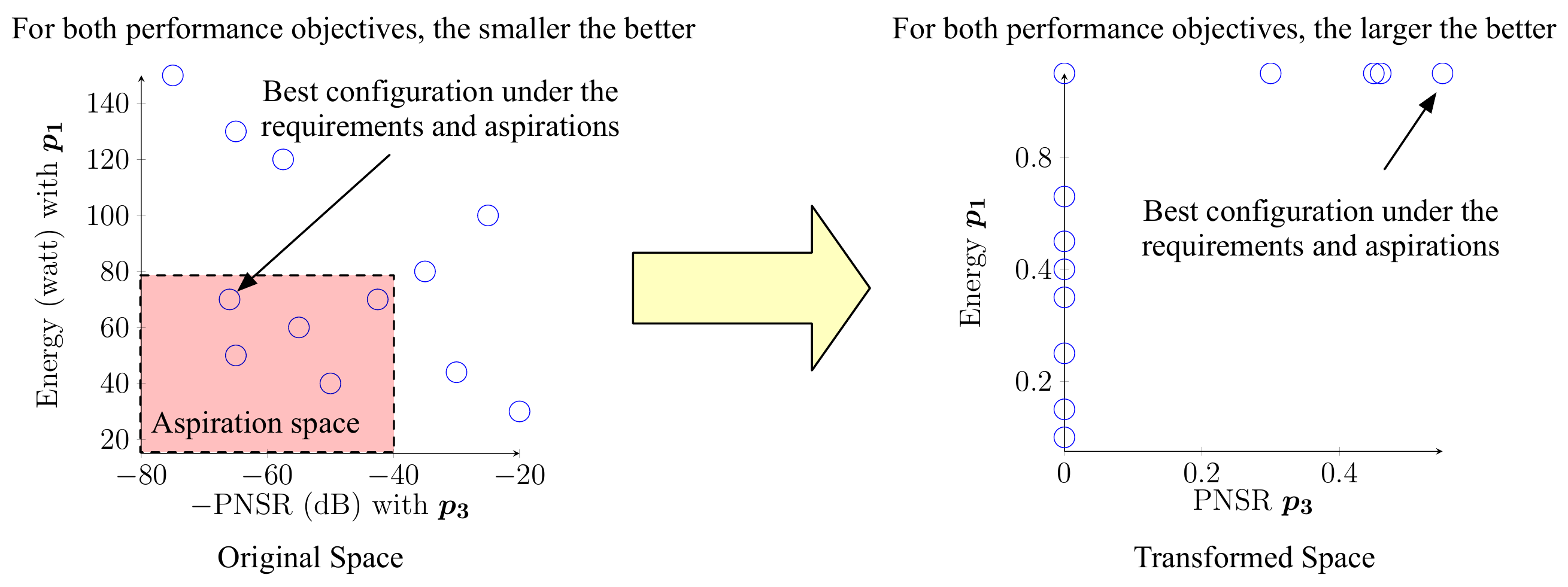}
\caption{Example of using the patterns for evaluating the goodness of configurations and guiding \texttt{PS-w} in the transformed space.}
\label{fig:quan}
\end{figure}

\subsection{Respecting Requirements and Aspiration in Software Configuration Tuning}

While the above requirement patterns are the key to evaluating the ``better'' or ``worse'' in the set of configurations produced by any Pareto optimizer and optimization model, they directly influence the behavior of \texttt{PS-w} (they correspond to the $p_n$ in Equation (1)) but not that of the \texttt{PS-w/o}.

%The above requirement patterns directly influence the behavior of \texttt{PS-w} (they correspond to the $p_n$ in Equation (1)) but not that of the \texttt{PS-w/o}. However, 

Most importantly, those patterns allow us to precisely quantify what is the best configuration(s) amongst the configurations produced by those two optimization models given a set of requirements and aspirations. Figure~\ref{fig:quan} shows an example of evaluating the configurations (and guiding \texttt{PS-w}) in a transformed space when taking the requirements and aspirations into account, i.e., energy usage with $\vect{p_1}$ and aspiration of 80 watts while PNSR with $\vect{p_3}$ and aspiration of 40dB. Here, we certainly prefer the points within the aspiration space in contrast to those outside. However, for those points within the aspiration space, we only prefer those with better PNSR while the energy usage is deemed as equivalent (due to the implication of $\vect{p_1}$ and  $\vect{p_3}$). 

\rev{The above is difficult to assess and quantify in the original space (Figure~\ref{fig:quan} left), since naturally the points that are non-dominated by each other (in the sense of the original objective values) are considered as equivalent when the requirements and aspiration are not involved. Therefore, the actual most preferred point (arrow highlighted) is not considered the best. In contrast, the evaluation becomes immediately obvious on what is the best point in the transformed space, where the energy and PNSR are converted by the equations for $\vect{p_1}$ and $\vect{p_3}$, respectively. Now, clearly, the most preferred point is the only non-dominated point therein (Figure~\ref{fig:quan} right).}

%% file: tables/req.tex
\begin{table}[t!]
%\vspace{0.05cm}
\caption{Performance requirements with aspirations.}
\label{tb:reqs}
\footnotesize
\centering
\begin{tabular}{lcl}\toprule

\textbf{Dataset}&\textbf{\# Requirements}&\textbf{Link}\\
\midrule

Do et al.~\cite{DBLP:conf/icsr/DoCB19}&52&\url{https://github.com/aqd14/ICSR-2019}\\
\rowcolor{steel!10}PROMISE~\cite{menzies2012promise}&48&\url{https://zenodo.org/record/268542}\\
PURE~\cite{DBLP:conf/re/FerrariSG17}&28&\url{https://zenodo.org/record/1414117}\\
\rowcolor{steel!10}Shaukat et al.~\cite{DBLP:conf/cse/ShaukatNZ18}&13&\url{https://zenodo.org/record/1209601}\\
Dalpiaz et al.~\cite{DBLP:conf/re/DalpiazDAC19}&10&\url{https://zeno do.org/record/3309669}\\

\bottomrule
\end{tabular}

\end{table}

%% file: tables/papers.tex
\begin{table}[t!]
\caption{Identified papers with aspiration quantification.}
\label{tb:papers}
\footnotesize
\centering
\begin{tabular}{lc|lc|lc}\toprule

\textbf{Venue}&\textbf{\# Papers}&\textbf{Venue}&\textbf{\# Papers}&\textbf{Venue}&\textbf{\# Papers}\\
\midrule

TSE (journal)&3&JSS (journal)&6&TAAS (journal)&3\\
\rowcolor{steel!10}ASE (journal)&2&ESE (journal)&1&ICPE (conference)&1\\
ICSE (conference)&1&FSE (conference)&3&ASE (conference)&1\\
\rowcolor{steel!10}SEAMS (symposium)&6&ICSA (conference)&1&MODELS (conference)&1\\

\bottomrule
\end{tabular}

\end{table}

%% file: tables/req-mapping.tex
\setlength{\tabcolsep}{1mm}
\footnotesize
\centering
\begin{tabular}{ll}\toprule

\textbf{Pattern}&\textbf{Implication}\\
\midrule
$\vect{p}_1$&$\mathcal{I}_1$, $\mathcal{I}_3$\\
\rowcolor{steel!10}$\vect{p}_2$&$\mathcal{I}_1$, $\mathcal{I}_2$, $\mathcal{I}_3$\\
$\vect{p}_3$&$\mathcal{I}_2$, $\mathcal{I}_3$\\

\bottomrule
\end{tabular}

%% file: study.tex
\section{Empirical Study Design}
\label{sec:emp}

As shown in Figure~\ref{fig:empirical}, our methodology consists of the following steps:

\begin{enumerate}[label=\textbf{Step~\arabic*:},leftmargin=1.5cm]

\item Assume that a requirement scenario has been negotiated by the software engineers and stakeholders, we quantify the requirements such that they are ready for the Pareto optimizers, i.e., in the forms of a combination of patterns from Section~\ref{sec:req} and their aspiration space. In particular, to form a requirement scenario, the given combination of the patterns is denoted as a two-dimensional vector $\mathbfcal{P}$, such that there is at least one that comes with a clear aspiration level, e.g., $\mathbfcal{P}=\{\vect{p}_0, \vect{p}_3\}$. In this work, we examine all possible combinations of the patterns (including $\vect{p_0}$). Under each combination, we also consider different aspiration spaces for our RQs; this will be further elaborated in Section~\ref{sec:aspiration}.

\begin{figure*}[t!]
\centering
\includegraphics[width=\textwidth]{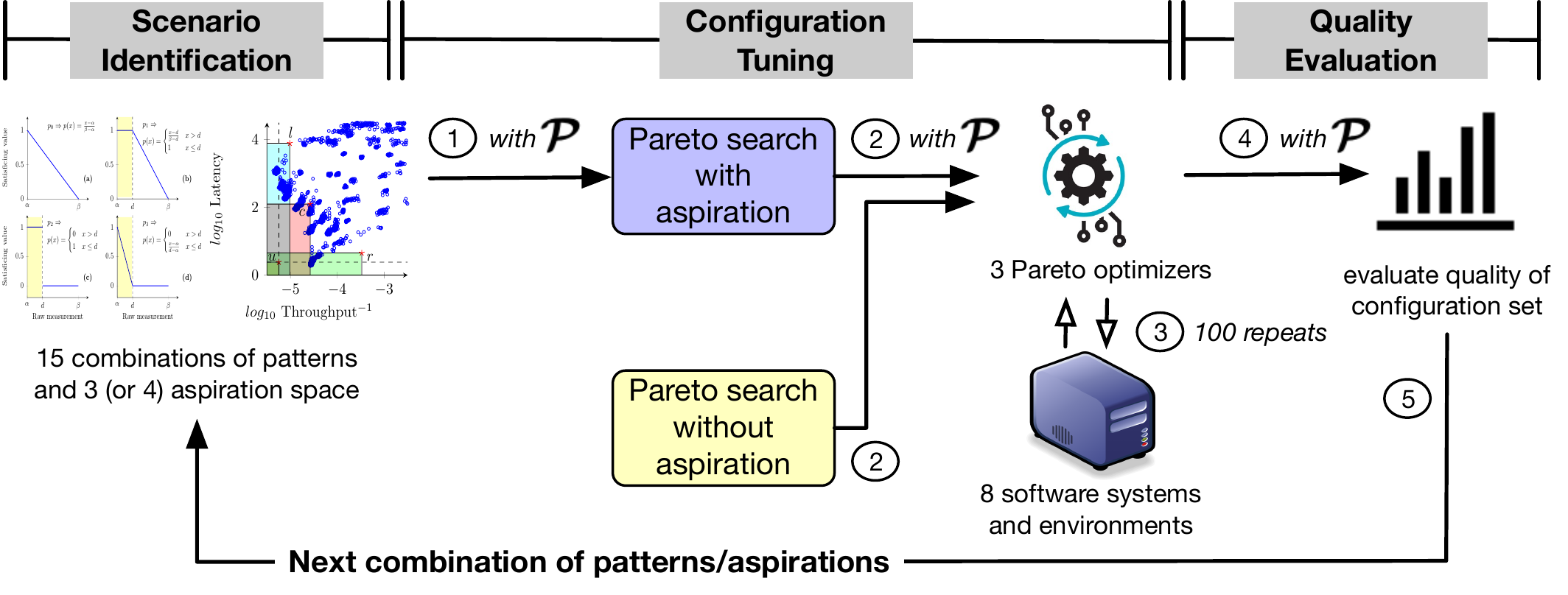}
\caption{Overview of the empirical study.}
\label{fig:empirical}
\end{figure*}

\item Run both \texttt{PS-w/o} and \texttt{PS-w} on different software systems. Particularly, when formulating the performance objectives, \texttt{PS-w/o} is steered by the raw measurements only\footnote{This is effectively identical to using $\vect{p_0}$ for all performance objectives.} while \texttt{PS-w} is designed to be guided by the given vector of patterns $\mathbfcal{P}$ as the new objectives. To ensure fairness, both optimization models are examined under the same optimizer and we consider three representative optimizers in this work, i.e., NSGA-II~\cite{Deb2002}, IBEA~\cite{DBLP:conf/ppsn/ZitzlerK04}, and MOEA/D~\cite{DBLP:journals/tec/ZhangL07}.

%We set $\mathbfcal{P}'=\mathbfcal{P}$ for \textbf{RQ1} and  \textbf{RQ2} while it can be of any combination of patterns for \textbf{RQ3}.

\item Measure the system as the search proceeds until the tuning budget has been exhausted; repeat 100 times.

\item Evaluate the set of configurations thereafter using $\mathbfcal{P}$ as part of the \textit{Quality Evaluation} phase.

\item Go back to Step 1 if there are more combinations of patterns and aspirations to examine.

\end{enumerate}

%assuming that a requirement scenario has been given by the performance engineer, i.e., a target combination of patterns from Section~\ref{sec:req} and their aspiration space (at step 1), we run both \texttt{PS-w/o} and \texttt{PS-w} on different software systems. Particularly, when formulating the performance attributes as objectives, \texttt{PS-w/o} is steered by the raw measurements only\footnote{This is effectively identical to using $\vect{p_0}$ as the objective for all attributes.} while \texttt{PS-w} is designed to be guided by the target patterns and their aspiration space. The produced set of configurations thereafter is evaluated in step 4. 

\input{tables/patterns}

It is worth noting that, although the patterns from Section~\ref{sec:req} are for single performance objective, they can be arbitrarily combined for the bi-objective software configuration tuning in the \textit{Scenario Identification} phase of Step 1~\cite{Calinescu2017Designing,DBLP:conf/wosp/MartensKBR10,Gerasimou2016Search,DBLP:journals/jss/CalinescuCGKP18}, as illustrated in Table~\ref{tb:com-pattern}.

In the \textit{Configuration Tuning} phase (Step 2 and 3), the patterns require normalization using the lower and/or upper bound (except for $\vect{p_0}$ and $\vect{p_2}$). However, since these are often unknown, we adopt a dynamic method wherein the raw measurements are normalized using the maximal and minimal values found so far as the tuning proceeds, which is common in SBSE for software configuration tuning~\cite{DBLP:conf/sigsoft/ShahbazianKBM20,DBLP:conf/ssbse/BowersFC18}. We record the raw measurements of each configuration throughout the tuning to efficiently utilize the tuning budget (Section~\ref{sec:budget}).

To mitigate stochastic bias, we repeat each experiment 100 runs. The study is conducted on a cluster of machines each with Intel i5 six cores CPU at 2.9GHz and 8GB memory, running numerous experiments in parallel over the course of five months ($24 \times 7$).

\subsection{Subject Software Systems}

\label{sec:sys}

We conduct our study on a set of real-world highly configurable software systems and environments that have been widely studied in existing work~\cite{DBLP:conf/sigsoft/JamshidiVKS18,nair2018finding,DBLP:conf/mascots/JamshidiC16,DBLP:conf/sigsoft/0001Chen}. These are selected according to the criteria below:

\begin{enumerate}

\item To ensure that the search landscape is not too trivial to be explored, the system should contain a mix of binary and enumerative configuration options.

\item A full exploration of the search space is infeasible, i.e., it cannot be done within 24 hours.

\item There are clear instructions on how to set up the benchmark under which the system will be measured.

\item If the same system of an environment has been used with a different set of configuration options, choose those with relatively higher complexity, i.e., larger search space and more configuration options. For example, \textsc{Storm} can be tuned under different workload benchmarks, and we choose \textsc{WordCount} and \textsc{RollingSort} as the two that satisfy the above criteria.

\end{enumerate}

We firstly eliminated \textsc{LLVM} from~\cite{nair2018finding}, as it violates Criterion (1). Similarly,  \textsc{sort-256} and \textsc{noc-CM-log} is also ruled out due to their rather small search space which can be exhaustively explored in 24 hours, i.e., Criterion (2). We cannot consider the system \textsc{SaC} as there is no clear instruction on under what benchmark it can be profiled, which violates Criterion (3). We also noticed that \textsc{Storm} and \textsc{Keras} (with DNN or LSTM) have been much more commonly used than others, but with different configuration options and environments. Therefore, according to Criterion (4), we use the settings that lead to a much larger search space and more options. As shown in Table~\ref{tb:sys}, the selected software systems come from diverse domains, e.g., video encoding, stream processing, and deep/machine learning, while having different performance objectives, scales, and search spaces. Their measurements are also expensive\footnote{Each measurement consists of 5 repeated samples and the median value is used.}, e.g., \textsc{XGBoost} needs 2,807 hours to explore less than 1\% of its search space. 

%In this work, we examine the two performance attributes for each software system as used in prior work~\cite{nair2018finding,Chen2018FEMOSAA}.

%Some of these involve a large space of configurations (e.g., \textsc{XGBoost}) while others are more tractable (e.g., \textsc{Storm/WC}). 

%We preprocessed the configuration options and their ranges for different environments based on domain understanding from the literature~\cite{DBLP:conf/sigsoft/JamshidiVKS18,nair2018finding,DBLP:conf/mascots/JamshidiC16} and the possible constraints, thereby only the most relevant ones and their ranges are kept. 

We keep the same performance objectives, configuration options, and their ranges as studied in the prior work that made use of them, e.g.,~\cite{DBLP:conf/sigsoft/JamshidiVKS18,nair2018finding,DBLP:conf/mascots/JamshidiC16,DBLP:conf/sigsoft/0001Chen}, since those have been shown
to be the key ones for the software systems under the related environment. As a result, although the software systems are the same, the actual search spaces are different, such as \textsc{Storm/WC} and \textsc{Storm/RS}. \rev{In particular, following what has been used in previous work, the environment/workload we consider are:}

\input{tables/sys-new}

\begin{itemize}
    \item \rev{\textbf{\textsc{Trimesh}:} we use the Shapenet dataset that contains 51,300 unique 3D models. In this work, we randomly sample 100 models as the standard benchmark.}
    
    \item \rev{\textbf{\textsc{x264}:} for this, the benchmark used is a standard video of 1GB size, which was chosen randomly.}
    
    \item \rev{\textbf{\textsc{Storm/WC}:} we use the \textsc{WordCount} as the benchmark. This is a typical simple streaming example where \textsc{Storm} is used to keep track of the words and their counts streaming in. \textsc{WordCount} generates a CPU-intensive workload.}
    
    \item \rev{\textbf{\textsc{Storm/RS}:} similar to \textsc{Storm/WC}, here we use the \textsc{RollingSort} as the benchmark. Unlike \textsc{WordCount}, \textsc{RollingSort} generates a memory intensive workload.}
    
    \item \rev{\textbf{\textsc{Keras/Adiac}:} we use the Deep Neural Network (DNN) from the \textsc{Keras} software and run it on the \textsc{Adiac} dataset. Generally, the dataset contains a task of automatic identification of diatoms (unicellular algae) among 31 classes with a training and testing size of 390 and 391, respectively.}
    
    \item \rev{\textbf{\textsc{Keras/DSR}:} we use the DNN from the \textsc{Keras} software and run it on the \textsc{DiatomSizeReduction} dataset. The dataset concerns the prediction of four types of diatoms with a training and testing size of 16 and 306, respectively.}
    
    \item \rev{\textbf{\textsc{Keras/SA}:} we use the DNN from the \textsc{Keras} software and run it on the \textsc{ShapesAll} dataset. Generally, the dataset aims to test contour/image and skeleton-based descriptors; there are 60 classes with a training and testing size of 600 each.}
    
     \item \rev{\textbf{\textsc{XGBoost}:} we use the \textsc{Covertype} dataset that contains 54 forest cover type from cartographic variables only. The size of the dataset is 581,012 and we follow a 70\%-30\% training and testing split.}
\end{itemize}

Indeed, the analyzed dataset and literature in Section~\ref{sec:req} may not specifically target the software systems considered in this work. However, the extracted implication and patterns are rather generic such that they can be applied to different cases. Further, some widely studied performance objectives (from both the dataset and literature) are overwhelmingly applicable. For example, latency- and throughput-related requirements (with different aspiration levels) are prevalent for a wide range of software~\cite{nair2018finding}. 

%The actual aspiration level/space may differ though.

% talk about the implication and patterns are generic

%\subsection{Requirements Combination and Aspiration Area}

\input{tables/aspir}

%\caption{Examples of aspiration space (shaded by different colors) in the log-transformed objective space for \textsc{Storm/RS}.}

\subsection{Aspiration Space}
\label{sec:aspiration}

% talk about changing aspiration level into area, and transfer the problem's objective space

% talk about why we need to record the original objective value for pareto search with aspiration

%\subsection{Aspiration Area}

To improve external validity, we consider aspiration levels that draw two types of aspiration space under two performance objectives: realistic and unrealistic ones. To that end, for each software system, we run all the Pareto optimizers for three hours each to obtain a landscape that contains an approximated Pareto front. We do so by ensuring that the obtained front is reasonably converged, i.e., increasing the budget only marginally changes the results. We then set the aspiration space based on such a front as summarized in Table~\ref{tb:aspir-levels}.

%Note that the aspiration does not apply to an attribute associated with $\vect{p_0}$.

\subsubsection{Realistic Aspiration Space}

For software configuration tuning with two performance objectives, we say an aspiration space is realistic if there is at least one configuration that can reach the aspiration levels of both performance objectives.

Using \textsc{Storm/RS} as an example in Figure~\ref{fig:aspir}, for the realistic ones under each combination of patterns, we set three aspiration space based on their positions in the objective space: left-shifted ($l$), right-shifted ($r$) and centered ($c$). In particular, $l$ is defined as using the value of the 20$th$ percentile for throughput and the value of the 80$th$ percentile for latency as their corresponding aspiration levels; similarly, $r$ uses the value of the 20$th$ percentile for latency and the value of the 80$th$ percentile for throughput; finally, $c$ uses the values of the 50$th$ percentile for both performance objectives. Clearly, despite covering diverse regions in the overall space of performance objectives, all those spaces contain at least one point (configuration). Note that the aspiration space is applicable to any combination of patterns with and without $\vect{p_0}$ (in Figure~\ref{fig:aspir}a and Figure~\ref{fig:aspir}b respectively), as long as there is a clear aspiration level for at least one performance objective. 

%\footnote{Those figures may be slightly vary depending on the system, as each of them come with a different Pareto front and we need to satisfy the definition of realistic aspiration space}

%Since $\vect{p_0}$ has no aspiration level, the aspiration does not applied to an attribute associated with $\vect{p_0}$$. For example, in Figure~\ref{fig:aspir}b, the aspiration space only applies for \textit{throughput} (with $\vect{p_1}$) but not \textit{latency}, which has a pattern of $\vect{p_0}$. 

%The how scenario implies that one prefer any possible \textit{throughput} within the aspiration space while as low as possible \textit{latency}.

%\begin{wrapfigure}{r}{0.5\linewidth}
%\centering
%\includegraphics[width=0.5\columnwidth]{figures/aspir.pdf}
%\caption{Aspiration space (highlighted in different colors) within the log-transformed objective space for \textsc{Storm/RS}.}
%\label{fig:aspir}
%\end{wrapfigure}

\subsubsection{Unrealistic Aspiration Space}

\begin{figure}[t!]
\centering
\includegraphics[width=0.8\columnwidth]{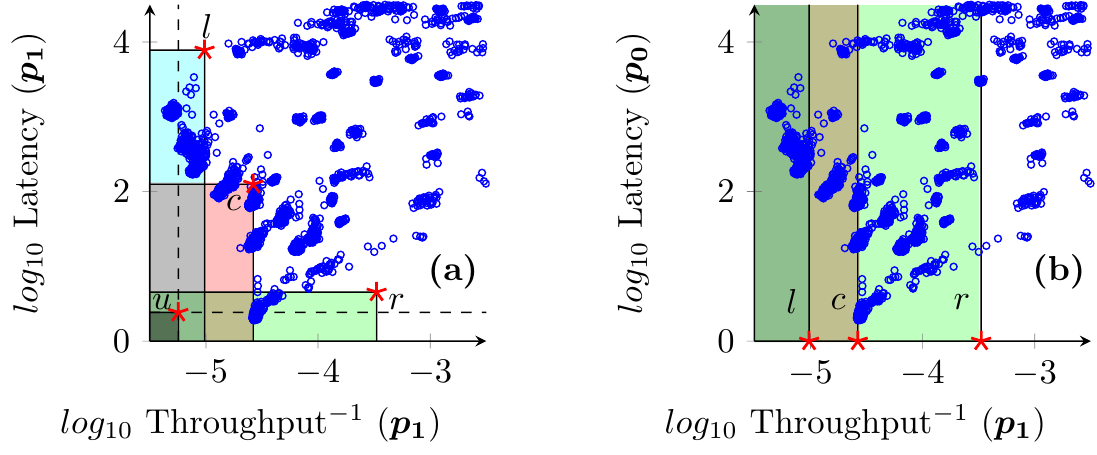}
\caption{Distant aspiration space (shaded by different colors) under different combinations of patterns for \textsc{Storm/RS}.}
\label{fig:aspir}
\end{figure}

Since the aspiration level/space is negotiated beforehand, it may be unrealistic. In this work, 
we refer to an unrealistic aspiration space as the situation 
wherein the aspiration levels of two performance objectives can be at most reached one at a time, but not both simultaneously. 
For example, in the case of two performance objectives from Figure~\ref{fig:aspir}a, $u$ is an unrealistic aspiration space such that the level is achievable for either of the two objectives individually (as indicated by the dashed lines), but not for both, as there is no point (configuration) residing in the space. As a result, it is not applicable when only one performance objective contains clear aspiration, e.g., in Figure~\ref{fig:aspir}b. To define such a space, we set the value of 5$th$ percentile of both performance objectives as the corresponding aspiration levels, which we have found as sufficient to create an unrealistic aspiration space for each system.

% We omit the cases where an aspiration space fails outside the bounds of all performance attributes with aspirations, as they are easy to detect and the impact on search is predictable. In contrast, the Pareto-unrealistic aspiration space is more difficult to notice and presume its implication to the search, rendering such as a highly probable case in the practice.

%it would simply weaken the search by ranking all configurations as equal or the pattern would be reduced to a similar form of $\vect{p_0}$

%As can be seen in Figure~\ref{fig:aspir}, $u$ represents one of those aspiration spaces that has no intersection with the approximated Pareto front, while still within the bounds of each individual dimension of the performance attributes (as can be seen from the dashed lines).

%As a result, only those combinations of patterns that do not involve $\vect{p_0}$ would be examined under an unrealistic aspiration space.

\subsection{Tuning Settings}

\subsubsection{Pareto Optimizer}

We consider three Pareto optimizers, i.e., NSGA-II~\cite{Deb2002}, IBEA~\cite{DBLP:conf/ppsn/ZitzlerK04}, and MOEA/D~\cite{DBLP:journals/tec/ZhangL07}, because:

\begin{itemize}
    \item They have been widely used for software configuration tuning in prior work~\cite{DBLP:conf/wosp/SinghBSH16,Calinescu2017Designing,DBLP:conf/wosp/MartensKBR10,Gerasimou2016Search,DBLP:journals/jss/CalinescuCGKP18,Chen2018FEMOSAA}.
    \item They are the representatives of three fundamentally different frameworks for Pareto search~\cite{Emmerich2018tutorial}. 
    %i.e., the dominance-based, indicator-based and decomposition-based.
    \item They can, but do not have to, rely on a surrogate model(s)~\cite{Chen2018FEMOSAA}, which greatly reduces the noise in our empirical study.
\end{itemize}

All the above optimizers are adopted for both \texttt{PS-w} and \texttt{PS-w/o} based on the implementations in jMetal~\cite{DBLP:journals/aes/DurilloN11}.

%\subsection{Settings}

\subsubsection{Tuning Budget}
\label{sec:budget}

%Since the measurement can be rather expensive for tuning software configuration and there is a large number of experiments we need to conduct, 

In this work, we set a budget of one hour for each run as commonly used for expensive SBSE problems~\cite{DBLP:conf/icse/LiX0WT20}. However, directly relying on the time as a termination criterion can suffer severe interference during the tuning as numerous experiments need to be run in parallel. To prevent this, for each software system, we did the following to convert the one-hour tuning budget into the number of unique measurements:

\begin{enumerate}

\item incrementally (100 each step) measuring distinct configurations on a dedicated machine using random sampling until the one-hour time budget is exhausted.

\item repeating the above 5 times and collect the number of measurements.

\item the median of the 5 repeats serves as the key termination criterion of the tuning thereafter (in Table~\ref{tb:settings}).

\end{enumerate}

%at first incrementally (100 each step) measured distinct configurations on a dedicated machine using random sampling until the time budget is exhausted. In this way, we collect the median number of measurements over 5 repeats (in Table~\ref{tb:settings}), which serves as the key termination criterion of the tuning thereafter.

%Since the search budget reflects the amount of measurements taken in one hour time, 

Note that in each run of the tuning, we cached the measurement of every distinct configuration for direct reuse. Hence, only the distinct configurations would consume the budget. 

%To avoid running forever, we set a pragmatic cap of 500 generations but it has never been hit.

%in case the tuning fails to explore any new configurations

\input{tables/alg}

\subsubsection{Parameters}
\label{sec:settings}

For the three optimizers in all cases, we apply the binary tournament, boundary mutation, and uniformed crossover, as used in prior work~\cite{Chen2018FEMOSAA,DBLP:journals/infsof/ChenLY19}. The mutation and crossover rates are set to 0.1 and 0.9, respectively, which also follows the most common setting for software configuration tuning~\cite{Chen2018FEMOSAA,DBLP:journals/infsof/ChenLY19}. Other specific settings for IBEA and MOEA/D are kept as default values, which have been shown to be effective~\cite{Chen2018FEMOSAA}.

%However, we could not find a common setting for the population size. Therefore, 

For each system, we pragmatically set the population size via:

%under the budget from Table~\ref{tb:settings} via the following pragmatic steps:
%Since the population size is often less than 100 [],
\begin{enumerate} % (one run each)
    \item \rev{examining different sizes in pilot runs under the budget in Table~\ref{tb:settings}, i.e., $\{10,20,...,$ $100\}$, over all optimizers, combinations of patterns and aspiration (on both \texttt{PS-w} and \texttt{PS-w/o}).}
    \item \rev{recording the average change rate of population over the last 10\% generations using $g={1\over k} \times \sum^k_{i=0} {c_i \over s}$, where $k$ is the number of the last 10\% generations; $c_i$ denotes the number of different configurations in the \textit{i}th generation compared with those in the \textit{i-1}th generation; $s$ is the population size.}
    \item \rev{the largest population size where $g \leq 0.1$ across all conditions (or 10, if no size satisfies the above constraints) will be used.}
    
    %the change rate over the last 10\% generations is less than 10\% 
\end{enumerate}

\rev{The results are also shown in Table~\ref{tb:settings}. In this way, we seek to reach a balance between convergence (smaller population change) and diversity (larger population size) under the given tuning budget. That is, increasing the budget will unlikely change the result. This has been practiced in~\cite{DBLP:journals/ase/GerasimouCT18,DBLP:conf/sigsoft/0001Chen}.}

\subsection{Analysis and Comparison}

\subsubsection{Metric}
\label{sec:metric}
%In particular, we make calculation on the raw measurements of the configurations.

%\mathbfcal{P} includes both patterns and aspiration space
To make a comparison and determine which optimization model is better in this work, we need to measure the ``best'' with two conditions in mind:

\begin{itemize}

\item \textbf{Condition 1:} The metric needs to be able to comprehensively compare the different sets of configurations as produced by the Pareto optimizers, covering diverse quality aspects, such as convergence and diversity.

\item \textbf{Condition 2:} The metric should be able to reflect the given requirement scenarios, i.e., taking the given patterns identified from Section~\ref{sec:req} into account when conducting the comparisons and evaluations.

\end{itemize}

To that end, we use Hypervolume (HV)~\cite{Zitzler1998,Zitzler2007a} as the basic metric to assess the quality of the configuration set produced in each run. In a nutshell, HV measures the volume between all points of a configuration set and a reference point (usually a nadir point); the larger the volume, the better convergence, and diversity that the set achieves. \rev{HV is chosen in this work because:}

\begin{itemize}

\item \rev{HV is a comprehensive metric that covers all quality aspects of a configuration set, 
i.e., convergence, uniformity, spread, and cardinality~\cite{li2019quality,9252185}, which meets \textbf{Condition 1}.} 

\item \rev{HV also does not require a reference Pareto front and is Pareto compliant\footnote{\rev{Generally speaking, 
	a quality indicator being Pareto compliant means 
	that its evaluation result does not conflict with the Pareto dominance relation between two solution sets. 
	More strictly,  
	if a solution set $A$ is \textit{better}~\cite{Zitzler2003} than $B$ 
	(i.e., for any solution in $B$, 
	there exists one solution in $A$ that covers (dominates or is equivalent to) it, 
	and there exists at least one solution in $A$ that is not covered by any solution in $B$), 
	then $A$ is always evaluated better than $B$ by the indicator.}}, which fits our case as the true Pareto front is unknown.} 

\item \rev{By following the guidelines proposed by \cite{li2018critical,9252185}, 
we landed on HV as the appropriate metric for our SBSE problem.}
\end{itemize}

%Indeed, other quality indicators such as IGD can also cover those aspects. However, according to Li et al.~\cite{9252185}, HV is more suitable in our case as it does not require a true Pareto front and is Pareto compliant. 

\begin{figure}[!t]
  \centering
   \begin{subfigure}[t]{0.53\textwidth}
        \centering
\includegraphics[width=\textwidth]{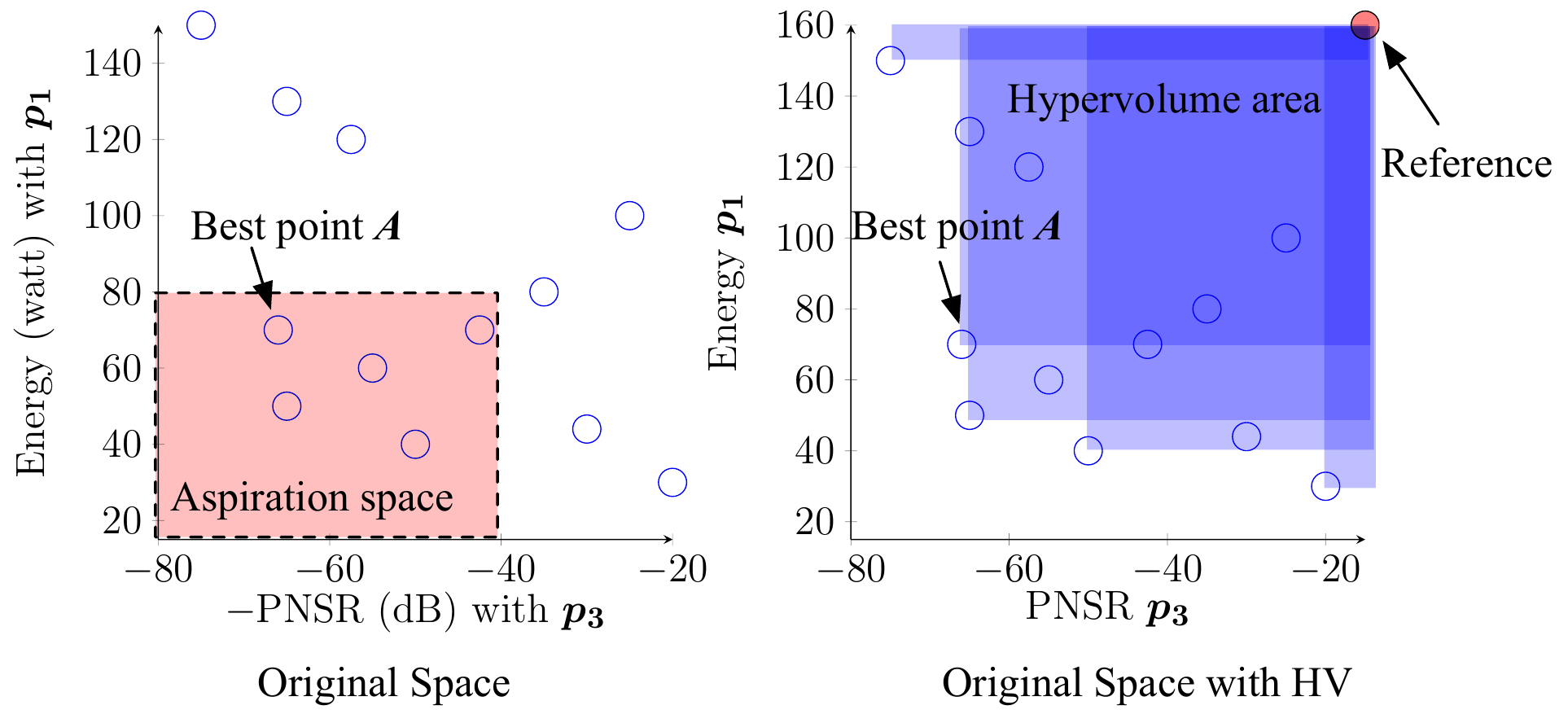}
    \subcaption{Evaluating with HV in the original space}
   \end{subfigure}
 ~
     \begin{subfigure}[t]{0.47\textwidth}
     \centering
\includegraphics[width=\textwidth]{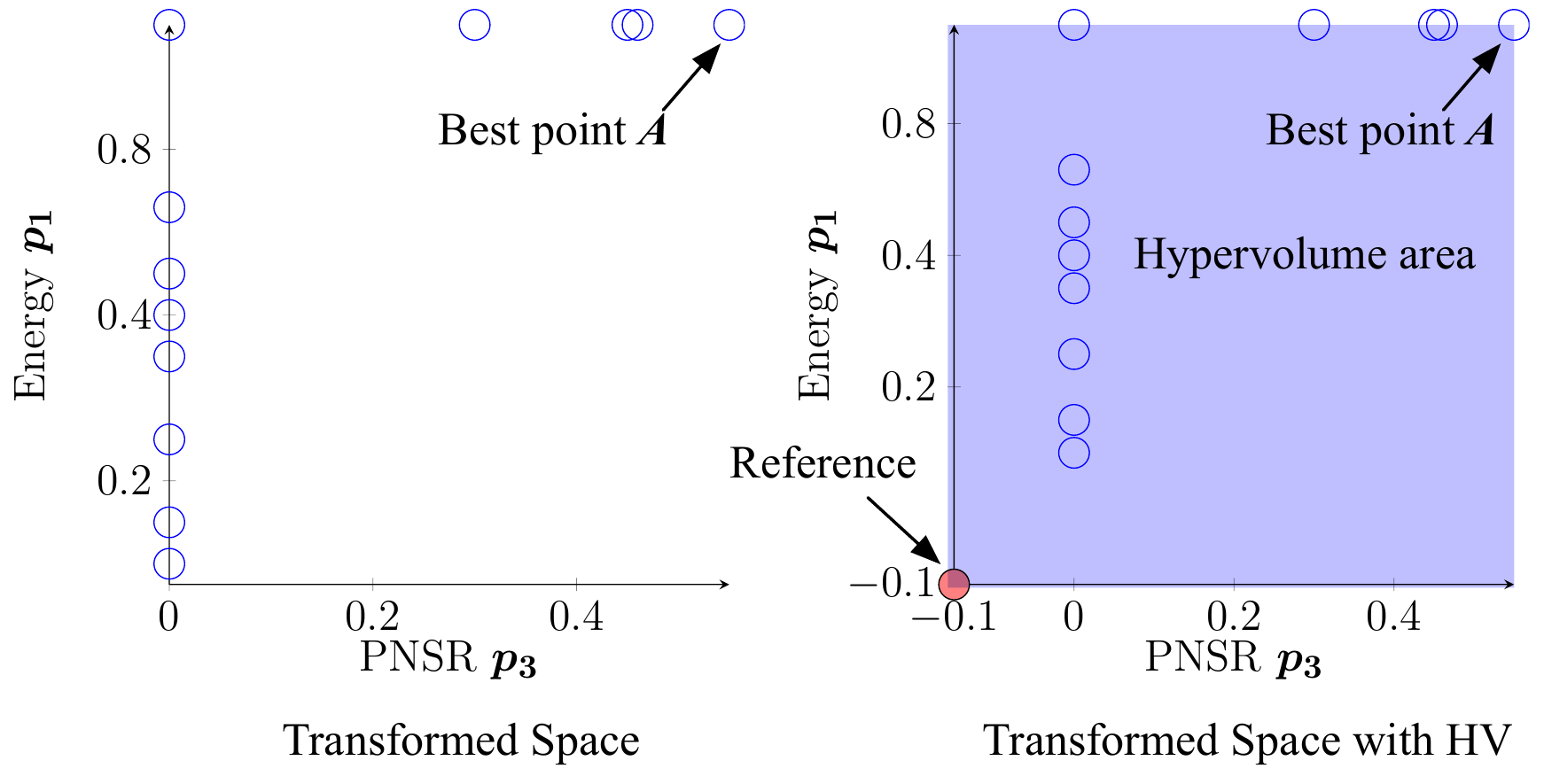}
    \subcaption{Evaluating with HV in the transformed space}
   \end{subfigure}
 
    \caption{Evaluation of HV with and without requirements/aspirations.}
      \label{fig:ahv}
  \end{figure}

Since we are interested in a requirement scenario that has a specific combination of patterns and aspiration space ($\mathbfcal{P}$) in the objective space, the original HV, which always favors the configurations that are close to the entire Pareto front, is no longer suitable. Therefore, we need to transfer these preferences into the HV following the guidance by Li \textit{et al.}~\cite{9252185} and leveraging the patterns and quantification from Section~\ref{sec:req} (for satisfying \textbf{Condition 2}). Using the same example from Section~\ref{sec:motivation}, as shown in Figure~\ref{fig:ahv}, the requirement scenario is that: \textit{the stakeholders prefer better PNSR and energy usage better than 80 watts, but any configurations better than 80 watts are equally preferred; willing to accept energy usage worse than 80 watts but do not accept PNSR worse than 40dB.} This means for any points in the aspiration space, the ones with better PNSR would be preferred more. \rev{Therefore, 
point $\vect{A}$ is the best based on the requirements and should contribute the most to the chosen metric. 
However, 
directly applying HV would make some configurations, which are less preferred to the requirements, contribute significantly to the HV value (Figure~\ref{fig:ahv}a). 
This would misleadingly evaluate some sets that have many non-preferred points to have a very good HV value. 
In contrast, 
when transferring the information of patterns before using HV (i.e., in the transformed space), 
the above requirements and aspirations can be better complied with, 
as $\vect{A}$ is certainly the one that contributes the most and other non-preferred points tend to have no or little contributions. (Figure~\ref{fig:ahv}b)}

To that end, we extend the HV in this work. Suppose that there are $m$ performance objectives (we have $m=2$ in this work) and $\mathbfcal{A}$ is a produced configuration set wherein the vector of a configuration's raw measurements is $\vect{\overline{x}_i}=\{x_1,x_2,...x_m\}$, we calculate HV based on the converted satisficing value of $\vect{\overline{x}_i}$ according to the given $\mathbfcal{P}$. We call it aspiration-aware HV (dubbed A-HV), which is formulated as:
\begin{equation}
	A\mhyphen HV(\mathbfcal{A}) = \lambda(\bigcup_{\vect{\overline{x}_i}\in \mathbfcal{A}} \{\vect{v}|\mathbfcal{P}(\vect{\overline{x}_i})\prec \vect{v} \prec \vect{r}\})
	\label{eq:HV}
\end{equation}
where $\lambda$ is the Lebesgue measure that quantifies the volume~\cite{Zitzler1998} as used in the original HV; $\vect{r}$ is the reference nadir point, which is often taken as the 1.1 times of the range of the nondominated set~\cite{9252185}, hence in our case, this would be $\{-0.1,-0.1\}$ as $\mathbfcal{P}(\vect{\overline{x}_i})$ converts the outputs to $[0,1]$. Like HV, a higher A-HV value is better. To ensure fair comparison with A-HV, we use the minimum and/or maximum values (of each performance objective) from all experiments for the posterior normalization in the patterns.

To enable more intuitive exposition, we report on the \% gain of the A-HV for considering requirements and aspirations in the tuning, i.e., \texttt{PS-w}, over that for \texttt{PS-w/o} on each run, which is defined as:
\begin{equation}
\text{\% Gain} =
{{{x_i - y_i} \over {y_i}}} \times 100
\end{equation}
whereby $x_i$ and $y_i$ are the A-HV value at the $i$th run for \texttt{PS-w} and \texttt{PS-w/o}, respectively, in their sorted lists. Clearly, a positive \% gain indicates that the aspirations are helpful (\texttt{PS-w} is better) while a negative value implies they are harmful (\texttt{PS-w/o} is better); zero gain means identical result.

\subsubsection{Statistical Validation}

%We leverage the following recommended methods for software engineering~\cite{DBLP:conf/icse/ArcuriB11,DBLP:journals/infsof/KampenesDHS07} to interpret the statistical significance of the results over 100 runs in each case: 

We use the standard methods to interpret the significance of the results over 100 runs in each case~\cite{DBLP:conf/icse/ArcuriB11,DBLP:journals/infsof/KampenesDHS07}:

\begin{itemize}
  
  \item \textbf{Wilcoxon test:} We apply the Wilcoxon test~\cite{Wilcoxon1945IndividualCB} with $a=0.05$~\cite{DBLP:conf/icse/ArcuriB11} to investigate the statistical significance of the A-HV comparisons over all 100 runs, as it is a non-parametric statistical test that makes little assumption about the data distribution and has been recommended in software engineering research for pair-wise comparisons~\cite{DBLP:conf/icse/ArcuriB11}.
  
  %A $p<0.05$ indicates significantly different magnitude. 
  
  %we say that the magnitude of differences in the comparisons are significant.
    
    %\item\textbf{Wilcoxon signed-rank test:} We apply the Wilcoxon signed-rank test~\cite{Wilcoxon1945IndividualCB} with $a=0.05$~\cite{ArcuriB11} to investigate the statistical significance of the A-HV comparisons over all 100 runs, as it is a non-parametric statistical test that makes a little assumption about the data distribution and has been recommended in software engineering research for pair-wise comparisons~\cite{ArcuriB11}.
    
    %To ensure the resulted differences are not generated from a trivial effect, 
    
    \item \textbf{$\mathbf{\hat{A}_{12}}$ effect size:} To ensure that a $p<0.05$ is not caused by a trivial amount of the samples, we apply $\hat{A}_{12}$~\cite{Vargha2000ACA} to measure the effect size. In this work, $\hat{A}_{12}>0.5$ denotes \texttt{PS-w} wins wherein it has better A-HV for more than 50\% of the runs. $\hat{A}_{12} \geq 0.6$ or $\hat{A}_{12} \leq 0.4$ indicate a non-trivial effect size. Since there are 100 runs (instead of the commonly used 30), we use a stricter interpretation by which $0.6 \leq \hat{A}_{12} < 0.7$ (0.3 < $\hat{A}_{12} \leq 0.4$), $0.7 \leq \hat{A}_{12} < 0.8$ (0.2 < $\hat{A}_{12} \leq 0.3$), and $\hat{A}_{12} \geq 0.8$ ($\hat{A}_{12} \leq 0.2$) indicate small, medium, and large effect, respectively.

\end{itemize}

%Note that both Wilcoxon signed-rank test and $A_{12}$ are also used in the Scott-Knott test for generating the clusters.

%% file: tables/patterns.tex
% the first point in convergence is either 50 measurement or the first number of measurement after the first generation, whichever that is greater
\begin{table}[t!]
\caption{The considered requirement scenarios (in terms of the combination of the patterns identified from Section~\ref{sec:req}) and their example interpretations. The interpretations are based on the assumption that the performance objectives are $\{latency, throughput\}$ with possible aspiration levels $d_1$ and $d_2$, respectively.}
\label{tb:com-pattern}
\footnotesize
\centering
\begin{tabular}{cp{11.6cm}}\toprule

\textbf{Possible} $\mathbfcal{P}$&\textbf{Example Interpretation}\\
\midrule

$\{\vect{p_0},\vect{p_1}\}$&Prefer better latency and throughput better than $d_2$, but any configurations better than $d_2$ are equally preferred; willing to accept throughput worse than $d_2$.\\

\rowcolor{steel!10}$\{\vect{p_1},\vect{p_0}\}$&Prefer better throughput and latency better than $d_1$, but any configurations better than $d_1$ are equally preferred; willing to accept latency worse than $d_1$.\\

$\{\vect{p_0},\vect{p_2}\}$&Prefer better latency and throughput better than $d_2$, but any configurations better than $d_2$ are equally preferred; do not accept throughput worse than $d_2$.\\

\rowcolor{steel!10}$\{\vect{p_2},\vect{p_0}\}$&Prefer better throughput and latency better than $d_1$, but any configurations better than $d_1$ are equally preferred; do not accept latency worse than $d_1$.\\

$\{\vect{p_0},\vect{p_3}\}$&Prefer better latency and throughput; do not accept throughput worse than $d_2$.\\

\rowcolor{steel!10}$\{\vect{p_3},\vect{p_0}\}$&Prefer better latency and throughput; do not accept latency worse than $d_1$.\\

$\{\vect{p_1},\vect{p_1}\}$&Prefer latency better than $d_1$ and throughput better than $d_2$, but any configurations better than $d_1$ and $d_2$ are equally preferred; willing to accept latency and throughput worse than $d_1$ and $d_2$, respectively.\\

\rowcolor{steel!10}$\{\vect{p_2},\vect{p_2}\}$&Prefer latency better than $d_1$ and throughput better than $d_2$, but any configurations better than $d_1$ and $d_2$ are equally preferred; do not accept latency and throughput worse than $d_1$ and $d_2$, respectively.\\

$\{\vect{p_3},\vect{p_3}\}$&Prefer better latency and throughput; do not accept latency and throughput worse than $d_1$ and $d_2$, respectively.\\

\rowcolor{steel!10}$\{\vect{p_1},\vect{p_2}\}$&Prefer latency better than $d_1$ and throughput better than $d_2$, but any configurations better than $d_1$ and $d_2$ are equally preferred; willing to accept latency worse than $d_1$ but do not accept throughput worse than $d_2$.\\

$\{\vect{p_2},\vect{p_1}\}$&Prefer latency better than $d_1$ and throughput better than $d_2$, but any configurations better than $d_1$ and $d_2$ are equally preferred; willing to accept throughput worse than $d_2$ but do not accept latency worse than $d_1$.\\

\rowcolor{steel!10}$\{\vect{p_1},\vect{p_3}\}$&Prefer better throughput and latency better than $d_1$, but any configurations better than $d_1$ are equally preferred; willing to accept latency worse than $d_1$ but do not accept throughput worse than $d_2$.\\

$\{\vect{p_3},\vect{p_1}\}$&Prefer better latency and throughput better than $d_2$, but any configurations better than $d_2$ are equally preferred; willing to accept throughput worse than $d_2$ but do not accept latency worse than $d_1$.\\

\rowcolor{steel!10}$\{\vect{p_2},\vect{p_3}\}$&Prefer better throughput and latency better than $d_1$, but any configurations better than $d_1$ are equally preferred; do not accept latency worse than $d_1$ nor throughput worse than $d_2$.\\

$\{\vect{p_3},\vect{p_2}\}$&Prefer better latency and throughput better than $d_2$, but any configurations better than $d_2$ are equally preferred; do not accept latency worse than $d_1$ nor throughput worse than $d_2$.\\

\bottomrule
\end{tabular}
\end{table}

%% file: tables/sys-new.tex
\begin{table}[t!]
\caption{Configurable software systems studied. We run all software systems under their standard benchmarks. \textsc{Storm} and \textsc{Keras} (with DNN) use two benchmarks and three dataset, respectively.}
\label{tb:sys}
\centering
\footnotesize
\begin{tabular}{lllcll}\toprule

\textbf{Software}&\textbf{Domain}&\textbf{Performance Objectives}&$\#$ \textbf{Options} &\textbf{Search Space}&\textbf{Used By}\\

\midrule

\textsc{Trimesh}&Mesh solver&Latency and $\#$ Iteration &13&239,260&~\cite{nair2018finding,DBLP:conf/sigsoft/0001Chen}\\

%&\makecell[l]{Latency\\Throughput}&3&1,343&\makecell[l]{Apache Storm under the Word \\Count on OpenNebula(1 CPU)}\\

\rowcolor{steel!10}\textsc{x264}&Video encoding&PSNR and Energy Usage&17&53,662&~\cite{nair2018finding,DBLP:conf/sigsoft/0001Chen}\\

\textsc{Storm/WC}&Stream processing&Latency and Throughput&6&2,880&~\cite{nair2018finding,DBLP:conf/sigsoft/0001Chen,DBLP:conf/mascots/JamshidiC16,DBLP:conf/sigsoft/JamshidiVKS18}\\

\rowcolor{steel!10}\textsc{Storm/RS}&Stream processing&Latency and Throughput&6&3,839&~\cite{nair2018finding,DBLP:conf/sigsoft/0001Chen,DBLP:conf/mascots/JamshidiC16,DBLP:conf/sigsoft/JamshidiVKS18}\\

\textsc{Keras/Adiac}&Deep learning&AUC and Inference Time&13&3.99$\times 10^{13}$&~\cite{DBLP:conf/sigsoft/JamshidiVKS18}\\

\rowcolor{steel!10}\textsc{Keras/DSR}&Deep learning&AUC and Inference Time&13&3.32$\times 10^{13}$&~\cite{DBLP:conf/sigsoft/JamshidiVKS18,DBLP:conf/sigsoft/0001Chen}\\

\textsc{Keras/SA}&Deep learning&AUC and Inference Time&13&2.66$\times 10^{13}$&~\cite{DBLP:conf/sigsoft/JamshidiVKS18}\\

\rowcolor{steel!10}\textsc{XGBoost}&Machine learning&Accuracy and Training Time&13&2.88$\times 10^{10}$&~\cite{DBLP:conf/sigsoft/JamshidiVKS18}\\

\bottomrule
\end{tabular}
%More details can be found at: \href{https://tinyurl.com/y8jczpeo}{\texttt{\textcolor{blue}{https://tinyurl.com/y8jczpeo}}}
\centering

\end{table}

%% file: tables/aspir.tex
\begin{table}[t!]
\caption{Aspiration levels and spaces for the configurable software systems studied (used for all combinations of patterns). $l$, $r$, $c$, and $u$ denote left-shifted, right-shifted, centered, and unrealistic aspirations, respectively.}
\label{tb:aspir-levels}
\setlength{\tabcolsep}{0.8mm}
\centering
\footnotesize
\begin{tabular}{lccccc}\toprule

\textbf{Software}&\textbf{Performance Objectives}&$l$&$r$&$c$&$u$\\

\midrule

\textsc{Trimesh}&$\{$Latency (s), \# Iterations$\}$&$\{81,4\}$&$\{461,15\}$&$\{135,7\}$&$\{37,501\}$\\

\rowcolor{steel!10}\textsc{x264}&$\{$PSNR (dB), Energy Usage (W)$\}$&$\{50,3680\}$&$\{37,462\}$&$\{46,1260\}$&$\{100,34\}$\\

\textsc{Storm/WC}&$\{$Throughput (msgs/m), Latency (ms)$\}$&$\{16473,15677\}$&$\{994,5\}$&$\{8982,101\}$&$\{34740,3\}$\\

\rowcolor{steel!10}\textsc{Storm/RS}&$\{$Throughput (msgs/m), Latency (ms)$\}$&$\{1.3\times 10^5,7819\}$&$\{3006,5\}$&$\{3.7\times 10^4,126\}$&$\{2.3\times 10^5,1.9\}$\\

\textsc{Keras/Adiac}&$\{$AUC, Inference Time (ms)$\}$&$\{0.030,44\}$&$\{0.017,0.05\}$&$\{0.028,3\}$&$\{0.292,0.03\}$\\

\rowcolor{steel!10}\textsc{Keras/DSR}&$\{$AUC, Inference Time (ms)$\}$&$\{0.307,123\}$&$\{0.107,0.12\}$&$\{0.300,25\}$&$\{0.581,0.031\}$\\

\textsc{Keras/SA}&$\{$AUC, Inference Time (ms)$\}$&$\{0.167,21\}$&$\{0.157,0.07\}$&$\{0.160,6\}$&$\{0.325,0.04\}$\\

\rowcolor{steel!10}\textsc{XGBoost}&$\{$Accuracy (\%), Training Time\;(s)$\}$&$\{80,42\}$&$\{54,3\}$&$\{72,8\}$&$\{92,1\}$\\

%\rowcolor{steel!10}\textsc{x264}&Video&PSNR and Energy Usage&17&53,662\\

%\textsc{Storm/WC}&SP&Latency and Throughput&6&2,880\\

%\rowcolor{steel!10}\textsc{Storm/RS}&SP&Latency and Throughput&6&3,839\\

%\textsc{Keras/Adiac}&ML&AUC and Inference Time&13&3.99$\times 10^{13}$\\

%\rowcolor{steel!10}\textsc{Keras/DSR}&ML&AUC and Inference Time&13&3.32$\times 10^{13}$\\

%\textsc{Keras/SA}&ML&AUC and Inference Time&13&2.66$\times 10^{13}$\\

%\rowcolor{steel!10}\textsc{XGBoost}&ML&Accuracy and Training Time&13&2.88$\times 10^{10}$\\

\bottomrule
\end{tabular}
\end{table}

%% file: tables/alg.tex
% the first point in convergence is either 50 measurement or the first number of measurement after the first generation, whichever that is greater
\begin{table}[t!]
\caption{Population size and measurement tuning budget.}
\label{tb:settings}
\footnotesize
\centering
\begin{tabular}{lcc|lcc}\toprule

\textbf{Software}&\textbf{Population Size}&\textbf{\# Measurements}&\textbf{Software}&\textbf{Population Size}&\textbf{\# Measurements}\\
\midrule

\textsc{Trimesh}&10&500&\textsc{x264}&50&1,500\\
\rowcolor{steel!10}\textsc{Storm/WC}&50&500&\textsc{Storm/RS}&30&700\\
\textsc{Keras/Adiac}&50&700&\textsc{Keras/DSR}&60&500\\
\rowcolor{steel!10}\textsc{Keras/SA}&60&500&\textsc{XGBoost}&30&300\\

\bottomrule
\end{tabular}

\end{table}

%% file: results.tex
\section{Results and Findings}
\label{sec:result}

In this section, we present the results of our empirical study and address the research questions posed in Section~\ref{sec:intr}. 

%All data and code are available at: \href{https://tinyurl.com/y8jczpeo}{\texttt{\textcolor{blue}{https://tinyurl.com/y8jczpeo}}}.

%\input{tables/quality-hard}

\subsection{RQ1: Which is Better under Requirements with Realistic Aspirations?}
\label{sec:realistic}

\subsubsection{Method} 

To answer \textbf{RQ1}, we compare \texttt{PS-w} and \texttt{PS-w/o} across 15 combinations of patterns, three realistic aspiration spaces ($l$, $r$, and $c$), three optimizers and eight subject systems, leading to $15 \times 3 \times 3 \times 8 = 1,080$ cases. Since we are interested in a pair-wise comparison of the A-HV under each case, the Wilcoxon test and $\hat{A}_{12}$ are used to verify the statistical significance over 100 runs. 

%Note that for \texttt{PS-w}, the combination of patterns that guides the tuning is identical to the target one from the scenario in a case, i.e., $\mathbfcal{P}' = \mathbfcal{P}$, reflecting what is required in the tuning process.

%\input{tables/count1}
\subsubsection{Results}

\begin{figure*}[t!]
\centering
\includegraphics[width=\textwidth]{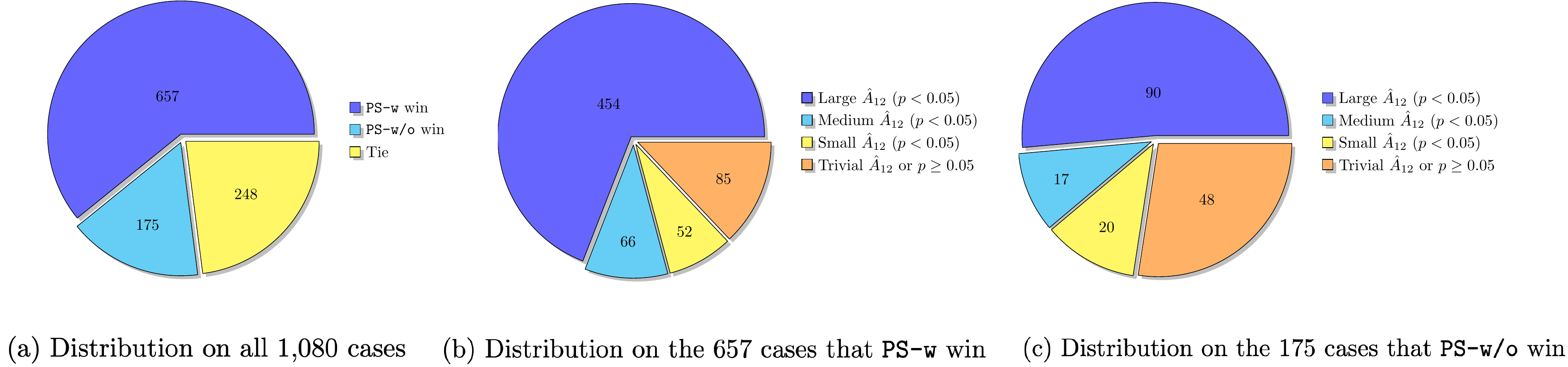}
\caption{Summary of the wins by \texttt{PS-w} and \texttt{PS-w/o} together with their detailed statistics validation results.}
\label{fig:rq1-pie}
\end{figure*}

As an overview, Figure~\ref{fig:rq1-pie} shows a summary of the $\hat{A}_{12}$ outcomes across the cases. Clearly, we see that \texttt{PS-w} performs overwhelmingly better than its \texttt{PS-w/o} counterpart. In particular, \texttt{PS-w} wins for 61\% (657/1080) of the cases and loses for 16\% (175/1080), while there is a 23\% (248/1080) tie. In other words, \texttt{PS-w} is better or similar for 84\% (905/1080) of the cases in contrast to the 39\% (423/1080) when using \texttt{PS-w/o}. Statistically, \texttt{PS-w} wins 572 cases with $\hat{A}_{12} \geq 0.6$ and $p<0.05$, while there are only 127 significant cases when it loses.

\input{tables/rq1}

%\begin{figure*}[htbp]
%\centering
%\includegraphics[width=\textwidth]{figures/gains.pdf}
%\caption{25th, 50th, and 75th percentile of the normalized \% gains on A-HV under realistic aspirations with $\hat{A}_{12}$ and statistical test. A row shows the result over 9 cases (3 aspiration space and 3 algorithms) with 100 runs each. 6(3):2(1) means that \texttt{PS-w} wins 6 cases with $\hat{A}_{12}>0.5$ (3 of them with $\hat{A}_{12}\geq 0.6$ and $p<0.05$), while \texttt{PS-w/o} wins on 2 cases with $\hat{A}_{12}<0.5$ (1 of which has $\hat{A}_{12}\leq 0.4$ and $p<0.05$); the remaining 1 case is a tie. The strategy with more statistically significant wins is shown in bold.}
%\label{fig:gains}
%\end{figure*}

To provide a more comprehensive view on the different systems and requirement scenarios, in Table~\ref{tb:rq1}, we see that \texttt{PS-w} performs considerably better in general, as it achieves reasonably well positive gains on the majority of the cases (up to 145\% improvement on A-HV in average) with generally more statistically significance wins. It is worth noting that we observed particularly high gains on \texttt{PS-w} under \textsc{Storm} (Table~\ref{tb:rq1}c and Table~\ref{tb:rq1}d). This is attributed to the highly diverse performance between configurations for the system, as what has been reported in prior work~\cite{DBLP:conf/sigsoft/0001Chen,DBLP:conf/mascots/JamshidiC16,nair2018finding}.

It is exciting to see that the superiority of \texttt{PS-w} is consistent across the given requirement patterns --- a clear sign to confirm that the requirements can offer important guidance to steer the tuning. However, as expected, when the scenario requires $\vect{p_1}$ or $\vect{p_0}$ only, \texttt{PS-w} and \texttt{PS-w/o} perform mostly identical (or very similar). As for the very few cases where \texttt{PS-w} is inferior to \texttt{PS-w/o}, the results can be the cause of some accidentally encountered local optima issues, which we will discuss in greater detail in what follow.

Therefore, we say:

%compared with when there is more search budget. Note that the above is consistent across different software systems.

%As a result, for \textbf{RQ1}, we say:

\begin{quotebox}
\noindent
\textit{\textbf{RQ1:} Given realistic aspiration space, \texttt{PS-w} is 84\% of the time similar or better than \texttt{PS-w/o} with considerable improvements, suggesting that the requirements and aspirations are beneficial for guiding the tuning in such a situation. Yet, the benefits can vary depending on some particular combinations of the patterns, i.e., it tends to be blurred when only the $\vect{p_1}$ and/or $\vect{p_0}$ is given.}
\end{quotebox}

%\begin{result}
%Given realistic aspiration space, \texttt{PS-w} is safer as it is considerably better or similar to \texttt{PS-w/o} for 84\% of the cases, including 23\% tie. Yet, \texttt{PS-w/o} tends to be better under limited tuning budget. 
%\end{result}

%\texttt{PS-w/o} can be a better option.
%as it converges faster in general. 
%\textcolor{red}{This is because \texttt{PS-w} can be guided to condense on local region according to the requirements and aspirations while emphasizing less on the diversity.}

\begin{figure}[!t]
  \centering
   \begin{subfigure}[t]{0.5\textwidth}
        \centering
\includegraphics[width=\textwidth]{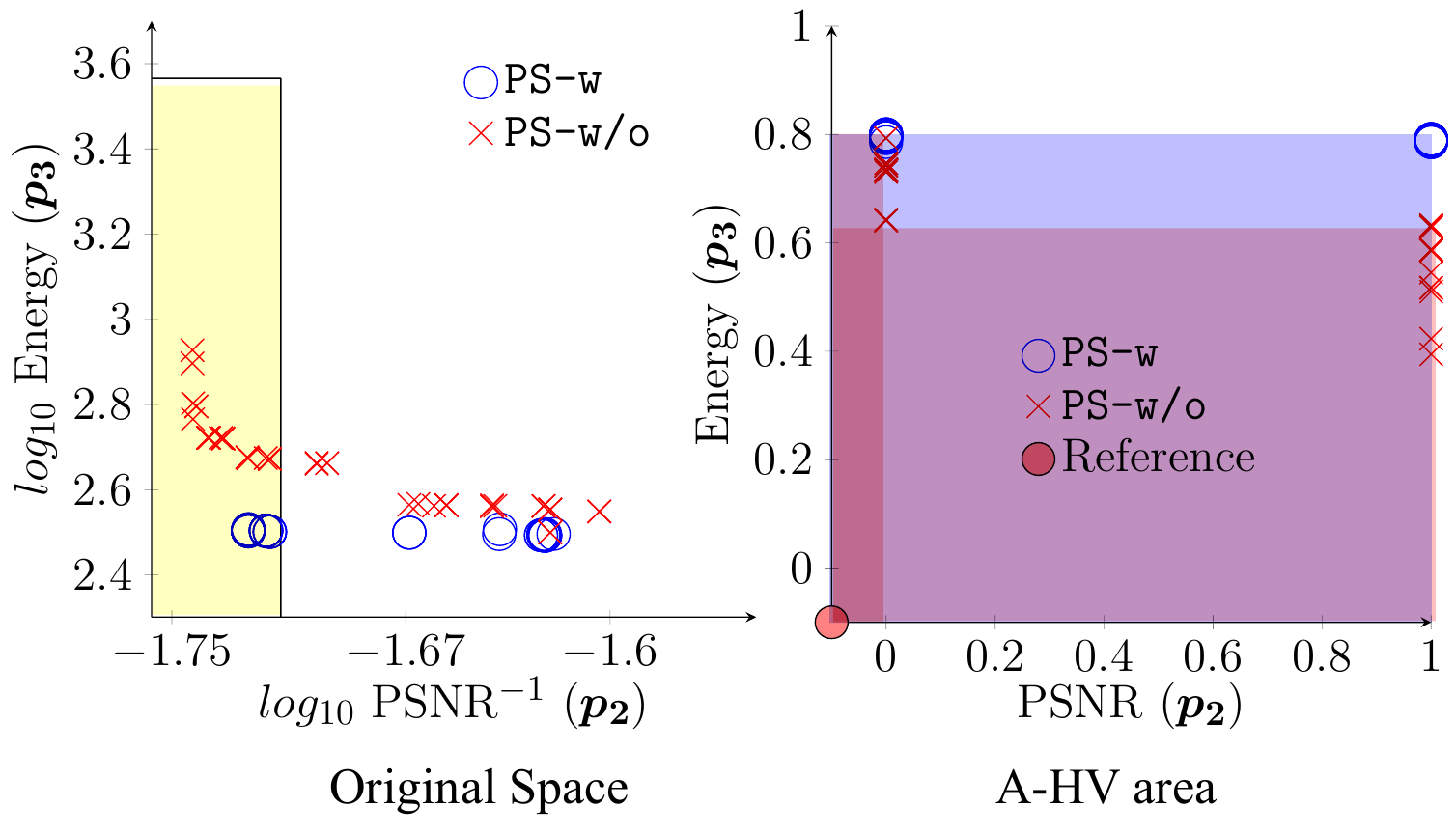}
    \subcaption{\textsc{x264}, \texttt{PS-w wins}}
   \end{subfigure}
 ~
     \begin{subfigure}[t]{0.5\textwidth}
     \centering
\includegraphics[width=\textwidth]{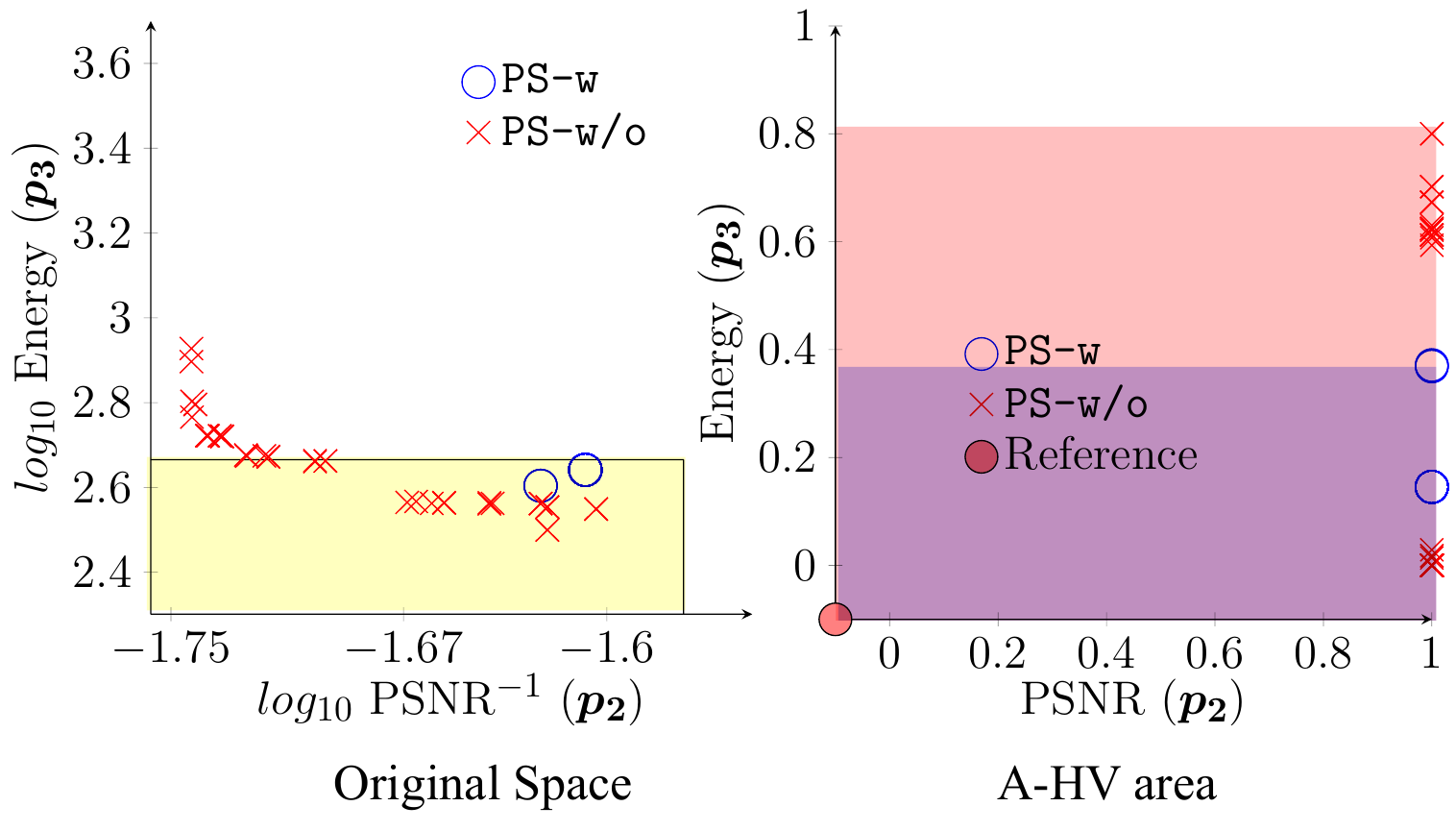}
    \subcaption{\textsc{x264}, \texttt{PS-w/o wins}}
   \end{subfigure}
 
    \caption{Example runs of the final configuration sets (with NSGA-II) under realistic aspiration space indicated by the shaded areas.}
     \label{fig:rq1-dis}
  \end{figure}

\subsubsection{Discussion}

To understand what causes the results under realistic aspiration space, in Figure~\ref{fig:rq1-dis} we show a common example from \textsc{x264}, where all PSNR values better than its aspiration are equally preferred and no worse results are acceptable ($\vect{p_2}$), while the energy usage is desired to be as low as possible, even if its aspiration has already been exceeded ($\vect{p_3}$). Figure~\ref{fig:rq1-dis}a is a superior case of \texttt{PS-w}, in which we see that the aspirations drive the tuning to focus more on the local regions within the objective space, hence the points of \texttt{PS-w} is much less spread than those of \texttt{PS-w/o} (as see in the \textit{Original Space}). Such a ``focused pressure'' is mostly sufficient to help find some more preferred regions by the scenario under a fixed tuning budget, hence the \texttt{PS-w} has better A-HV than \texttt{PS-w/o} (larger volume, as seen in the \textit{A-HV area}). 

However, \texttt{PS-w} is not always beneficial. As reported by Chen and Li~\cite{DBLP:conf/sigsoft/0001Chen}, Nair \textit{et al.}~\cite{nair2018finding}, and the others~\cite{DBLP:conf/mascots/JamshidiC16,DBLP:conf/icse/HaZ19}, configurable software systems are known to exhibit a high degree of sparsity, i.e., the close configurations can also have radically different performance, thus only a small amount of them may achieve certain performance range, causing rather sparse objective points (e.g., Figure~\ref{fig:aspir}). \rev{For example, switching the \textit{wait\_strategy} in \textsc{Storm} can have dramatic impacts on the performance, despite that it is merely a single change on an option. This is because the \textit{wait\_strategy} conserves CPU usage depending on whether the wait is a fixed interval or is progressively determined based on the length of the queue at runtime, therefore it has a large impact on latency and throughput. However, in the tuning, it is represented as a single configuration option with a value chosen from $\{0,1,2,3\}$ where each value represents a distinct wait strategy.} The presence of high sparsity exacerbates the problem of local optima traps --- some undesired regions that are difficult to escape from by an optimizer. Occasionally, searching focally under high sparsity does cause \texttt{PS-w} to overemphasize the less desired local optima, which harms the results. This is why there are some cases where the \texttt{PS-w} show no advantage, as illustrated in Figure~\ref{fig:rq1-dis}b where the points of \texttt{PS-w} are too densely populated compared with those of the \texttt{PS-w/o} (as see in the \textit{Original Space}), causing the volume covered by \texttt{PS-w} is smaller than that of \texttt{PS-w/o} and smaller A-HV (as see in the \textit{A-HV area}).

It is interesting to observe that under certain combinations of patterns, i.e., with $\vect{p_0}$ and/or $\vect{p_1}$ only, both optimization models perform similarly. This makes sense, as in those cases the requirements would create similar discriminative power between configurations to that of \texttt{PS-w/o} (which is essentially guided by $\{\vect{p_0}, \vect{p_0}\}$), generating configurations that are equally preferred under the given needs.

%In general, it is less difficult for the aspirations to overemphasize certain configurations under intractable search space. This is why \texttt{PS-w} can lead to some large improvements for ML systems.

%The x- and y-axis denote the \# measurements and A-HV, respectively.

% talk a bit more about those cases of tie?

\subsection{RQ2: How do Different Aspirations Influence the Comparisons?}

\subsubsection{Method} 

To understand \textbf{RQ2}, we follow the procedure used for \textbf{RQ1}, but with particular focus on the results with respect to the three aspiration spaces used (i.e., $l$, $c$, and $r$). 

\begin{figure*}[t!]
\centering
\includegraphics[width=\textwidth]{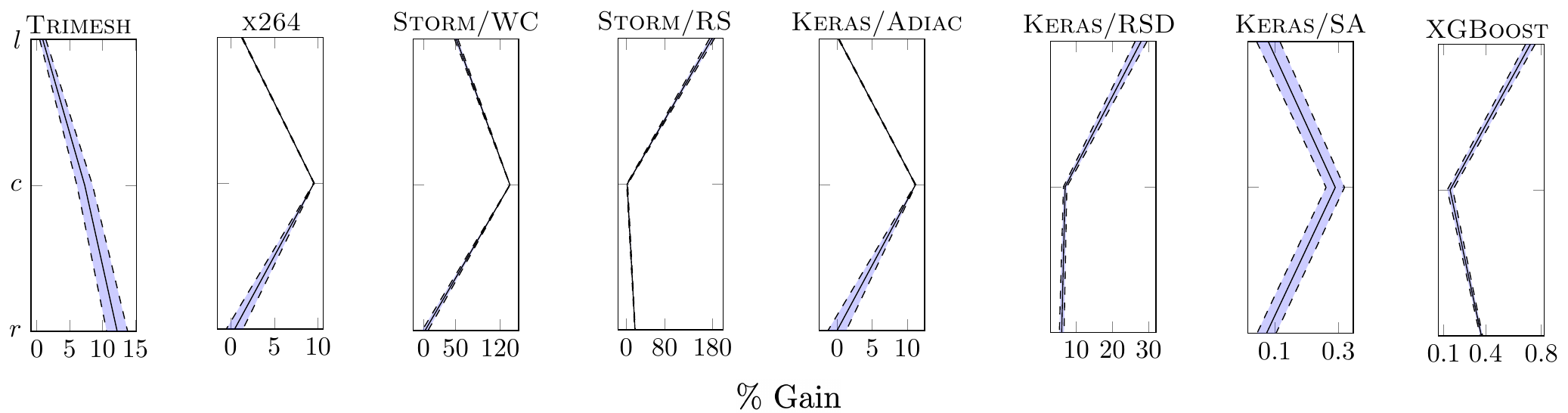}
\caption{Sensitivity of the \% gain on \texttt{PS-w} over \texttt{PS-w/o} to different positions of the realistic aspiration space. Each point is the average and standard error over all combinations of patterns and optimizers. $l$, $c$, and $r$ denote left-shifted, centered, and right-shifted position in the performance landscape, respectively.}
\label{fig:sen}
\end{figure*}

\subsubsection{Results} 

Figure~\ref{fig:sen} plots the sensitivity of A-HV to the different aspiration spaces. While the overall conclusion is consistent with that for \textbf{RQ1} over different patterns and systems, we see that there is often a strong bias on the gains for a certain position of the aspiration spaces. For example, on \textsc{x264} and \textsc{Keras/Adiac}, the improvement of \texttt{PS-w} is particularly high for aspiration space located at the centered area of the objective space. In contrast, the gain is particularly high on \textit{left-shifted} aspiration space under \texttt{Storm/RS} and \textit{centered} space for \texttt{Storm/WC}, which is possible depending on the landscape of a system (as we will discuss next). Indeed, some aspiration spaces can easily cause the \texttt{PS-w} to be trapped at the local optima, making its improvements over \texttt{PS-w/o} blurred. For example, on \textsc{Storm/WC} with \textit{right-shifted} aspiration space, this effect is largely detrimental and hence severely influence the benefits of \texttt{PS-w}.

In summary, we have:

\begin{quotebox}
\noindent
%\textit{\textbf{RQ2:} The given (realistic) aspiration space is an important factor that determines the improvement of \texttt{PS-w} over \texttt{PS-w/o}. }
\textit{\textbf{RQ2:} The improvement of \texttt{PS-w} over \texttt{PS-w/o} is often largely biased to certain position of the aspiration space in the performance landscape, e.g., centered or left-shifted. Yet, \texttt{PS-w} still performs more advantageously in general.}
\end{quotebox}

\begin{figure}[t!]
\centering
\includegraphics[width=0.5\columnwidth]{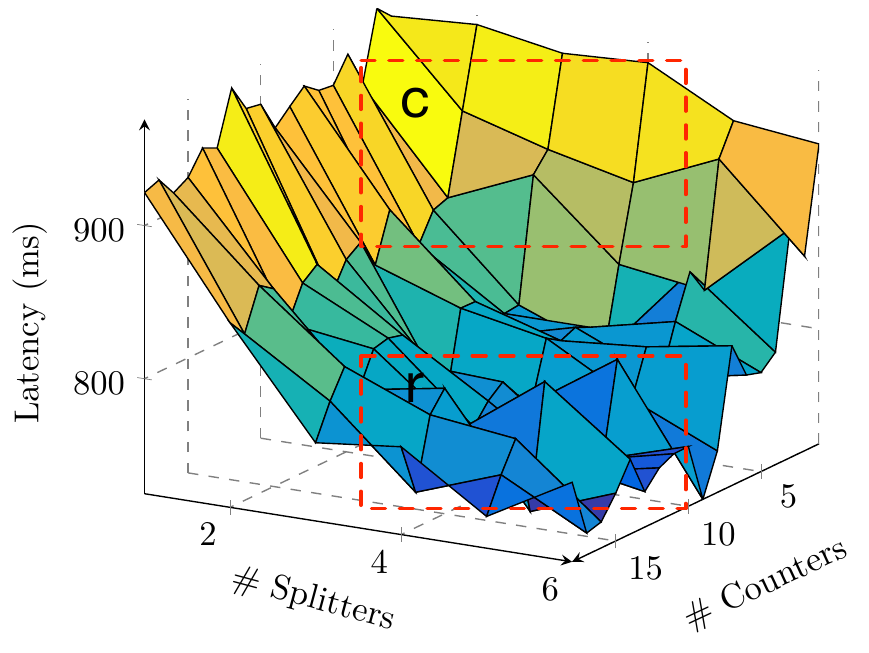}
\caption{A projected landscape of the performance objective Latency with respect to configuration options \texttt{Splitters} and \texttt{Counters} for \textsc{Storm/WC}. $c$ and $r$ denote centered and right-shifted aspiration space, respectively. Note that the aspirations spaces are bounded because the throughput objective is also considered; it is however not showed here for simpler exposition.}
\label{fig:rq2-dis}
\end{figure}

\subsubsection{Discussion}

As discussed for \textbf{RQ1}, the main reason that \texttt{PS-w} can perform better than \texttt{PS-w/o} is due to the ``focused search pressure''. However, this may not be always helpful if the tuning encounters complex local optima that are difficult for the optimizer to escape from. The high sensitivity of the gains to the positions of aspiration space suggests that the local optima can be distributed unevenly across the landscape. If the aspiration space covers many local optima regions, then certainly the gains of \texttt{PS-w} would be marginal.

For example, in Figure~\ref{fig:rq2-dis}, clearly the aspiration space $c$ (which covers the requirements for latency and throughput) would be bounded on some regions in the landscape with a much more smooth surface for the latency. However, for $r$, the region becomes highly rugged and steep, which involves some very difficult local optima. Unfortunately, we did not see consistent patterns of such a sensitivity across the configurable software systems, which makes sense as the performance landscape of those systems can be very different too.

\begin{figure*}[t!]
\centering
\includegraphics[width=\textwidth]{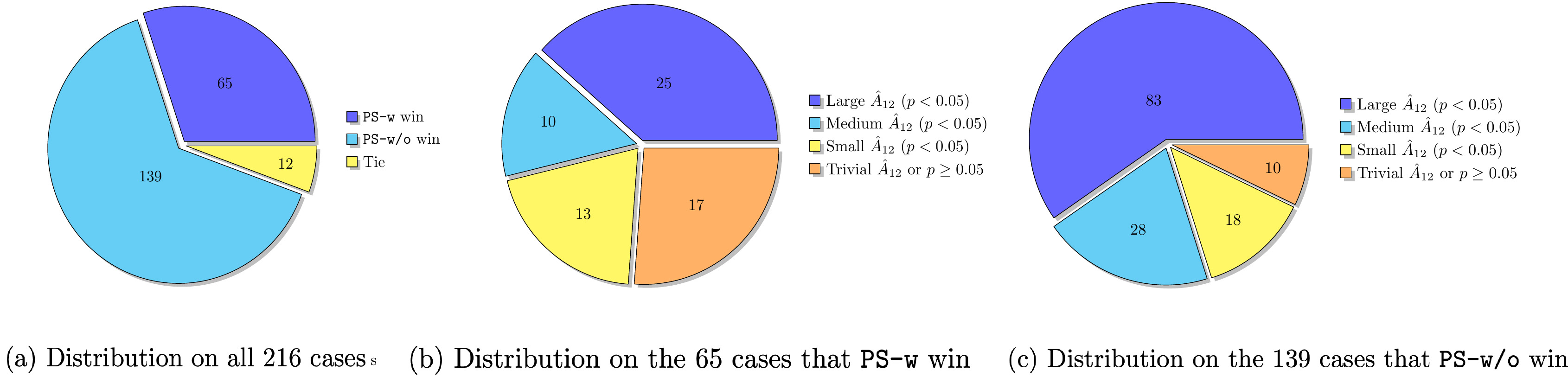}
\caption{Summary of the wins by \texttt{PS-w} and \texttt{PS-w/o} together with their detailed statistics validation results.}
\label{fig:rq3-pie}
\end{figure*}

\subsection{RQ3: What if the Aspirations are Unrealistic?}

\input{tables/rq3}

\subsubsection{Method} 

%As mentioned, since it is always possible that the aspiration space specified in a prior is (Pareto-)unrealistic (i.e., $u$ from Figure), for \textbf{RQ2}, we seek to understand how the two types of Pareto search perform in such situation. To that end

To investigate \textbf{RQ3}, we omit the scenarios with $\vect{p_0}$ as they cannot create an unrealistic aspiration space. This has left us with nine combinations of patterns, which, together with three Pareto optimizers and eight subjects, provide $9 \times 3 \times 8 = 216$ cases. All other settings are identical to those for \textbf{RQ1}.

\subsubsection{Results}

As the summary from Figure~\ref{fig:rq3-pie}, we see that \texttt{PS-w/o} is generally better across all the cases, as it wins on 64\% (139/216) while loses on 30\% (65/216). There is also a 6\% (12/216) tie. This means that \texttt{PS-w/o} is better or similar on 70\% (151/216) of the cases against the 36\% (77/216) for \texttt{PS-w}. Among these, \texttt{PS-w/o} wins 129 cases with $\hat{A}_{12} \leq 0.4$ and $p<0.05$ comparing with 41 of such cases when it loses.

Similar results can be confirmed in Table~\ref{tb:rq3} when inspecting specific system and requirement scenarios. Albeit there is a limited number of cases where \texttt{PS-w} is still advantageous, it is more common to show no improvement at all or even cause fairly negative gains, which could be up to an average of $-$243\%. It also has overall much less statistically significant wins across the cases. Particularly, we found that under $\{\vect{p_1}, \vect{p_1}\}$ on all systems, the two optimization models perform similarly but \texttt{PS-w} tends to obtain more wins. This is because such a combination pattern is the only case where the unrealism of aspiration does not lead to too many incomparable configurations.

Overall, we conclude that:

%\input{tables/quality-mix-all}

%In summary, our answer for \textbf{RQ2} is:

\begin{quotebox}
\noindent
\textit{\textbf{RQ3:} When the aspiration space is unrealistic, \texttt{PS-w/o} is safer as it is similar or reasonably better than \texttt{PS-w} for 70\% of the time, meaning that the requirements and aspirations are more harmful for guiding the tuning in this case. Yet, the only exception applied to $\{\vect{p_1}, \vect{p_1}\}$.}
\end{quotebox}

%\begin{result}
%Given unrealistic aspiration space, \texttt{PS-w/o} is safer as it is reasonably better or similar to \texttt{PS-w} for 70\% cases with 6\% tie. 
%\end{result}
%The same applied even under limited search budget. 

%\textcolor{red}{This is because the guidance in \texttt{PS-w} is weak as the aspiration space is hard to reach.}

\subsubsection{Discussion}

%all results of latency better than the aspiration are equally preferred but there is a tolerance if it goes worse ($\vect{p_1}\}$), while better throughput is preferred and no worse result than the aspiration is acceptable ($\vect{p_3}\}$). Here,

\begin{figure}[!t]
  \centering
   \begin{subfigure}[t]{0.5\textwidth}
        \centering
\includegraphics[width=\textwidth]{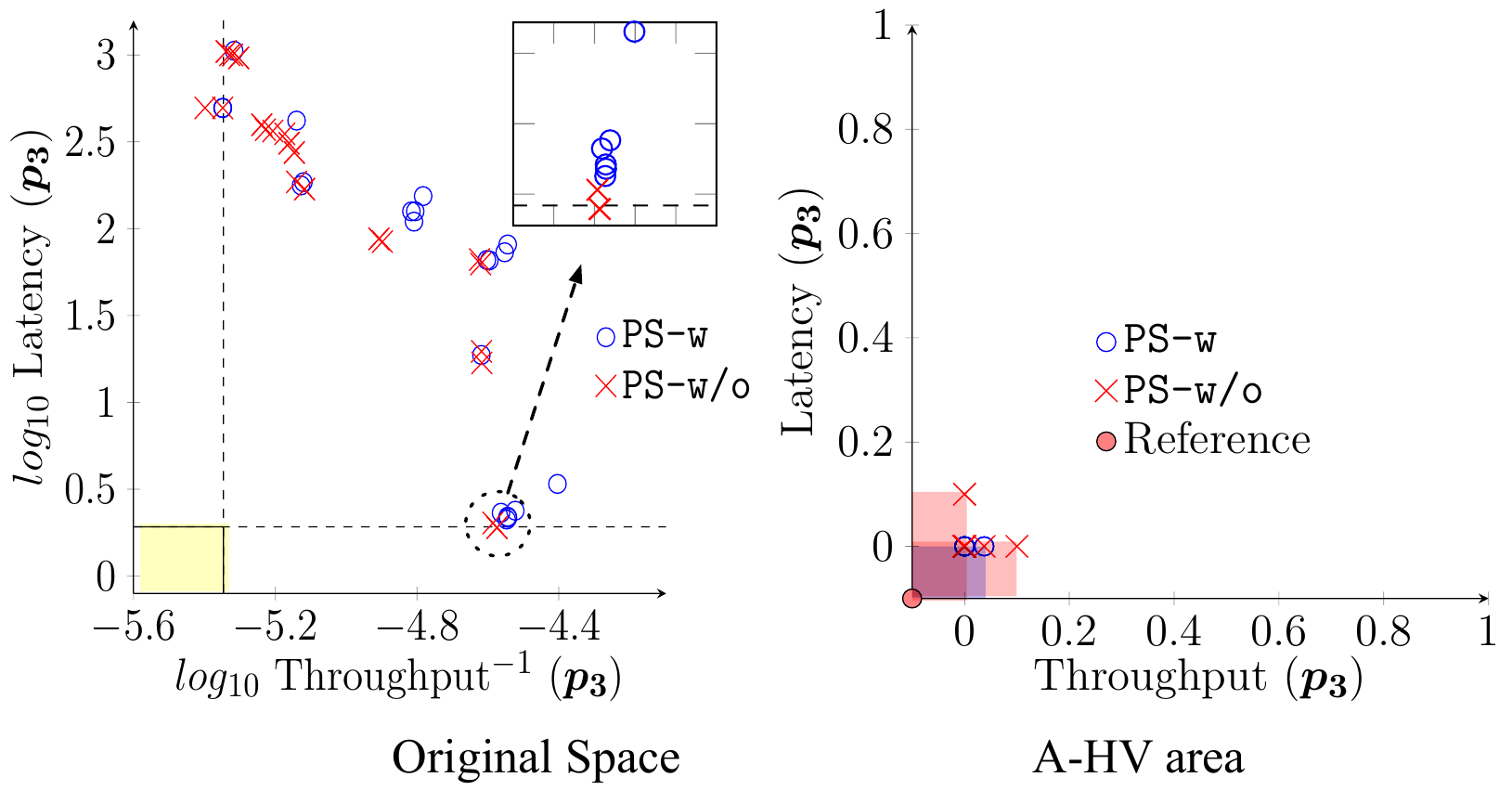}
    \subcaption{\textsc{Storm/RS}, \texttt{PS-w/o wins}}
   \end{subfigure}
 ~
     \begin{subfigure}[t]{0.5\textwidth}
     \centering
\includegraphics[width=\textwidth]{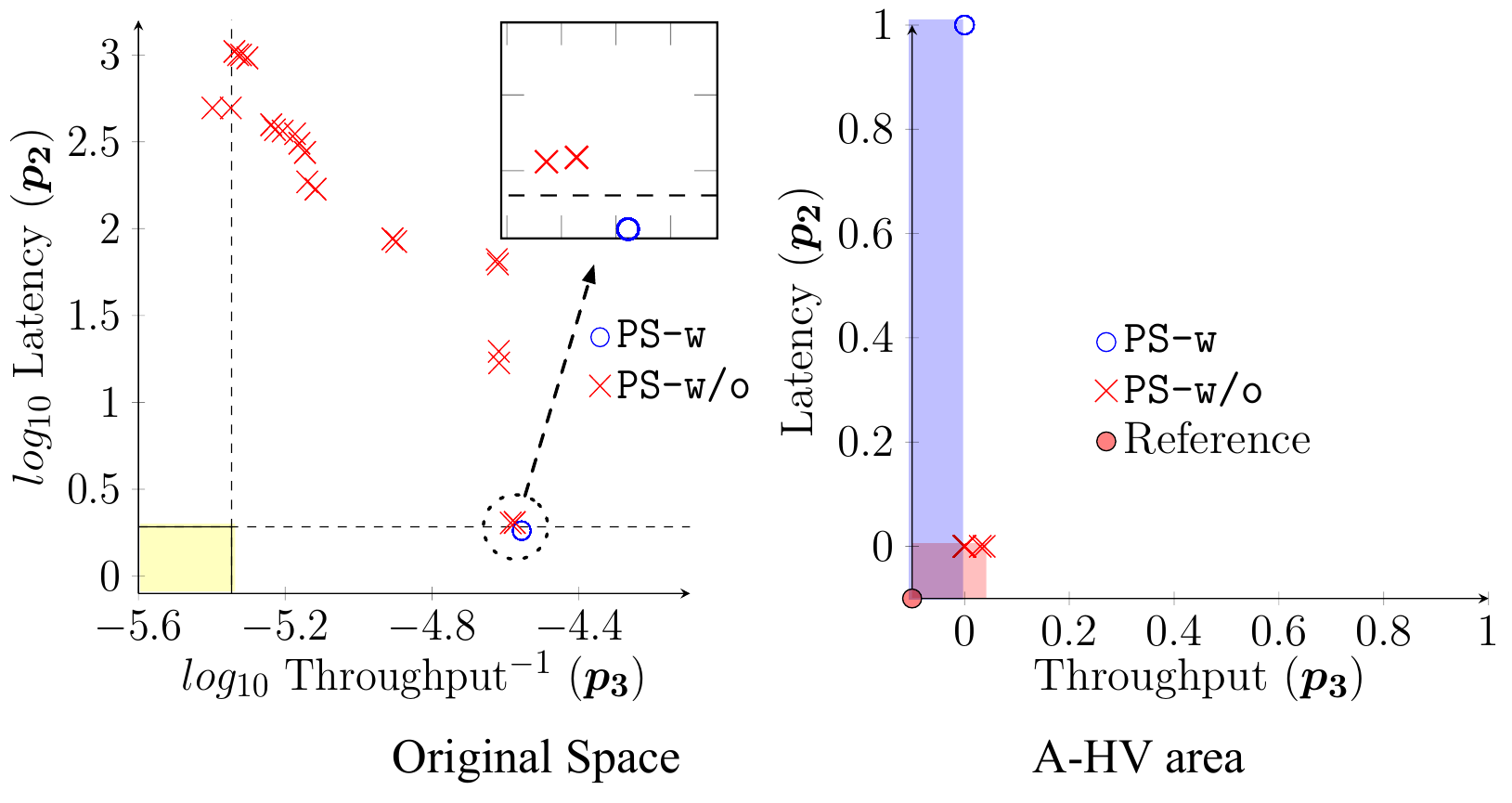}
    \subcaption{\textsc{Storm/RS}, \texttt{PS-w wins}}
   \end{subfigure}
 
    \caption{Example runs of the final configuration sets (with NSGA-II) under unrealistic aspiration space indicated by the shaded areas.}
     \label{fig:rq3-dis}
  \end{figure}

Given unrealistic aspiration space, the most common cases are similar to the \textsc{Storm/RS} example in Figure~\ref{fig:rq3-dis}, where the \texttt{PS-w} is commonly inferior to \texttt{PS-w/o} when the diversity tends to be high (Figure~\ref{fig:rq3-dis}a), but sometimes superior to \texttt{PS-w/o} under limited diversity (Figure~\ref{fig:rq3-dis}b). This is because in most of the cases, after being transformed using the requirements with unrealistic aspirations,  \texttt{PS-w} tends to find too many incomparable configurations from the beginning (as in the cases other than $\{\vect{p_1},\vect{p_1}\}$, most configurations are fully unsatisfied on at least one performance objective), implying that the guidance provided by an unrealistic aspiration space is dramatically weakened. Such an incomparability, although may prompt slightly better diversity to escape from the local regions, can often severely harm the tendency towards more preferred configurations that reach/exceed the aspirations, leading to worse A-HV (the smaller volume) than \texttt{PS-w/o} in Figure~\ref{fig:rq3-dis}a. This is because no selection pressure (i.e., discriminative power) can be generated in such a case. It is also the reason why \texttt{PS-w} is not deteriorated by the unrealistic aspirations under $\{\vect{p_1},\vect{p_1}\}$, which can still ensure that the configurations are comparable. Yet sometimes (Figure~\ref{fig:rq3-dis}b), such a high incomparability does help \texttt{PS-w} to find a good configuration by chance (e.g., better than the aspiration of latency), which is more desired than those of \texttt{PS-w/o}, leading to better HV (the larger volume). Hence the \texttt{PS-w} remains better for certain cases, despite that the tuning would be easily trapped at that configuration due to the high sparsity. 

%Intuitively, the smaller the search space, the more likely it that a good configuration may be found by chance. This is why the aspirations can be more harmful to the ML systems, as they are often intractable.

%can be guided to concentrate on certain region according to the requirements and aspiration, while focusing less on the diversity.

%This is also the reason why no one tends to have faster A-HV convergence.

\subsection{RQ4: Does the Given Tuning Resource Important?}
\label{sec:rq4}

\subsubsection{Method} 

To understand the resource efficiency of both optimization models in \textbf{RQ4}, for each system, we use the following procedure:

\begin{enumerate}

\item Plot the trajectories of A-HV along with the number of measurements for both \texttt{PS-w} and \texttt{PS-w/o}, where each point is the average of all requirement patterns, aspiration spaces, and optimizers.

\item Identify a baseline, $b$, taken as the smallest number of measurements that the baseline model consumes to achieve its best A-HV (say $T$).

\item For the other model, find the smallest number of measurements, denoted as $m$, at which the average A-HV is equivalent to or better than $T$.

\item Calculate the speedup over the baseline model, i.e., $s = {b \over m}$, according to the metric used by Gao \textit{et al.}~\cite{DBLP:conf/icse/GaoZ0LY21}. 

\end{enumerate}

Since we found that the generally better optimization model differs depending on the realism of the aspiration space, we use \texttt{PS-w/o} and \texttt{PS-w} as the baseline for realistic and unrealistic aspiration situations, respectively.

\begin{figure*}[t!]
\centering
\includegraphics[width=\textwidth]{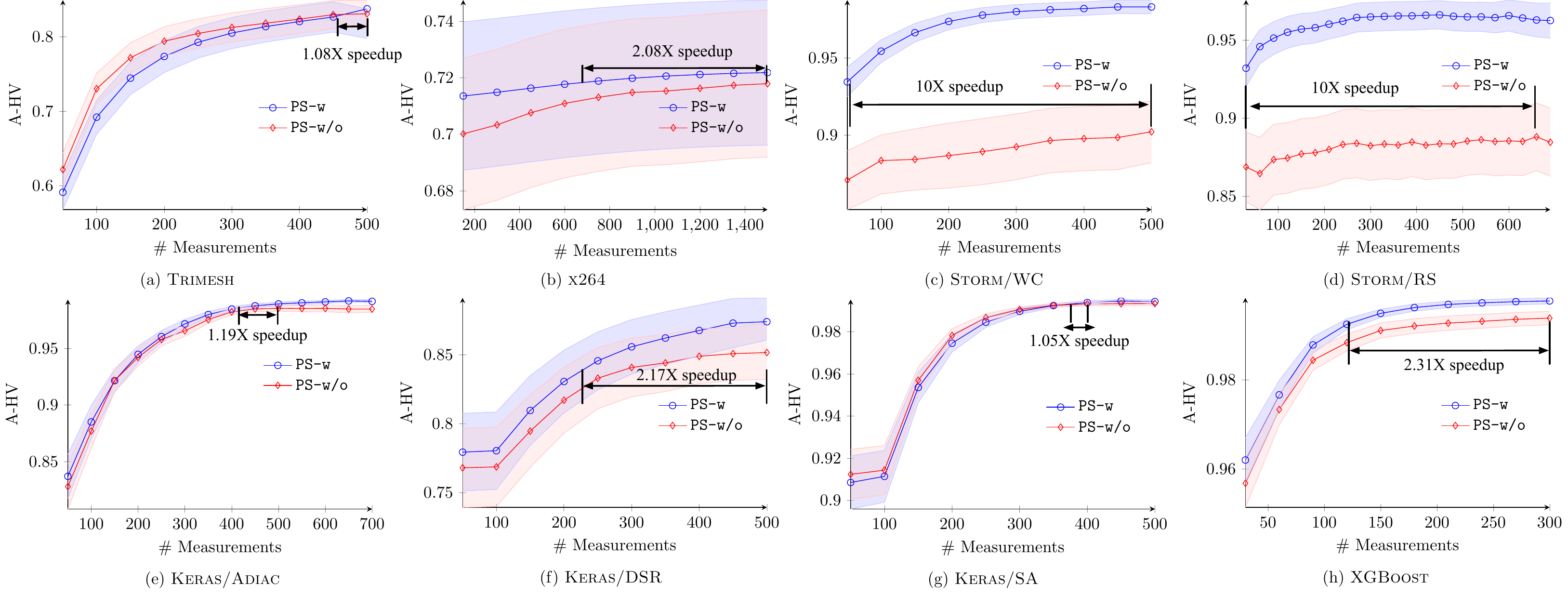}
\caption{Speedup on \texttt{PS-w} over \texttt{PS-w/o} under realistic aspirations (each point is the average and standard error over all combinations of patterns, aspiration space, optimziers and their runs).}
\label{fig:sp1}
\end{figure*}

\begin{figure*}[t!]
\centering
\includegraphics[width=\textwidth]{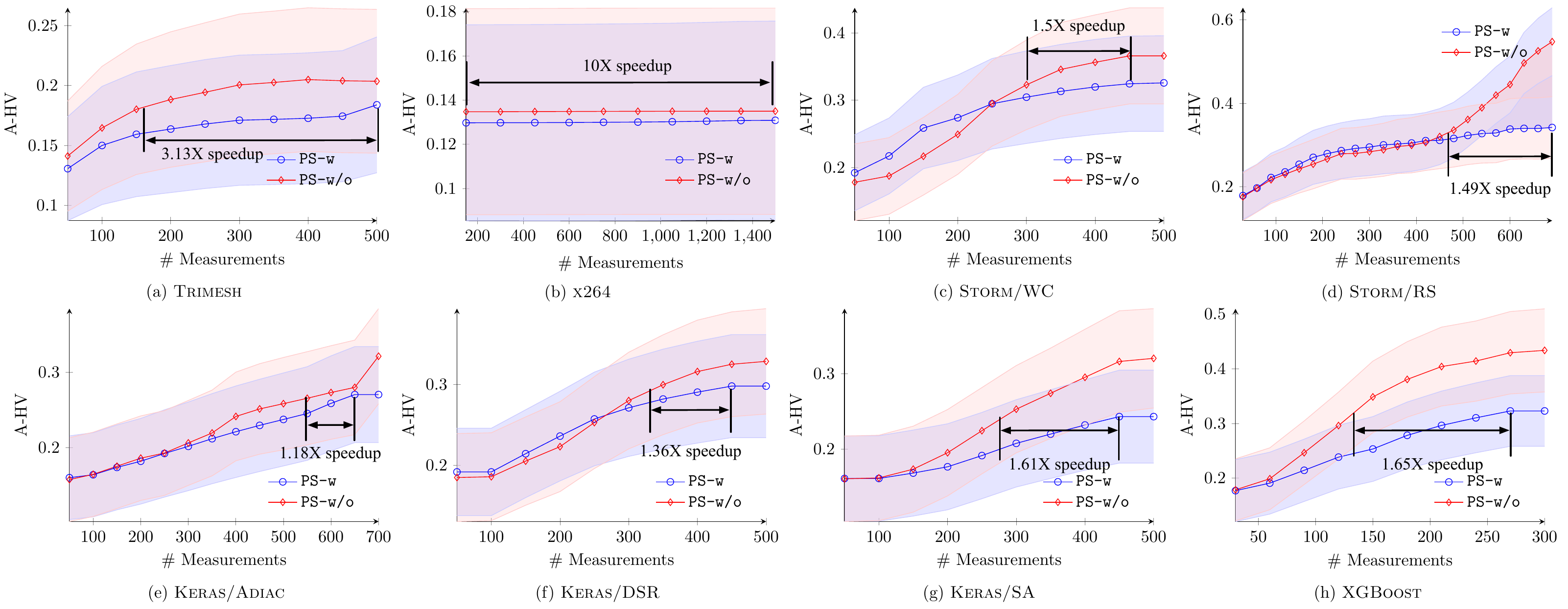}
\caption{Speedup on \texttt{PS-w/o} over \texttt{PS-w} under unrealistic aspirations (each point is the average and standard error over all combinations of patterns, optimziers and their runs).}
\label{fig:sp2}
\end{figure*}

\subsubsection{Results}

From the results under realistic aspirations as shown in Figure~\ref{fig:sp1}, we see that \texttt{PS-w} outperforms \texttt{PS-w/o} throughout the trajectories over different configurable systems, which further strengthen our findings for \textbf{RQ1}. The improvement in resource efficiency has been remarkable: there is a speedup between $1.05 \times$ and $10 \times$. In contrast, when the given aspirations are unrealistic (Figure~\ref{fig:sp2}), \texttt{PS-w/o} is much more resource-efficient, as it enables a speedup from $1.18 \times$ to $10 \times$. This again complies with the findings for \textbf{RQ3}. However, under unrealistic aspirations, the advantages of \texttt{PS-w/o} may not be obvious at the early stage of the tuning; on some systems (e.g., Figure~\ref{fig:sp2}c and Figure~\ref{fig:sp2}f), it is even inferior to the \texttt{PS-w} until around 250 configurations have been measured. 

In summary, we found that:

\begin{quotebox}
\noindent
%\textit{\textbf{RQ4:} The resource (tuning budget) is less likely to affect the choice between  \texttt{PS-w} and \texttt{PS-w/o} under realistic aspirations, but it may be an influential factor when the given aspirations are unrealistic.}
\textit{\textbf{RQ4:} Under realistic aspirations, \texttt{PS-w} often obtains consistently better A-HV than \texttt{PS-w/o} throughout the trajectory and with a speedup up to $10\times$. When the aspirations are unrealistic, in contrast, the two optimization models are competitive in the early stage of tuning but soon \texttt{PS-w/o} would lead to better results with considerably high speedup.}
\end{quotebox}

\subsubsection{Discussion}

Under realistic aspirations, the reasons that \texttt{PS-w} has better A-HV throughout the trajectory with remarkably high speedup are two folds: firstly, as what we have already discussed for \textbf{RQ1}, the guidance provided by the requirements and aspirations are often helpful to enable the tuning to be more focus-driven, hence better utilizing the resources to explore the more promising area. Secondly, \texttt{PS-w/o} would waste the valuable tuning budget to explore those configurations that it favors, but would never be preferred under the given requirements, since it is naturally interested in the whole Pareto front. Therefore, the above difference enables \texttt{PS-w} to be a particularly attractive model for some systems, such as \textsc{Storm}, where the performance of diverse configurations can be radically different.

The situation is completely different when the given aspirations are unrealistic and it is mainly due to the high incomparability in \texttt{PS-w} as mentioned for \textbf{RQ3} --- many of the configurations are incomparable when transformed using the requirements with unrealistic aspirations. It has been shown that this situation can cause severe issues for any Pareto optimizer~\cite{Li2014a}, as the resources would have been spent mainly on exploration. However, such an incomparability can occasionally be helpful to explore some preferred configurations by chance, especially at the early stage of the tuning where the \texttt{PS-w/o} has yet explored enough space to pursue the Pareto front. As such, we see that at the beginning \texttt{PS-w} performs similarly to \texttt{PS-w/o} and, sometimes, even better.

%% file: tables/rq1.tex
% CS Table for RQ2
\begin{table*}[t!]
\caption{Comparing \texttt{PS-w} and \texttt{PS-w/o} under realistic requirements and aspirations over 100 runs. \quartexp{0}{20}{10}{20} and \bquartexp{0}{20}{10}{20} denote the average (Avg) and standard error (SE) of the positive and negative \% gain, respectively.  \tquartexp{0}{20}{10}{20}  means zero gain overall. The column ``\texttt{PS-w}'' and ``\texttt{PS-w/o}'' show the number of cases that the corresponding optimization model wins. 9 (6) means one wins on 9 cases within which 6 shows statistical significance, i.e., $\hat{A}_{12} \geq 0.6$ (or $\hat{A}_{12} \leq 0.4$) and $p<0.05$ (each combination of requirement patterns has 9 cases in total, as there are 3 aspiration space and 3 optimizers). The \setlength{\fboxsep}{1.5pt}\colorbox{steel!30}{blue cells} denote \texttt{PS-w} wins more while \setlength{\fboxsep}{1.5pt}\colorbox{red!10}{red cells} mean it loses more.}
    \label{tb:rq1}
    \footnotesize
    \setlength{\tabcolsep}{1mm}
  \begin{center}
    \begin{adjustbox}{max width = 1\textwidth}

    \begin{tabular}{cc@{}cc@{}cc}
        \begin{tabular}{lcccll}
        & \textbf{\texttt{PS-w}} & \textbf{\texttt{PS-w/o}} & \textbf{Tie} &  \multicolumn{2}{c}{\textbf{Avg (SE) of A-HV Gain}} \\
            \hline
\rowcolor{steel!30}${\{\vect{p_0}\textit{,}\vect{p_1}\}}$&8 (7)&1 (1)&0&1.1\% (0.3\%)& \quart{21.5}{0.6}{21.8}{100}\\ 
\rowcolor{steel!30}${\{\vect{p_1}\textit{,}\vect{p_0}\}}$&7 (7)&2 (2)&0&-4.6\% (1.4\%)& \bquart{9.5}{2.7}{10.9}{100}\\ 
\rowcolor{steel!30}${\{\vect{p_0}\textit{,}\vect{p_2}\}}$&5 (5)&4 (4)&0&36.4\% (10.4\%)& \quart{79.8}{20.2}{89.9}{100}\\ 
\rowcolor{steel!30}${\{\vect{p_2}\textit{,}\vect{p_0}\}}$&7 (7)&2 (2)&0&-3.5\% (2.4\%)& \bquart{10.6}{4.7}{12.9}{100}\\ 
\rowcolor{steel!30}${\{\vect{p_0}\textit{,}\vect{p_3}\}}$&6 (5)&3 (2)&0&29.0\% (7.2\%)& \quart{68.8}{13.9}{75.7}{100}\\ 
\rowcolor{steel!30}${\{\vect{p_3}\textit{,}\vect{p_0}\}}$&8 (8)&1 (1)&0&2.9\% (0.6\%)& \quart{24.7}{1.2}{25.3}{100}\\ 
\rowcolor{steel!30}${\{\vect{p_1}\textit{,}\vect{p_1}\}}$&3 (0)&2 (1)&4&-4.6\% (1.4\%)& \bquart{9.5}{2.7}{10.8}{100}\\ 
\rowcolor{red!10}${\{\vect{p_2}\textit{,}\vect{p_2}\}}$&1 (0)&5 (1)&3&-7.6\% (5.1\%)& \bquart{0.0}{9.9}{4.9}{100}\\ 
\rowcolor{steel!30}${\{\vect{p_3}\textit{,}\vect{p_3}\}}$&6 (5)&3 (3)&0&17.8\% (7.0\%)& \quart{47.3}{13.5}{54.1}{100}\\ 
${\{\vect{p_1}\textit{,}\vect{p_2}\}}$&3 (0)&3 (1)&3&-3.5\% (5.2\%)& \bquart{8.0}{10.1}{13.0}{100}\\ 
\rowcolor{red!10}${\{\vect{p_2}\textit{,}\vect{p_1}\}}$&2 (0)&4 (1)&3&-6.2\% (1.6\%)& \bquart{6.3}{3.1}{7.8}{100}\\ 
\rowcolor{steel!30}${\{\vect{p_1}\textit{,}\vect{p_3}\}}$&6 (6)&3 (3)&0&6.0\% (5.2\%)& \quart{26.3}{10.0}{31.3}{100}\\ 
\rowcolor{steel!30}${\{\vect{p_3}\textit{,}\vect{p_1}\}}$&9 (9)&0 (0)&0&3.1\% (1.3\%)& \quart{24.4}{2.6}{25.7}{100}\\ 
\rowcolor{steel!30}${\{\vect{p_2}\textit{,}\vect{p_3}\}}$&6 (6)&3 (3)&0&-1.4\% (5.9\%)& \bquart{11.3}{11.5}{17.0}{100}\\ 
\rowcolor{red!10}${\{\vect{p_3}\textit{,}\vect{p_2}\}}$&4 (4)&5 (3)&0&28.1\% (8.2\%)& \quart{66.1}{15.7}{74.0}{100}\\

%&&&&& \squart{-28\%}{21\%}{69\%}{100}\\ 

        \end{tabular} & 
        &
       \begin{tabular}{lcccllll}
                 & \textbf{\texttt{PS-w}} & \textbf{\texttt{PS-w/o}} & \textbf{Tie} &  \multicolumn{2}{c}{\textbf{Avg (SE) of A-HV Gain}}  \\
            \hline
\rowcolor{steel!30}${\{\vect{p_0}\textit{,}\vect{p_1}\}}$&9 (9)&0 (0)&0&0.2\% (0.0\%)& \quart{1.7}{0.2}{1.8}{100}\\ 
\rowcolor{steel!30}${\{\vect{p_1}\textit{,}\vect{p_0}\}}$&6 (6)&2 (2)&1&0.1\% (0.0\%)& \quart{0.1}{0.1}{0.1}{100}\\ 
\rowcolor{steel!30}${\{\vect{p_0}\textit{,}\vect{p_2}\}}$&8 (6)&1 (1)&0&4.9\% (4.2\%)& \quart{24.2}{36.9}{42.6}{100}\\ 
\rowcolor{steel!30}${\{\vect{p_2}\textit{,}\vect{p_0}\}}$&6 (5)&3 (3)&0&0.1\% (0.0\%)& \quart{0.0}{0.1}{0.1}{100}\\ 
\rowcolor{red!10}${\{\vect{p_0}\textit{,}\vect{p_3}\}}$&3 (2)&6 (4)&0&7.2\% (2.1\%)& \quart{54.0}{18.8}{63.4}{100}\\ 
\rowcolor{steel!30}${\{\vect{p_3}\textit{,}\vect{p_0}\}}$&6 (6)&3 (3)&0&2.1\% (0.3\%)& \quart{17.4}{2.7}{18.8}{100}\\ 
\rowcolor{steel!30}${\{\vect{p_1}\textit{,}\vect{p_1}\}}$&1 (0)&0 (0)&8&0.1\% (0.0\%)& \quart{0.0}{0.0}{0.0}{100}\\ 
\rowcolor{steel!30}${\{\vect{p_2}\textit{,}\vect{p_2}\}}$&1 (0)&0 (0)&8&5.0\% (4.5\%)& \quart{24.3}{39.1}{43.8}{100}\\ 
\rowcolor{red!10}${\{\vect{p_3}\textit{,}\vect{p_3}\}}$&3 (3)&6 (6)&0&3.6\% (1.1\%)& \quart{26.7}{9.8}{31.6}{100}\\ 
\rowcolor{steel!30}${\{\vect{p_1}\textit{,}\vect{p_2}\}}$&1 (0)&0 (0)&8&5.0\% (4.5\%)& \quart{24.3}{39.1}{43.8}{100}\\ 
\rowcolor{steel!30}${\{\vect{p_2}\textit{,}\vect{p_1}\}}$&1 (0)&0 (0)&8&0.1\% (0.0\%)& \quart{0.0}{0.0}{0.0}{100}\\ 
\rowcolor{steel!30}${\{\vect{p_1}\textit{,}\vect{p_3}\}}$&5 (5)&4 (3)&0&10.0\% (2.8\%)& \quart{75.5}{24.5}{87.7}{100}\\ 
\rowcolor{steel!30}${\{\vect{p_3}\textit{,}\vect{p_1}\}}$&9 (9)&0 (0)&0&2.8\% (0.3\%)& \quart{22.9}{2.9}{24.3}{100}\\ 
\rowcolor{steel!30}${\{\vect{p_2}\textit{,}\vect{p_3}\}}$&6 (6)&3 (3)&0&9.3\% (2.8\%)& \quart{69.5}{24.7}{81.9}{100}\\ 
\rowcolor{steel!30}${\{\vect{p_3}\textit{,}\vect{p_2}\}}$&8 (6)&1 (1)&0&6.3\% (3.5\%)& \quart{39.7}{30.9}{55.1}{100}\\

        \end{tabular} &
        &
       \begin{tabular}{lcccllll}
                  & \textbf{\texttt{PS-w}} & \textbf{\texttt{PS-w/o}} & \textbf{Tie} &  \multicolumn{2}{c}{\textbf{Avg (SE) of A-HV Gain}}  \\
            \hline
\rowcolor{steel!30}${\{\vect{p_0}\textit{,}\vect{p_1}\}}$&8 (8)&1 (1)&0&0.1\% (0.0\%)& \quart{0.0}{0.0}{0.0}{100}\\ 
\rowcolor{steel!30}${\{\vect{p_1}\textit{,}\vect{p_0}\}}$&8 (8)&1 (1)&0&0.1\% (0.0\%)& \quart{0.0}{0.0}{0.0}{100}\\ 
\rowcolor{steel!30}${\{\vect{p_0}\textit{,}\vect{p_2}\}}$&8 (8)&0 (0)&1&82.6\% (17.2\%)& \quart{48.4}{11.3}{54.1}{100}\\ 
\rowcolor{steel!30}${\{\vect{p_2}\textit{,}\vect{p_0}\}}$&6 (6)&3 (3)&0&-0.1\% (0.0\%)& \bquart{0.0}{0.0}{0.0}{100}\\ 
\rowcolor{steel!30}${\{\vect{p_0}\textit{,}\vect{p_3}\}}$&7 (6)&2 (1)&0&90.4\% (17.4\%)& \quart{53.5}{11.4}{59.2}{100}\\ 
\rowcolor{steel!30}${\{\vect{p_3}\textit{,}\vect{p_0}\}}$&9 (9)&0 (0)&0&67.4\% (13.4\%)& \quart{39.8}{8.7}{44.1}{100}\\ 
\rowcolor{steel!30}${\{\vect{p_1}\textit{,}\vect{p_1}\}}$&1 (1)&0 (0)&8&0.1\% (0.0\%)& \quart{0.0}{0.0}{0.0}{100}\\ 
\rowcolor{steel!30}${\{\vect{p_2}\textit{,}\vect{p_2}\}}$&1 (1)&0 (0)&8&83.0\% (17.4\%)& \quart{48.7}{11.4}{54.3}{100}\\ 
\rowcolor{steel!30}${\{\vect{p_3}\textit{,}\vect{p_3}\}}$&9 (8)&0 (0)&0&142.1\% (18.8\%)& \quart{86.9}{12.3}{93.0}{100}\\ 
\rowcolor{steel!30}${\{\vect{p_1}\textit{,}\vect{p_2}\}}$&1 (1)&0 (0)&8&83.0\% (17.4\%)& \quart{48.7}{11.4}{54.3}{100}\\ 
\rowcolor{steel!30}${\{\vect{p_2}\textit{,}\vect{p_1}\}}$&1 (1)&0 (0)&8&0.1\% (0.0\%)& \quart{0.0}{0.0}{0.0}{100}\\ 
\rowcolor{steel!30}${\{\vect{p_1}\textit{,}\vect{p_3}\}}$&7 (7)&2 (2)&0&91.5\% (17.6\%)& \quart{54.1}{11.5}{59.9}{100}\\ 
\rowcolor{steel!30}${\{\vect{p_3}\textit{,}\vect{p_1}\}}$&9 (9)&0 (0)&0&67.9\% (13.4\%)& \quart{40.1}{8.8}{44.5}{100}\\ 
\rowcolor{steel!30}${\{\vect{p_2}\textit{,}\vect{p_3}\}}$&6 (6)&3 (3)&0&91.5\% (17.6\%)& \quart{54.1}{11.5}{59.9}{100}\\ 
\rowcolor{steel!30}${\{\vect{p_3}\textit{,}\vect{p_2}\}}$&8 (8)&1 (0)&0&142.9\% (19.6\%)& \quart{87.2}{12.8}{93.6}{100}\\

        \end{tabular}  \\
            (a). \textsc{Trimesh}&& (b). \textsc{x264} && (c). \textsc{Storm/WC}

        \\
        
        \\
        
            \begin{tabular}{lcccll}
        & \textbf{\texttt{PS-w}} & \textbf{\texttt{PS-w/o}} & \textbf{Tie} &  \multicolumn{2}{c}{\textbf{Avg (SE) of A-HV Gain}} \\
            \hline
\rowcolor{steel!30}${\{\vect{p_0}\textit{,}\vect{p_1}\}}$&8 (8)&1 (1)&0&0.1\% (0.0\%)& \quart{0.0}{0.0}{0.0}{100}\\ 
\rowcolor{steel!30}${\{\vect{p_1}\textit{,}\vect{p_0}\}}$&6 (5)&3 (2)&0&-0.1\% (0.0\%)& \bquart{0.0}{0.0}{0.0}{100}\\ 
\rowcolor{steel!30}${\{\vect{p_0}\textit{,}\vect{p_2}\}}$&8 (7)&1 (1)&0&1.0\% (2.0\%)& \quart{0.0}{1.2}{0.6}{100}\\ 
\rowcolor{steel!30}${\{\vect{p_2}\textit{,}\vect{p_0}\}}$&8 (7)&1 (0)&0&114.8\% (19.9\%)& \quart{65.1}{12.4}{71.3}{100}\\ 
\rowcolor{steel!30}${\{\vect{p_0}\textit{,}\vect{p_3}\}}$&5 (5)&4 (4)&0&1.5\% (1.9\%)& \quart{0.3}{1.2}{0.9}{100}\\ 
\rowcolor{steel!30}${\{\vect{p_3}\textit{,}\vect{p_0}\}}$&9 (9)&0 (0)&0&145.6\% (18.5\%)& \quart{84.6}{11.5}{90.4}{100}\\ 
\rowcolor{steel!30}${\{\vect{p_1}\textit{,}\vect{p_1}\}}$&2 (1)&0 (0)&7&0.1\% (0.0\%)& \quart{0.0}{0.0}{0.0}{100}\\ 
\rowcolor{steel!30}${\{\vect{p_2}\textit{,}\vect{p_2}\}}$&2 (1)&0 (0)&7&117.0\% (20.2\%)& \quart{66.4}{12.5}{72.6}{100}\\ 
\rowcolor{steel!30}${\{\vect{p_3}\textit{,}\vect{p_3}\}}$&8 (8)&1 (1)&0&95.5\% (14.3\%)& \quart{54.8}{8.9}{59.3}{100}\\ 
\rowcolor{steel!30}${\{\vect{p_1}\textit{,}\vect{p_2}\}}$&2 (1)&0 (0)&7&1.0\% (2.0\%)& \quart{0.0}{1.2}{0.6}{100}\\ 
\rowcolor{steel!30}${\{\vect{p_2}\textit{,}\vect{p_1}\}}$&2 (1)&0 (0)&7&116.0\% (20.1\%)& \quart{65.8}{12.5}{72.0}{100}\\ 
\rowcolor{steel!30}${\{\vect{p_1}\textit{,}\vect{p_3}\}}$&5 (4)&4 (3)&0&1.5\% (1.9\%)& \quart{0.3}{1.2}{0.9}{100}\\ 
\rowcolor{steel!30}${\{\vect{p_3}\textit{,}\vect{p_1}\}}$&9 (9)&0 (0)&0&151.4\% (19.4\%)& \quart{88.0}{12.0}{94.0}{100}\\ 
\rowcolor{steel!30}${\{\vect{p_2}\textit{,}\vect{p_3}\}}$&8 (7)&1 (0)&0&112.5\% (19.3\%)& \quart{63.9}{12.0}{69.9}{100}\\ 
\rowcolor{steel!30}${\{\vect{p_3}\textit{,}\vect{p_2}\}}$&8 (7)&1 (1)&0&139.5\% (19.2\%)& \quart{80.6}{11.9}{86.6}{100}\\

%&&&&& \squart{-28\%}{21\%}{69\%}{100}\\ 

        \end{tabular} & 
        &
       \begin{tabular}{lcccllll}
                 & \textbf{\texttt{PS-w}} & \textbf{\texttt{PS-w/o}} & \textbf{Tie} &  \multicolumn{2}{c}{\textbf{Avg (SE) of A-HV Gain}}  \\
            \hline
\rowcolor{steel!30}${\{\vect{p_0}\textit{,}\vect{p_1}\}}$&6 (5)&3 (2)&0&-0.1\% (0.0\%)& \bquart{11.4}{0.2}{11.5}{100}\\ 
\rowcolor{steel!30}${\{\vect{p_1}\textit{,}\vect{p_0}\}}$&9 (9)&0 (0)&0&0.1\% (0.0\%)& \quart{11.9}{0.0}{11.9}{100}\\ 
\rowcolor{steel!30}${\{\vect{p_0}\textit{,}\vect{p_2}\}}$&7 (7)&2 (1)&0&5.0\% (4.5\%)& \quart{30.5}{30.2}{45.6}{100}\\ 
\rowcolor{steel!30}${\{\vect{p_2}\textit{,}\vect{p_0}\}}$&9 (9)&0 (0)&0&0.1\% (0.0\%)& \quart{11.9}{0.0}{11.9}{100}\\ 
\rowcolor{steel!30}${\{\vect{p_0}\textit{,}\vect{p_3}\}}$&6 (4)&3 (2)&0&8.3\% (4.5\%)& \quart{53.0}{30.3}{68.1}{100}\\ 
\rowcolor{red!10}${\{\vect{p_3}\textit{,}\vect{p_0}\}}$&2 (1)&7 (6)&0&-1.2\% (0.6\%)& \bquart{1.6}{3.8}{3.5}{100}\\ 
\rowcolor{steel!30}${\{\vect{p_1}\textit{,}\vect{p_1}\}}$&1 (0)&0 (0)&8&0.1\% (0.0\%)& \quart{11.9}{0.0}{11.9}{100}\\ 
\rowcolor{steel!30}${\{\vect{p_2}\textit{,}\vect{p_2}\}}$&1 (0)&0 (0)&8&5.0\% (4.5\%)& \quart{30.6}{30.1}{45.6}{100}\\ 
\rowcolor{steel!30}${\{\vect{p_3}\textit{,}\vect{p_3}\}}$&6 (5)&2 (1)&1&10.6\% (5.0\%)& \quart{66.6}{33.4}{83.3}{100}\\ 
\rowcolor{steel!30}${\{\vect{p_1}\textit{,}\vect{p_2}\}}$&1 (0)&0 (0)&8&5.0\% (4.5\%)& \quart{30.6}{30.1}{45.6}{100}\\ 
\rowcolor{steel!30}${\{\vect{p_2}\textit{,}\vect{p_1}\}}$&1 (0)&0 (0)&8&0.1\% (0.0\%)& \quart{11.9}{0.0}{11.9}{100}\\ 
\rowcolor{steel!30}${\{\vect{p_1}\textit{,}\vect{p_3}\}}$&9 (9)&0 (0)&0&8.4\% (4.5\%)& \quart{53.5}{30.3}{68.6}{100}\\ 
\rowcolor{steel!30}${\{\vect{p_3}\textit{,}\vect{p_1}\}}$&6 (4)&3 (2)&0&-1.4\% (0.6\%)& \bquart{0.0}{4.3}{2.2}{100}\\ 
\rowcolor{steel!30}${\{\vect{p_2}\textit{,}\vect{p_3}\}}$&9 (9)&0 (0)&0&9.6\% (4.5\%)& \quart{61.8}{30.4}{77.0}{100}\\ 
\rowcolor{steel!30}${\{\vect{p_3}\textit{,}\vect{p_2}\}}$&6 (5)&3 (1)&0&5.5\% (5.0\%)& \quart{32.2}{33.6}{49.0}{100}\\

        \end{tabular} &
        &
       \begin{tabular}{lcccllll}
                  & \textbf{\texttt{PS-w}} & \textbf{\texttt{PS-w/o}} & \textbf{Tie} &  \multicolumn{2}{c}{\textbf{Avg (SE) of A-HV Gain}}  \\
            \hline
\rowcolor{steel!30}${\{\vect{p_0}\textit{,}\vect{p_1}\}}$&6 (3)&3 (2)&0&0.2\% (0.1\%)& \quart{0.3}{0.1}{0.3}{100}\\ 
\rowcolor{steel!30}${\{\vect{p_1}\textit{,}\vect{p_0}\}}$&8 (8)&1 (0)&0&0.1\% (0.0\%)& \quart{0.0}{0.0}{0.0}{100}\\ 
\rowcolor{steel!30}${\{\vect{p_0}\textit{,}\vect{p_2}\}}$&6 (5)&2 (1)&1&1.1\% (1.9\%)& \quart{0.3}{3.6}{2.1}{100}\\ 
\rowcolor{steel!30}${\{\vect{p_2}\textit{,}\vect{p_0}\}}$&8 (7)&1 (0)&0&9.1\% (6.0\%)& \quart{11.8}{11.4}{17.5}{100}\\ 
\rowcolor{steel!30}${\{\vect{p_0}\textit{,}\vect{p_3}\}}$&6 (4)&3 (2)&0&1.3\% (1.9\%)& \quart{0.7}{3.6}{2.5}{100}\\ 
\rowcolor{steel!30}${\{\vect{p_3}\textit{,}\vect{p_0}\}}$&6 (5)&3 (1)&0&22.6\% (5.3\%)& \quart{38.3}{10.1}{43.3}{100}\\ 
\rowcolor{steel!30}${\{\vect{p_1}\textit{,}\vect{p_1}\}}$&2 (1)&0 (0)&7&0.1\% (0.0\%)& \quart{0.0}{0.0}{0.0}{100}\\ 
\rowcolor{steel!30}${\{\vect{p_2}\textit{,}\vect{p_2}\}}$&2 (1)&0 (0)&7&18.1\% (7.3\%)& \quart{27.5}{14.0}{34.5}{100}\\ 
\rowcolor{steel!30}${\{\vect{p_3}\textit{,}\vect{p_3}\}}$&8 (7)&1 (1)&0&23.9\% (3.6\%)& \quart{42.4}{6.8}{45.8}{100}\\ 
\rowcolor{steel!30}${\{\vect{p_1}\textit{,}\vect{p_2}\}}$&2 (1)&0 (0)&7&1.0\% (2.0\%)& \quart{0.0}{3.8}{2.0}{100}\\ 
\rowcolor{steel!30}${\{\vect{p_2}\textit{,}\vect{p_1}\}}$&2 (1)&0 (0)&7&9.1\% (6.0\%)& \quart{11.8}{11.4}{17.5}{100}\\ 
\rowcolor{steel!30}${\{\vect{p_1}\textit{,}\vect{p_3}\}}$&9 (8)&0 (0)&0&1.5\% (2.0\%)& \quart{0.9}{3.8}{2.8}{100}\\ 
\rowcolor{steel!30}${\{\vect{p_3}\textit{,}\vect{p_1}\}}$&8 (8)&1 (1)&0&49.1\% (6.4\%)& \quart{87.8}{12.2}{93.9}{100}\\ 
\rowcolor{steel!30}${\{\vect{p_2}\textit{,}\vect{p_3}\}}$&8 (8)&1 (1)&0&22.6\% (7.3\%)& \quart{36.2}{14.0}{43.2}{100}\\ 
\rowcolor{steel!30}${\{\vect{p_3}\textit{,}\vect{p_2}\}}$&9 (8)&0 (0)&0&44.6\% (6.9\%)& \quart{78.6}{13.3}{85.2}{100}\\

        \end{tabular}  \\
            (d). \textsc{Storm/RS}&& (e). \textsc{Keras/Adiac} && (f). \textsc{Keras/DSR}
            \\
            \\

            \begin{tabular}{lcccll}
        & \textbf{\texttt{PS-w}} & \textbf{\texttt{PS-w/o}} & \textbf{Tie} &  \multicolumn{2}{c}{\textbf{Avg (SE) of A-HV Gain}} \\
            \hline
\rowcolor{steel!30}${\{\vect{p_0}\textit{,}\vect{p_1}\}}$&5 (5)&3 (2)&1&0.1\% (0.0\%)& \quart{45.3}{1.1}{45.8}{100}\\ 
\rowcolor{steel!30}${\{\vect{p_1}\textit{,}\vect{p_0}\}}$&9 (9)&0 (0)&0&0.1\% (0.0\%)& \quart{45.0}{0.0}{45.0}{100}\\ 
\rowcolor{red!10}${\{\vect{p_0}\textit{,}\vect{p_2}\}}$&4 (3)&5 (4)&0&0.1\% (0.0\%)& \quart{44.2}{1.1}{44.7}{100}\\ 
\rowcolor{steel!30}${\{\vect{p_2}\textit{,}\vect{p_0}\}}$&9 (9)&0 (0)&0&0.1\% (0.0\%)& \quart{45.0}{0.0}{45.0}{100}\\ 
\rowcolor{steel!30}${\{\vect{p_0}\textit{,}\vect{p_3}\}}$&6 (5)&3 (3)&0&0.5\% (0.1\%)& \quart{77.8}{4.9}{80.2}{100}\\ 
\rowcolor{steel!30}${\{\vect{p_3}\textit{,}\vect{p_0}\}}$&6 (3)&3 (0)&0&-0.5\% (0.3\%)& \bquart{0.0}{22.4}{11.2}{100}\\ 
${\{\vect{p_1}\textit{,}\vect{p_1}\}}$&0 (0)&0 (0)&9&0.0\% (0.0\%)& \tquart{44.9}{0.0}{44.9}{100}\\ 
${\{\vect{p_2}\textit{,}\vect{p_2}\}}$&0 (0)&0 (0)&9&0.0\% (0.0\%)& \tquart{44.9}{0.0}{44.9}{100}\\ 
\rowcolor{steel!30}${\{\vect{p_3}\textit{,}\vect{p_3}\}}$&6 (6)&2 (1)&1&0.4\% (0.1\%)& \quart{72.4}{10.6}{77.7}{100}\\ 
${\{\vect{p_1}\textit{,}\vect{p_2}\}}$&0 (0)&0 (0)&9&0.0\% (0.0\%)& \tquart{44.9}{0.0}{44.9}{100}\\ 
${\{\vect{p_2}\textit{,}\vect{p_1}\}}$&0 (0)&0 (0)&9&0.0\% (0.0\%)& \tquart{44.9}{0.0}{44.9}{100}\\ 
\rowcolor{steel!30}${\{\vect{p_1}\textit{,}\vect{p_3}\}}$&9 (9)&0 (0)&0&0.7\% (0.1\%)& \quart{94.1}{5.0}{96.6}{100}\\ 
\rowcolor{steel!30}${\{\vect{p_3}\textit{,}\vect{p_1}\}}$&4 (3)&3 (1)&2&-0.3\% (0.3\%)& \bquart{10.1}{23.3}{21.7}{100}\\ 
\rowcolor{steel!30}${\{\vect{p_2}\textit{,}\vect{p_3}\}}$&9 (9)&0 (0)&0&0.7\% (0.1\%)& \quart{94.9}{5.1}{97.4}{100}\\ 
\rowcolor{red!10}${\{\vect{p_3}\textit{,}\vect{p_2}\}}$&4 (3)&5 (4)&0&-0.2\% (0.3\%)& \bquart{23.2}{19.3}{32.8}{100}\\

%&&&&& \squart{-28\%}{21\%}{69\%}{100}\\ 

        \end{tabular} & 
        &
       \begin{tabular}{lcccllll}
                 & \textbf{\texttt{PS-w}} & \textbf{\texttt{PS-w/o}} & \textbf{Tie} &  \multicolumn{2}{c}{\textbf{Avg (SE) of A-HV Gain}}  \\
            \hline
\rowcolor{steel!30}${\{\vect{p_0}\textit{,}\vect{p_1}\}}$&8 (5)&0 (0)&1&0.1\% (0.0\%)& \quart{9.9}{0.6}{10.1}{100}\\ 
\rowcolor{steel!30}${\{\vect{p_1}\textit{,}\vect{p_0}\}}$&9 (9)&0 (0)&0&0.1\% (0.0\%)& \quart{4.2}{0.0}{4.2}{100}\\ 
\rowcolor{steel!30}${\{\vect{p_0}\textit{,}\vect{p_2}\}}$&8 (3)&0 (0)&1&0.1\% (0.0\%)& \quart{9.8}{0.5}{10.1}{100}\\ 
\rowcolor{steel!30}${\{\vect{p_2}\textit{,}\vect{p_0}\}}$&9 (9)&0 (0)&0&0.1\% (0.0\%)& \quart{4.2}{0.0}{4.3}{100}\\ 
\rowcolor{red!10}${\{\vect{p_0}\textit{,}\vect{p_3}\}}$&4 (4)&5 (5)&0&-0.1\% (0.0\%)& \bquart{0.0}{1.1}{0.6}{100}\\ 
\rowcolor{steel!30}${\{\vect{p_3}\textit{,}\vect{p_0}\}}$&6 (5)&3 (1)&0&1.7\% (0.2\%)& \quart{77.0}{8.6}{81.3}{100}\\ 
${\{\vect{p_1}\textit{,}\vect{p_1}\}}$&0 (0)&0 (0)&9&0.0\% (0.0\%)& \tquart{4.2}{0.0}{4.2}{100}\\ 
${\{\vect{p_2}\textit{,}\vect{p_2}\}}$&0 (0)&0 (0)&9&0.0\% (0.0\%)& \tquart{4.2}{0.0}{4.2}{100}\\ 
\rowcolor{steel!30}${\{\vect{p_3}\textit{,}\vect{p_3}\}}$&5 (3)&4 (4)&0&0.3\% (0.1\%)& \quart{17.7}{4.6}{20.0}{100}\\ 
${\{\vect{p_1}\textit{,}\vect{p_2}\}}$&0 (0)&0 (0)&9&0.0\% (0.0\%)& \tquart{4.2}{0.0}{4.2}{100}\\ 
${\{\vect{p_2}\textit{,}\vect{p_1}\}}$&0 (0)&0 (0)&9&0.0\% (0.0\%)& \tquart{4.2}{0.0}{4.2}{100}\\ 
\rowcolor{steel!30}${\{\vect{p_1}\textit{,}\vect{p_3}\}}$&9 (9)&0 (0)&0&0.1\% (0.0\%)& \quart{7.1}{1.0}{7.6}{100}\\ 
\rowcolor{steel!30}${\{\vect{p_3}\textit{,}\vect{p_1}\}}$&9 (5)&0 (0)&0&1.9\% (0.2\%)& \quart{90.2}{9.8}{95.1}{100}\\ 
\rowcolor{steel!30}${\{\vect{p_2}\textit{,}\vect{p_3}\}}$&9 (9)&0 (0)&0&0.1\% (0.0\%)& \quart{6.3}{1.4}{7.0}{100}\\ 
\rowcolor{steel!30}${\{\vect{p_3}\textit{,}\vect{p_2}\}}$&8 (4)&0 (0)&1&1.9\% (0.2\%)& \quart{89.2}{9.8}{94.1}{100}\\

        \end{tabular} &
        &

       \\
            (g). \textsc{Keras/SA}&& (h). \textsc{XGBoost} &&
            \\
       
    \end{tabular}
  \end{adjustbox}
   \end{center}
\end{table*}

%% file: tables/rq3.tex
% CS Table for RQ2
\begin{table*}[t!]
\caption{Comparing \texttt{PS-w} and \texttt{PS-w/o} under unrealistic requirements and aspirations over 100 runs. Formats are the same as Table~\ref{tb:rq1}.}
    \label{tb:rq3}
    \footnotesize
       \setlength{\tabcolsep}{1mm}
  \begin{center}
    \begin{adjustbox}{max width = 1\textwidth}

    \begin{tabular}{cc@{}cc@{}cc}
    \begin{tabular}{lcccll}
        & \textbf{\texttt{PS-w}} & \textbf{\texttt{PS-w/o}} & \textbf{Tie} &  \multicolumn{2}{c}{\textbf{Avg (SE) of A-HV Gain}} \\
            \hline
\rowcolor{steel!30}${\{\vect{p_1}\textit{,}\vect{p_1}\}}$&2 (2)&1 (1)&0&-1.1\% (0.7\%)& \bquart{58.2}{1.1}{58.7}{100}\\ 
\rowcolor{red!10}${\{\vect{p_2}\textit{,}\vect{p_2}\}}$&0 (0)&3 (3)&0&-1.3\% (22.8\%)& \bquart{40.6}{35.6}{58.4}{100}\\ 
\rowcolor{red!10}${\{\vect{p_3}\textit{,}\vect{p_3}\}}$&0 (0)&3 (2)&0&-30.7\% (5.3\%)& \bquart{8.3}{8.2}{12.4}{100}\\ 
\rowcolor{red!10}${\{\vect{p_1}\textit{,}\vect{p_2}\}}$&1 (1)&2 (1)&0&-18.1\% (8.0\%)& \bquart{25.9}{12.5}{32.2}{100}\\ 
\rowcolor{red!10}${\{\vect{p_2}\textit{,}\vect{p_1}\}}$&1 (1)&2 (2)&0&15.7\% (17.2\%)& \quart{71.5}{26.9}{84.9}{100}\\ 
\rowcolor{red!10}${\{\vect{p_1}\textit{,}\vect{p_3}\}}$&1 (1)&2 (1)&0&-24.6\% (4.5\%)& \bquart{18.6}{7.1}{22.1}{100}\\ 
\rowcolor{steel!30}${\{\vect{p_3}\textit{,}\vect{p_1}\}}$&2 (2)&1 (1)&0&0.1\% (0.1\%)& \quart{60.6}{0.1}{60.7}{100}\\ 
\rowcolor{red!10}${\{\vect{p_2}\textit{,}\vect{p_3}\}}$&0 (0)&3 (2)&0&12.3\% (26.1\%)& \quart{59.2}{40.8}{79.6}{100}\\ 
\rowcolor{red!10}${\{\vect{p_3}\textit{,}\vect{p_2}\}}$&0 (0)&3 (2)&0&-36.0\% (5.5\%)& \bquart{0.0}{8.6}{4.3}{100}\\

%&&&&& \squart{-28\%}{21\%}{69\%}{100}\\ 

        \end{tabular} & 
        &
       \begin{tabular}{lcccllll}
                 & \textbf{\texttt{PS-w}} & \textbf{\texttt{PS-w/o}} & \textbf{Tie} &  \multicolumn{2}{c}{\textbf{Avg (SE) of A-HV Gain}}  \\
            \hline
\rowcolor{red!10}${\{\vect{p_1}\textit{,}\vect{p_1}\}}$&1 (1)&2 (2)&0&-3.8\% (0.4\%)& \bquart{0.0}{10.8}{5.4}{100}\\ 
${\{\vect{p_2}\textit{,}\vect{p_2}\}}$&0 (0)&0 (0)&3&0.0\% (0.0\%)& \tquart{100.0}{0.0}{100.0}{100}\\ 
${\{\vect{p_3}\textit{,}\vect{p_3}\}}$&0 (0)&0 (0)&3&0.0\% (0.0\%)& \tquart{100.0}{0.0}{100.0}{100}\\ 
\rowcolor{red!10}${\{\vect{p_1}\textit{,}\vect{p_2}\}}$&1 (1)&2 (2)&0&-2.7\% (0.3\%)& \bquart{28.7}{7.4}{32.4}{100}\\ 
\rowcolor{red!10}${\{\vect{p_2}\textit{,}\vect{p_1}\}}$&1 (1)&2 (2)&0&-0.8\% (0.1\%)& \bquart{80.3}{2.2}{81.5}{100}\\ 
\rowcolor{red!10}${\{\vect{p_1}\textit{,}\vect{p_3}\}}$&1 (1)&2 (2)&0&-2.7\% (0.3\%)& \bquart{28.9}{7.3}{32.6}{100}\\ 
\rowcolor{red!10}${\{\vect{p_3}\textit{,}\vect{p_1}\}}$&1 (1)&2 (2)&0&-0.7\% (0.1\%)& \bquart{81.2}{2.2}{82.3}{100}\\ 
${\{\vect{p_2}\textit{,}\vect{p_3}\}}$&0 (0)&0 (0)&3&0.0\% (0.0\%)& \tquart{100.0}{0.0}{100.0}{100}\\ 
${\{\vect{p_3}\textit{,}\vect{p_2}\}}$&0 (0)&0 (0)&3&0.0\% (0.0\%)& \tquart{100.0}{0.0}{100.0}{100}\\

        \end{tabular} &
        &
       \begin{tabular}{lcccllll}
                  & \textbf{\texttt{PS-w}} & \textbf{\texttt{PS-w/o}} & \textbf{Tie} &  \multicolumn{2}{c}{\textbf{Avg (SE) of A-HV Gain}}  \\
            \hline
\rowcolor{steel!30}${\{\vect{p_1}\textit{,}\vect{p_1}\}}$&2 (1)&1 (1)&0&-0.0\% (0.0\%)& \bquart{44.9}{0.0}{44.9}{100}\\ 
\rowcolor{steel!30}${\{\vect{p_2}\textit{,}\vect{p_2}\}}$&2 (2)&1 (1)&0&70.0\% (29.8\%)& \quart{67.0}{11.9}{72.9}{100}\\ 
\rowcolor{red!10}${\{\vect{p_3}\textit{,}\vect{p_3}\}}$&1 (1)&2 (1)&0&-95.5\% (224.5\%)& \bquart{0.0}{89.9}{6.7}{100}\\ 
\rowcolor{steel!30}${\{\vect{p_1}\textit{,}\vect{p_2}\}}$&2 (0)&1 (1)&0&53.0\% (30.4\%)& \quart{60.1}{12.2}{66.2}{100}\\ 
\rowcolor{red!10}${\{\vect{p_2}\textit{,}\vect{p_1}\}}$&1 (1)&2 (2)&0&-90.4\% (218.8\%)& \bquart{1.2}{87.6}{8.8}{100}\\ 
\rowcolor{steel!30}${\{\vect{p_1}\textit{,}\vect{p_3}\}}$&2 (2)&1 (1)&0&119.1\% (37.0\%)& \quart{85.2}{14.8}{92.6}{100}\\ 
\rowcolor{red!10}${\{\vect{p_3}\textit{,}\vect{p_1}\}}$&1 (1)&2 (2)&0&-52.7\% (136.9\%)& \bquart{17.5}{54.8}{23.8}{100}\\ 
\rowcolor{steel!30}${\{\vect{p_2}\textit{,}\vect{p_3}\}}$&2 (1)&1 (1)&0&74.2\% (30.0\%)& \quart{68.6}{12.0}{74.6}{100}\\ 
\rowcolor{steel!30}${\{\vect{p_3}\textit{,}\vect{p_2}\}}$&2 (0)&1 (1)&0&21.2\% (21.7\%)& \quart{49.1}{8.7}{53.4}{100}\\

        \end{tabular}  \\
            (a). \textsc{Trimesh}&& (b). \textsc{x264} && (c). \textsc{Storm/WC}

        \\
        
        \\
        
       \begin{tabular}{lcccll}
        & \textbf{\texttt{PS-w}} & \textbf{\texttt{PS-w/o}} & \textbf{Tie} &  \multicolumn{2}{c}{\textbf{Avg (SE) of A-HV Gain}} \\
            \hline
\rowcolor{steel!30}${\{\vect{p_1}\textit{,}\vect{p_1}\}}$&2 (2)&1 (1)&0&0.0\% (0.0\%)& \quart{50.0}{0.0}{50.0}{100}\\ 
\rowcolor{red!10}${\{\vect{p_2}\textit{,}\vect{p_2}\}}$&1 (1)&2 (2)&0&-243.4\% (541.0\%)& \bquart{0.0}{100.0}{5.0}{100}\\ 
\rowcolor{red!10}${\{\vect{p_3}\textit{,}\vect{p_3}\}}$&1 (1)&2 (2)&0&-147.7\% (336.0\%)& \bquart{19.0}{62.1}{22.7}{100}\\ 
\rowcolor{red!10}${\{\vect{p_1}\textit{,}\vect{p_2}\}}$&1 (1)&2 (2)&0&6.2\% (23.3\%)& \quart{49.0}{4.3}{51.2}{100}\\ 
\rowcolor{steel!30}${\{\vect{p_2}\textit{,}\vect{p_1}\}}$&2 (2)&1 (1)&0&195.3\% (44.2\%)& \quart{82.0}{8.2}{86.1}{100}\\ 
\rowcolor{red!10}${\{\vect{p_1}\textit{,}\vect{p_3}\}}$&1 (1)&2 (2)&0&-10.3\% (19.1\%)& \bquart{46.3}{3.5}{48.1}{100}\\ 
\rowcolor{steel!30}${\{\vect{p_3}\textit{,}\vect{p_1}\}}$&2 (1)&1 (1)&0&81.9\% (28.0\%)& \quart{62.6}{5.2}{65.1}{100}\\ 
\rowcolor{red!10}${\{\vect{p_2}\textit{,}\vect{p_3}\}}$&1 (1)&2 (2)&0&-214.2\% (478.1\%)& \bquart{5.8}{88.4}{10.4}{100}\\ 
\rowcolor{red!10}${\{\vect{p_3}\textit{,}\vect{p_2}\}}$&1 (1)&2 (2)&0&-139.8\% (320.2\%)& \bquart{20.4}{59.2}{24.2}{100}\\

%&&&&& \squart{-28\%}{21\%}{69\%}{100}\\ 

        \end{tabular} & 
        &
       \begin{tabular}{lcccllll}
                 & \textbf{\texttt{PS-w}} & \textbf{\texttt{PS-w/o}} & \textbf{Tie} &  \multicolumn{2}{c}{\textbf{Avg (SE) of A-HV Gain}}  \\
            \hline
\rowcolor{red!10}${\{\vect{p_1}\textit{,}\vect{p_1}\}}$&1 (1)&2 (2)&0&-0.0\% (0.0\%)& \bquart{46.3}{0.0}{46.3}{100}\\ 
\rowcolor{red!10}${\{\vect{p_2}\textit{,}\vect{p_2}\}}$&0 (0)&3 (3)&0&-56.6\% (5.0\%)& \bquart{0.9}{3.8}{2.8}{100}\\ 
\rowcolor{red!10}${\{\vect{p_3}\textit{,}\vect{p_3}\}}$&0 (0)&3 (3)&0&-42.4\% (5.1\%)& \bquart{11.8}{3.9}{13.8}{100}\\ 
\rowcolor{red!10}${\{\vect{p_1}\textit{,}\vect{p_2}\}}$&1 (1)&2 (2)&0&1.4\% (20.4\%)& \quart{39.6}{15.6}{47.4}{100}\\ 
\rowcolor{steel!30}${\{\vect{p_2}\textit{,}\vect{p_1}\}}$&2 (2)&1 (1)&0&55.8\% (28.3\%)& \quart{78.3}{21.7}{89.1}{100}\\ 
\rowcolor{red!10}${\{\vect{p_1}\textit{,}\vect{p_3}\}}$&0 (0)&3 (2)&0&-9.7\% (10.8\%)& \bquart{34.7}{8.3}{38.8}{100}\\ 
\rowcolor{red!10}${\{\vect{p_3}\textit{,}\vect{p_1}\}}$&1 (0)&2 (2)&0&-26.4\% (4.6\%)& \bquart{24.3}{3.5}{26.0}{100}\\ 
\rowcolor{red!10}${\{\vect{p_2}\textit{,}\vect{p_3}\}}$&0 (0)&3 (3)&0&-45.4\% (5.1\%)& \bquart{9.5}{3.9}{11.5}{100}\\ 
\rowcolor{red!10}${\{\vect{p_3}\textit{,}\vect{p_2}\}}$&0 (0)&3 (3)&0&-57.9\% (4.8\%)& \bquart{0.0}{3.7}{1.8}{100}\\

        \end{tabular} &
        &
       \begin{tabular}{lcccllll}
                  & \textbf{\texttt{PS-w}} & \textbf{\texttt{PS-w/o}} & \textbf{Tie} &  \multicolumn{2}{c}{\textbf{Avg (SE) of A-HV Gain}}  \\
            \hline

\rowcolor{steel!30}${\{\vect{p_1}\textit{,}\vect{p_1}\}}$&2 (1)&1 (1)&0&0.1\% (0.1\%)& \quart{42.7}{0.0}{42.7}{100}\\ 
\rowcolor{red!10}${\{\vect{p_2}\textit{,}\vect{p_2}\}}$&1 (1)&2 (2)&0&-47.3\% (122.7\%)& \bquart{0.0}{85.2}{9.8}{100}\\ 
\rowcolor{red!10}${\{\vect{p_3}\textit{,}\vect{p_3}\}}$&1 (1)&2 (2)&0&-33.6\% (89.6\%)& \bquart{11.5}{62.2}{19.3}{100}\\ 
\rowcolor{red!10}${\{\vect{p_1}\textit{,}\vect{p_2}\}}$&0 (0)&3 (3)&0&-1.7\% (26.9\%)& \bquart{33.3}{18.7}{41.4}{100}\\ 
\rowcolor{steel!30}${\{\vect{p_2}\textit{,}\vect{p_1}\}}$&2 (1)&1 (0)&0&53.3\% (24.2\%)& \quart{71.2}{16.8}{79.6}{100}\\ 
\rowcolor{red!10}${\{\vect{p_1}\textit{,}\vect{p_3}\}}$&1 (1)&2 (2)&0&-2.4\% (18.5\%)& \bquart{34.5}{12.9}{40.9}{100}\\ 
\rowcolor{steel!30}${\{\vect{p_3}\textit{,}\vect{p_1}\}}$&2 (1)&1 (0)&0&69.8\% (25.8\%)& \quart{82.1}{17.9}{91.1}{100}\\ 
\rowcolor{steel!30}${\{\vect{p_2}\textit{,}\vect{p_3}\}}$&2 (1)&1 (1)&0&5.9\% (11.0\%)& \quart{42.9}{7.7}{46.7}{100}\\ 
\rowcolor{red!10}${\{\vect{p_3}\textit{,}\vect{p_2}\}}$&1 (1)&2 (2)&0&-33.3\% (93.2\%)& \bquart{10.2}{64.7}{19.5}{100}\\

        \end{tabular}  \\
            (d). \textsc{Storm/RS}&& (e). \textsc{Keras/Adiac} && (f). \textsc{Keras/DSR}
            \\
            \\

       \begin{tabular}{lcccll}
        & \textbf{\texttt{PS-w}} & \textbf{\texttt{PS-w/o}} & \textbf{Tie} &  \multicolumn{2}{c}{\textbf{Avg (SE) of A-HV Gain}} \\
            \hline
\rowcolor{steel!30}${\{\vect{p_1}\textit{,}\vect{p_1}\}}$&3 (2)&0 (0)&0&-1.0\% (0.0\%)& \bquart{82.1}{0.1}{82.1}{100}\\ 
\rowcolor{red!10}${\{\vect{p_2}\textit{,}\vect{p_2}\}}$&0 (0)&3 (3)&0&-64.6\% (4.6\%)& \bquart{3.5}{5.4}{6.2}{100}\\ 
\rowcolor{red!10}${\{\vect{p_3}\textit{,}\vect{p_3}\}}$&0 (0)&3 (3)&0&-39.1\% (4.7\%)& \bquart{33.3}{5.6}{36.1}{100}\\ 
\rowcolor{red!10}${\{\vect{p_1}\textit{,}\vect{p_2}\}}$&0 (0)&3 (3)&0&6.7\% (17.0\%)& \quart{80.0}{20.0}{90.0}{100}\\ 
\rowcolor{red!10}${\{\vect{p_2}\textit{,}\vect{p_1}\}}$&0 (0)&3 (3)&0&-46.2\% (5.2\%)& \bquart{24.7}{6.1}{27.8}{100}\\ 
\rowcolor{red!10}${\{\vect{p_1}\textit{,}\vect{p_3}\}}$&0 (0)&3 (3)&0&4.9\% (15.3\%)& \quart{78.9}{18.0}{87.9}{100}\\ 
\rowcolor{red!10}${\{\vect{p_3}\textit{,}\vect{p_1}\}}$&1 (0)&2 (2)&0&-26.6\% (4.3\%)& \bquart{48.3}{5.1}{50.9}{100}\\ 
\rowcolor{red!10}${\{\vect{p_2}\textit{,}\vect{p_3}\}}$&0 (0)&3 (3)&0&-67.7\% (4.3\%)& \bquart{0.0}{5.0}{2.5}{100}\\ 
\rowcolor{red!10}${\{\vect{p_3}\textit{,}\vect{p_2}\}}$&0 (0)&3 (3)&0&-38.3\% (5.1\%)& \bquart{34.1}{6.0}{37.1}{100}\\

%&&&&& \squart{-28\%}{21\%}{69\%}{100}\\ 

        \end{tabular} & 
        &
       \begin{tabular}{lcccllll}
                 & \textbf{\texttt{PS-w}} & \textbf{\texttt{PS-w/o}} & \textbf{Tie} &  \multicolumn{2}{c}{\textbf{Avg (SE) of A-HV Gain}}  \\
            \hline
\rowcolor{steel!30}${\{\vect{p_1}\textit{,}\vect{p_1}\}}$&2 (1)&1 (1)&0&-1.0\% (0.0\%)& \bquart{36.1}{0.0}{36.1}{100}\\ 
\rowcolor{red!10}${\{\vect{p_2}\textit{,}\vect{p_2}\}}$&0 (0)&3 (3)&0&-58.3\% (4.4\%)& \bquart{5.0}{2.3}{6.2}{100}\\ 
\rowcolor{red!10}${\{\vect{p_3}\textit{,}\vect{p_3}\}}$&0 (0)&3 (3)&0&-68.4\% (3.8\%)& \bquart{0.0}{1.9}{1.0}{100}\\ 
\rowcolor{red!10}${\{\vect{p_1}\textit{,}\vect{p_2}\}}$&0 (0)&3 (3)&0&-51.0\% (5.1\%)& \bquart{8.6}{2.6}{9.9}{100}\\ 
\rowcolor{red!10}${\{\vect{p_2}\textit{,}\vect{p_1}\}}$&1 (0)&2 (1)&0&106.3\% (36.4\%)& \quart{81.3}{18.7}{90.7}{100}\\ 
\rowcolor{red!10}${\{\vect{p_1}\textit{,}\vect{p_3}\}}$&0 (0)&3 (3)&0&-60.3\% (3.9\%)& \bquart{4.1}{2.0}{5.1}{100}\\ 
\rowcolor{steel!30}${\{\vect{p_3}\textit{,}\vect{p_1}\}}$&2 (0)&1 (1)&0&-0.0\% (0.0\%)& \bquart{36.1}{0.0}{36.1}{100}\\ 
\rowcolor{red!10}${\{\vect{p_2}\textit{,}\vect{p_3}\}}$&0 (0)&3 (3)&0&-63.2\% (3.7\%)& \bquart{2.7}{1.9}{3.7}{100}\\ 
\rowcolor{red!10}${\{\vect{p_3}\textit{,}\vect{p_2}\}}$&0 (0)&3 (3)&0&-64.8\% (4.7\%)& \bquart{1.6}{2.4}{2.8}{100}\\

        \end{tabular}  &
        &

       \\
            (g). \textsc{Keras/SA}&& (h). \textsc{XGBoost} && 
            \\
       
    \end{tabular}
  \end{adjustbox}
   \end{center}
\end{table*}

%% file: lessons.tex
\section{Lessons Learned}
\label{sec:lessons}

In this section, we discuss how our findings can be useful for the practitioners in the field in light of the lessons learned and future opportunities discovered.

\begin{implbox}
   \raisebox{-0.05ex}{}
   \tcblower
   \textit{\textbf{Lesson 1:} The choice on whether to exploit aspirations for guiding the tuning is primarily dependent on their realism.}
\end{implbox}

It is interesting to find that we cannot draw the conclusion to choose between \texttt{PS-w} and \texttt{PS-w/o} arbitrarily for software configuration tuning with two performance objectives, as opposed to what has been overwhelmingly assumed in existing work. Instead, from \textbf{RQ1}, \textbf{RQ3}, and \textbf{RQ4}, we discovered that the realism of the given aspirations is crucial to the choice: \texttt{PS-w} is more beneficial for realistic aspirations while \texttt{PS-w/o} is safer when the aspirations are unrealistic (given that the tuning budget is also sufficient). This raises the importance of understanding whether the given requirements and aspirations can be realistic, or the assumption therein, prior to choosing the right optimization model for tuning software configuration with two performance objectives.

\begin{implbox}
   \raisebox{-0.05ex}{}
   \tcblower
   \textit{\textbf{Lesson 2:} Little combinations of patterns can change the decision on whether to incorporate aspiration in the tuning, but it can influence the benefit/detriment of aspiration-guided tuning.}
\end{implbox}

Although from \textbf{RQ1}, we noticed that the benefits of \texttt{PS-w} is blurred when the given combination of patterns contain $\vect{p_1}$ and/or $\vect{p_0}$ only, this does not change the decision as \texttt{PS-w} remains outperform its \texttt{PS-w/o} counterpart. The only definitive case is when the aspirations are unrealistic, \texttt{PS-w} should be chosen under a patterns of $\{\vect{p_1}, \vect{p_1}\}$. Therefore, we envisage that the sensitivity of given patterns to the choice between \texttt{PS-w} and \texttt{PS-w/o} is marginal and we have discovered other more important factors. However, we do see that the extent of improvement/degradation from \textit{PS-w} can be sensitive to the given patterns.

\begin{implbox}
   \raisebox{-0.05ex}{}
   \tcblower
   \textit{\textbf{Lesson 3:} The positions of realistic aspiration space in the objective space can largely affect the benefits brought by considering aspirations within tuning, but it is less likely to influence the choice.}
\end{implbox}

An unexpected discovery from \textbf{RQ2} is that, when given realistic aspirations, the position of the aspiration space can largely influence the benefits of \texttt{PS-w}. While this is unlikely to affect the choice between \texttt{PS-w} and \texttt{PS-w/o}, it does raise the need to systematically analyze the correlation between the aspiration space and the configuration landscape of the system, particularly on the likelihood of covering some difficult local optima and their implication.

\begin{implbox}
   \raisebox{-0.05ex}{}
   \tcblower
   \textit{\textbf{Lesson 4:} The given tuning budget has marginal impact to the choice when the aspirations are realistic. However, it can be an important factor to consider under unrealistic aspirations.}
\end{implbox}

According to \textbf{RQ4} we have also revealed that, given realistic aspirations, the choice between \texttt{PS-w} and \texttt{PS-w/o} is marginally sensitive to the tuning budget, but it can be influenced by the budget when the aspirations are unrealistic. This adds an extra layer of consideration for unrealistic aspirations. In this case, what we observed, in general, is that for a small tuning budget, the benefit of  \texttt{PS-w/o} is much less justified, hence using either of the two optimization models may not lead to significantly different results. However, given sufficient budget, \texttt{PS-w/o} is likely to dominate its \texttt{PS-w} counterparts. Unfortunately, with the current evidence, it remains very difficult to precisely quantify how ``small'' or ``large'' the tuning budget is required to make such a distinction.

The above lessons not only reveal the important factors for the practitioners to consider when choosing \texttt{PS-w} and \texttt{PS-w/o} for bi-objective software configuration tuning but also hint at a few future research opportunities in this regard. These are:

\begin{itemize}

\item \textbf{Landscape analysis for configurable software systems:} We have found that the realism of aspiration space, its position in the objective landscape, and tuning budget can be the key factors to consider when choosing between \texttt{PS-w} and \texttt{PS-w/o}. All of those are relevant to the landscape analysis of the configurable system itself. Indeed, by systematically analyzing any collected data, we are able to obtain more knowledge about the above factors, and hence make more informed decisions on whether to incorporate requirements into the tuning.

\item \textbf{Requirement-robust optimizer for configuration tuning:} The realism of the aspiration is certainly the key factor in the choice between \texttt{PS-w} and \texttt{PS-w/o}. However, it may not be always possible to obtain such knowledge in advance, leaving uncertainty to the decision. In this regard, it would be desirable to combine the strength of \texttt{PS-w} and \texttt{PS-w/o} to design an optimizer that is robust to such an uncertainty in the requirements. Again, the landscape analysis from the previous opportunity can provide insights into the designs.

\item \textbf{Rigorous analysis of requirement patterns and their relationships to the tuning:} Although we see little implication of the requirement patterns to the choice between \texttt{PS-w} and \texttt{PS-w/o}, it is important to better understand why they work more diversely on some of the patterns and how exactly they can affect the performance of \texttt{PS-w}. In fact, on the theoretical side, the quantification from Section~\ref{sec:req} provides the foundation of theoretical reasoning for switching between patterns, which is important in the topic of requirement relaxation. \rev{For example, this can be achieved in two aspects:}

\begin{itemize}

\item With the quantification of the patterns, one can formally show the relations between them. For example, since all points in $\vect{p_1}$ have a higher satisficing value than those of $\vect{p_3}$, we can say that $\vect{p_1}$ is a ``relaxed'' form of $\vect{p_3}$.

\item \rev{Similarly, we can quantify the relationships between a pattern with two different aspiration levels.}

\end{itemize}

\rev{With the above understanding, we allow the software engineers to achieve more explainability in terms of the given requirements during the tuning. For example, once the tuning completes, one would know how to relax or tighten the requirements, such that the most preferred configuration can be found under the requirements. This can be a unified process that combines both requirement negotiations and the tuning itself.}

\item \rev{\textbf{Interactive configuration tuning:} On the empirical side, our findings provide a few insights on what to do under different circumstances during interactive tuning. For example, if the software engineers find that the tuning never (or rarely) produces configurations that satisfy the requirements/aspirations under \texttt{PS-w}, then one can immediately switch to \texttt{PS-w/o} instead before concerns about the suitability of the underlying optimizer. Similarly, one can influence the results produced by \texttt{PS-w} (or \texttt{PS-w/o}) by changing the position of the aspiration space.} 

\end{itemize}

%% file: threats.tex
\section{Threats to Validity}
\label{sec:threats}

As with other empirical studies in software engineering, our work may contain threats to \textbf{construct validity} in the following aspects:

\begin{itemize}

\item \textbf{Metric:} Pareto search produces a set of configurations, and thus the comparisons need to work on a set rather than a single configuration. We used HV, which is a comprehensive metric for evaluating solution sets, following the methodology proposed by Li \textit{et al.}~\cite{9252185}.  Since there can be different given sets of requirements with aspirations, the configuration sets ought to be compared under such a scenario. To that end, we extend the HV to explicitly consider the patterns of requirements, as discussed in Section~\ref{sec:metric}.

\item \textbf{Statistics:} The stochastic nature of the Pareto optimizers can raise threats to the stability of results. To mitigate such, we repeat the experiments 100 runs and use Wilcoxon test along with $\hat{A}_{12}$ to verify all pairwise comparisons. All the above methods have been recommended and widely used for Software Engineering research~\cite{DBLP:conf/icse/ArcuriB11}.
 
 %For multiple comparisons, we use Scott-Knott test to statistically rank them.
 
\end{itemize}

Two factors may form threats to \textbf{internal validity} in our study:

\begin{itemize}

    \item \textbf{Tuning budget:} Given the size of our study, we set a one-hour budget for each case, which is a common setting for expensive problems in SBSE~\cite{DBLP:conf/icse/LiX0WT20}. To mitigate the interference of our experiments, this is then converted into the number of unique measurements following systematic steps (Section~\ref{sec:budget}). We have also analyzed the trajectories of A-HV, in Section~\ref{sec:rq4}, showing what would happen if a smaller budget is used. Admittedly, investigating a larger tuning budget may affect some of the results, but confirming this would need even more computational resources and time (due to the expensive tuning), which we will plan as part of future work.

    \item \textbf{Optimizer setting:} In this work, we follow what has been shown to be effective for a SBSE problem in the literature, as our aim is to compare the most common practices. The only part we could not have found for sure is the population size, which is highly problem-dependent. To tackle this, we have followed carefully designed criteria (Section~\ref{sec:settings}) that strike a balance between reasonable convergence and the time required under the tuning budget. However, we do agree that exploring alternative parameter settings can be a thread that requires further exploration, which we leave as part of future work.
    
    %We have found that the parameter settings tend to be appropriate and the sensitivity of Pareto search to the parameters is marginal, as discussed in Section~\ref{sec:sen-p}.

\end{itemize}

Threats to \textbf{external validity} can come from various sources, including:

\begin{itemize}
    \item \textbf{Software systems:} In this work, we select the eight most representative systems/environments from existing work on software configuration tuning based on carefully codified rules (Section~\ref{sec:sys}). Those subject systems come from diverse domains and with different scales, performance objectives, and search spaces. \rev{A worth noting point is that the requirements extracted include those for more complex systems, such as Cyber-Physical systems, while the subjects we examined are mainly software systems. This does not severely invalid our conclusion because the extracted implication and patterns are rather generic such that they can be applied to different cases while there exist some performance attributes that are of relevance to a wide range of systems, e.g., latency- and throughput-related requirements (with different aspiration levels)~\cite{nair2018finding}. Nonetheless, we agree that this list of the studied systems is not exhaustive and we may miss some particular situations that can only become clear for more complex systems. Experimenting with more systems that are of diverse types may prove fruitful. A relevant point is that we did not examine our results on highly complex software systems that cut across the software and hardware layers. In those cases, the interaction between cross-layered configuration options can be more complex, leading to some different configuration landscapes~\cite{DBLP:conf/eurosys/IqbalKJRJ22}. Therefore, examining those highly complex systems may provide new insights and further consolidate our findings.}
    
    \rev{It is worth noting that it can be particularly attractive to relate the results with respect to the different types of software systems. However, unfortunately, we have not yet observed consistent patterns in the results according to the domain of systems, hence unable to draw a general conclusion thereupon. This can be attributed to two reasons:}

\begin{itemize}

\item \rev{The workload and benchmark under which each of the systems runs are rather different, creating a distinct configuration landscape.}

\item \rev{Because of the above, the appropriate aspirations (levels) used are also different even for systems that are of the same domain.}

\end{itemize}

\rev{Again, using even more software systems may help us to achieve such, which we certainly plan to do for future work. However, this does not invalidate the conclusions drawn regarding the comparison between \texttt{PS-w} and \texttt{PS-w/o}.}

       \item \rev{\textbf{Configuration options:} The discretization level of each configuration option can have a non-trivial impact on the tuning. In this work, we use exactly the same configuration options and their values as used in previous work~\cite{DBLP:conf/sigsoft/JamshidiVKS18,nair2018finding,DBLP:conf/mascots/JamshidiC16,DBLP:conf/sigsoft/0001Chen}. However, it is necessary to note that changing the discretization level may disclose new insights, which we will seek to investigate as part of future work.}

     \item \textbf{Requirement patterns and aspiration space:} To emulate real-world requirement scenarios, as shown in Section~\ref{sec:req}, we capture the implications and how they are quantified by surveying the relevant datasets and papers. This has enabled us to concentrate on four patterns that cover a wide range of situations, leading to 15 combinations of the patterns. As for the aspiration space, we cover both realistic and unrealistic aspirations, and for the former, we set three types of aspiration space including two skewed spaces and a more balanced one. Yet, admittedly, unintentionally ignored cases are always possible.

    \item \textbf{Optimizers:} In this study, three common Pareto optimizers based on evolutionary search are used, 
    each of which is a distinct representative of its own kind. \rev{Admittedly, 
    	there are other popular multi-objective optimization approaches used in SBSE, 
    	such as exact methods and Bayesian optimization methods.}
    	
    \rev{When the given multi-objective optimization problems are of special characteristics 
    	(e.g., linearity in both objective functions and constraints and the scale is small or moderate),
    	then exact methods (e.g., integer linear programming) can be very good choices, 
    	where solutions of the Pareto front can be iteratively found by specifying different weights or desirable/tolerant values.
    	Such optimization problems have been commonly seen in the next release problem, 
    	and well-established exact methods, such as $\epsilon$-constraint and augmented Tchebycheff methods, 
    	have shown promising results~\cite{Dominguez2019,Veerapen2015}.}
    
    \rev{Another popular kind of optimizers, 
    		particularly used in software configuration tuning,
    	is Bayesian optimization.
   		Recently, 
   		there are multi-objective Bayesian optimizers (e.g.,~\cite{Iqbal2020}) 
   		which search for the whole Pareto front of the problem.
  		Compared to evolutionary algorithms, 
  		Bayesian optimizers are usually more sample efficient~\cite{DBLP:conf/mascots/JamshidiC16,Jamshidi2017}.
  		However, 
  		one issue with such approaches is that they may not be as straightforward as evolutionary algorithms to incorporate the stakeholders' performance aspirations 
  		(e.g., for Bayesian optimization this may need a careful design of the acquisition function).
  		Different incorporation ideas may lead to different results, 
  		thus likely affecting the reliability of the conclusions drawn from the direct comparison between Pareto search with and without performance aspirations.}
    
   	\rev{In addition, 
   		it is necessary to point out that the conclusions drawn from multi-objective evolutionary algorithms 
   		may not apply to other optimization approaches.
   		Optimizers, 
   		which are guided ``heavily'' by the aspirations, 
   		may find them quickly if they are realistic, 
   		but may end up with undesirable solutions if unrealistic. 
   		Optimizers, 
   		which can strike a good balance between exploitation and exploration (under limited budgets) like Bayesian optimization, 
   		may bring different results, 
   		though it depends on the incorporation of the aspirations in the optimization process. 
  		Consequently, 
  		it is desirable to investigate different optimization approaches to study the generalizability of our findings,
  		particularly those with fruitful theoretical results 
  		(e.g., convergence rate regarding the response surface's smoothness as well as the regret bounds in Bayesian optimization), 
  		which may help support our empirical conclusions.
  	    This will be an important part of our future work.}

    \item \textbf{Number of objectives:} Our study covers the case of two performance objectives for software configuration tuning, which, as we have found from our review in Section~\ref{sec:req}, tends to be the most common situation when multiple objectives are considered. The results may not be generalizable to higher dimension cases of the objectives though. Extending the study to more objectives can be part of future work, but there would also be exponentially increasing factors to consider, e.g., the number of pattern combinations. Our results from this work serve as the very first step to raising the importance of studying whether to use aspiration to guide software configuration tuning with more than one performance objective.
    
    %However, regardless whichever optimization model used, we anticipate that the benefits of Pareto search would be blurred. This is because it has been well studied in the evolutionary computation community that the performance of many well-established multi-objective evolutionary algorithms falls rapidly with the increase of the number of objectives,particularly for Pareto-based search algorithms (e.g., NSGA-II)which only use the Pareto dominance relation to distinguish between solutions with respect to their convergence~\cite{Wagner2007,Li2013b}.Recent studies even show that mainstream algorithms like NSGA-II, MOEA/D, and IBEA even completely fail on some four-objective problems~\cite{Li2018a}. 

\end{itemize}

Overall, the above settings have provided us with more than 1,000 cases to generalize our findings in this study.

%% file: related.tex
\section{Related Work}
\label{sec:related}

Here, we discuss the related work in light of the purpose of our empirical study.

\subsection{\texttt{PS-w/o} for Software Configuration Tuning}

Search-based approaches for software configuration tuning have been commonly studied under a wide range of optimizers, such as random search~\cite{DBLP:conf/sigsoft/OhBMS17}, hill climbing~\cite{DBLP:conf/www/XiLRXZ04}, genetic algorithm~\cite{DBLP:conf/sigsoft/ShahbazianKBM20}, and ant colony optimization~\cite{Chen2017Self}. In the presence of more than one objective, Pareto search has been shown to be highly effective. Among others, Chen \textit{et al.}~\cite{Chen2018FEMOSAA} and Singh \textit{et al.}~\cite{DBLP:conf/wosp/SinghBSH16} leverage different multi-objective evolutionary algorithms to search the Pareto optimal configurations. Nair \textit{et al.}~\cite{nair2018finding} also aim for the same, but their approach applies Bayesian optimization wherein the two performance objectives are handled similarly to MOEA/D. More recently, Zhu \textit{et al.}~\cite{DBLP:conf/icse/GaoZ0LY21} also propose an extended Bayesian optimization approach to reach a given performance aspiration while considering both latency and resource consumption; however, such information has not been used to explicitly guide the search.

The above work has one thing in common: they have ignored the aspirations in the search process. The assumption therein is that the concept of optimization can obtain whatever best configuration that satisfies any given patterns of requirements.

\subsection{\texttt{PS-w} for Software Configuration Tuning}

In contrast, Calinescu \textit{et al.}~\cite{Calinescu2017Designing,DBLP:journals/ase/GerasimouCT18} explicitly quantifies aspirations as part of the objectives (as $\vect{p_1}$) to guide the Pareto search. Martens \textit{et al.}~\cite{DBLP:conf/wosp/MartensKBR10} also bear similar idea, but their pattern matches with $\vect{p_3}$. Ghanbari \textit{et al.}~\cite{DBLP:journals/fgcs/GhanbariSLI12} has also been relying aspiration to guide the tuning, and they assume a smoother curve over the requirement patterns. However, there has been no study that justifies the importance of aspirations in guiding the Pareto search for bi-objective software configuration tuning. Indeed, a recent discussion paper from Fekry \textit{et al.}~\cite{DBLP:conf/icdcs/FekryCPRH19} commented that studying the aspirations
for guiding the optimizers and measuring its effectiveness is an important future challenge for
software configuration tuning.

These are typical examples of the \texttt{PS-w} optimization model, such that the performance requirements are precisely quantified as part of the search and tuning process. It is also worth noting that the actual requirements patterns used can vary depending on the assumption, and none of the existing work for \texttt{PS-w} has considered all the patterns we summarized in our study.

\subsection{Tuning with or without Surrogate}

From another perspective, existing search-based approaches, regardless the number of performance objectives considered, can be classified as model-based (e.g., \texttt{EvoChecker}~\cite{DBLP:journals/ase/GerasimouCT18}, \texttt{FLASH}~\cite{nair2018finding}, and \texttt{BOCA}~\cite{DBLP:conf/icse/0003XC021}) and measurement-based (e.g., \texttt{FEMOSAA}~\cite{Chen2018FEMOSAA}, \texttt{eQual}~\cite{DBLP:conf/sigsoft/ShahbazianKBM20}, and \texttt{Plato}~\cite{DBLP:conf/icac/RamirezKCM09}), by which the former relies on surrogate models~\cite{DBLP:journals/tse/ChenB17} to guide the search while the latter do so via direct measurements from the software. The key difference between those two are the landscape upon which the search is conducted: the measurement-based approaches do so directly on the configuration landscape while the model-based approaches searches in a surrogate landscape, which is an approximation of the true configuration landscape. 

%The model-based methods seek to mitigate the expensiveness challenge of software configuration tuning, but doing so may lead to compromise in the accuracy.

This work focuses on tuning by directly measuring the systems without using the surrogate, because for the following reasons:

\begin{itemize}
    \item We seek to avoid the noises caused by the surrogate models as they would inevitably introduce errors, which, as demonstrated by Zhu \textit{et al.}~\cite{DBLP:conf/cloud/ZhuLGBMLSY17}, can severely affect the search and tuning behavior.
    \item In fact, we treat these two categories as complementary rather than alternative. For example, a measurement-based approach that works well can be also applied in Bayesian optimization, which is model-based, to search for the acquisition. This means that our findings on whether aspirations matter are also applicable therein.
    
    %of comparing \texttt{PS-w} and \texttt{PS-w/o} on the measurement-based manner can be also generalized to the model-based case.  
\end{itemize}

%For the empirical comparison between Pareto search with and without aspiration, 

\subsection{General Multi-objective Optimization}

Conceptually, 
our work can be relevant to the theme of preference-driven multi-objective optimization. 
In this regard, as surveyed by Wang \textit{et al.}~\cite{wang2017mini}, 
the preferences on the objective values (i.e., the performance requirements in this work) can form the following categories:

\begin{itemize}

\item \textbf{Weights:} In this case, 
a weight vector representing the relative importance of the objectives is given. 
Most commonly, this would convert the multi-objective problem into a single one via some form of aggregation, e.g., weighted sum. Indeed, both the weights and the aspirations in this work are some forms of preferences. However, they are very different because the weights need to be specified in-between the performance objectives, representing an explicit trade-off~\cite{DBLP:conf/sigsoft/ShahbazianKBM20}. The aspiration level, in contrast, serves as the expectation for a single objective, which is often easier to specify, and no explicit trade-off is required. \rev{In fact, using the weight can be thought as a special case of the kind of preference we consider in this work: in the case of weight, the best configuration is typically a particular point on the Pareto front. In contrast, under the requirements and aspirations considered, there are often more than one best point, including a proportion of the points on the Pareto front.} Interestingly, given a set of weights, it has been shown that the Pareto search (which runs without the weights) can generally find better configurations than the search guided by the weights~\cite{10.1145/3514233}.

\item \textbf{Objective relation:} This refers to the case where a full or partial rank of the objectives has been provided. For example, one may prefer to satisfy the requirement of throughput first before considering latency. The combinations of patterns and aspirations in this work, in contrast, have no direct ranking between the objectives.

\item \textbf{Area of objective space:} One may provide a rough notion of a particularly preferred area in the objective space, e.g., knee points or extreme points. Compared with the combinations of patterns from Section~\ref{sec:req}, this preference is vague --- the knee or extreme points are relative among the solutions found, while the patterns are still guided by a clearly defined aspiration space.

\item \textbf{Reference points~\cite{DBLP:conf/cec/YuJO19,9066927}:} Here, a vector representing the expectations of the objectives are given. In this regard, 
the concept is indeed similar to the aspiration levels we discussed in this work. 
However, a major difference is that reference point-based multi-objective optimization is always under the assumption that Pareto optimality needs to be considered first, i.e., the Pareto optimal solutions close to the reference point (along certain direction specified by the decision-maker) are preferred~\cite{Auger2009}. 
Moreover, 
additional parameters are required to specify the spread of the preferred solutions~\cite{9066927}. 
In contrast, 
the combinations of patterns we discovered for software configuration tuning can be rather different from the above, 
as the solutions in (outside) the aspiration space may be equally preferred (unpreferred) 
while do not favor the solutions close to the aspiration vector. 
Further, no other parameters are needed besides the aspiration levels.

%From that perspective, the notion of reference points in general multi-objective optimization can be regarded as a specific instance of the patterns we summarized here.

\end{itemize}

Therefore, all the above forms of preferences differ from the requirement aspirations and patterns used for software configuration tuning, as we summarized in Section~\ref{sec:req}. This is important as the summarized patterns are derived from empirical findings for the characteristics of the problems --- they may not be generalizable to other problems but are significant to software configuration tuning.

An empirical study on the importance of considering preferences in the search also exists from the general optimization community~\cite{9066927,Yu2016}. However, they differ from our work in two aspects: 

\begin{itemize}
    \item They focus on reference points, which, as we discussed, are rather different from the patterns considered in this work.
   
    \item The studied optimizers are specialized algorithms that take specific forms of preferences into account. In contrast, our study considers classic optimizers which are equipped with the requirement patterns to guide the search, as what has been commonly used for software configuration tuning. 
    
    %which is compatible to a wide range of classic Pareto search algorithms without being restricted by their internal design.
    %\item The preference-driven Pareto search has not yet been studied for multi-objective software configuration tuning.
\end{itemize}

In summary, this work is, to the best of our knowledge, the first empirical study to understand whether, when, and why aspirations matter for guiding bi-objective software configuration tuning, according to the characteristics of requirement patterns and practice summarized for the problem.

%, providing new insights for future research.

% search based without aspiration, starting with single objective 
% search based with aspiration, starting with single objective
%(say with model and without)
% comparing weight vs Pareto, comparing reference point search, from SBSE or EC field

%% file: conclusion.tex
\section{Conclusion}
\label{sec:con}

In this paper, we conduct a comprehensive empirical study, which offers an in-depth understanding of whether performance aspirations matter to bi-objective tuning for configurable software systems. Our study covers 15 combinations of patterns that quantify the aspirations, four types of aspiration space, three search algorithms, and eight software systems/environments, leading to 1,296 cases of investigation. The results challenge the belief of \textit{``it does not matter whether to use aspiration in the tuning''} and reveal that:

\begin{itemize}
    \item the aspirations are generally helpful in guiding the tuning when they are realistic. 
    \item but, they can often be harmful to the tuning when these aspirations are unrealistic.
    \item the different patterns and position of the aspiration space do not change the above conclusion, but can affect the extent of benefits/detriments generated.
    
    %\item exhibit less significant and divergent impact for ML software.
    \item when the aspirations are realistic, the tuning budget has marginal implication on their benefits during tuning. In contrast, it is an important factor when the aspirations are unrealistic.
\end{itemize}

Our findings in this work provide useful insights for the practitioners in this particular field of research, particularly on the factors/information needed for answering the question of \textit{``when aspirations should be considered during the tuning?''}.  

We outline some exciting future research opportunities:

\begin{itemize}

\item To better understand the likelihood of realism and position of the given aspirations, we shall analyze the landscape of the configurable software systems even based on some limited samples.

\item To cater for the uncertainty of given requirements and aspirations (e.g., possible to be unrealistic), we should investigate requirement-robust optimizers for tuning software configuration.

\item To explain the impact of requirements patterns for guiding the tuning, we ought to provide the foundation to rigorously analyze their relationships and how they may be able to switch to one another for achieving relaxation.

\end{itemize}

%From this work, we hope to spark a dialogue on systematically choose the right optimization model for the recent trend of tuning software configuration with two performance objectives, and even beyonds, sheding the lights on future research in this area.
%Our findings suggest several future directions on this topic for multi-objective software configuration tuning, including a hybrid approach that combines \texttt{PS-w} and \texttt{PS-w/o} and formally provable ``relaxation'' between the patterns of requirements with aspirations. 

%% file: main.bbl
%%% -*-BibTeX-*-
%%% Do NOT edit. File created by BibTeX with style
%%% ACM-Reference-Format-Journals [18-Jan-2012].

\begin{thebibliography}{87}

%%% ====================================================================
%%% NOTE TO THE USER: you can override these defaults by providing
%%% customized versions of any of these macros before the \bibliography
%%% command.  Each of them MUST provide its own final punctuation,
%%% except for \shownote{}, \showDOI{}, and \showURL{}.  The latter two
%%% do not use final punctuation, in order to avoid confusing it with
%%% the Web address.
%%%
%%% To suppress output of a particular field, define its macro to expand
%%% to an empty string, or better, \unskip, like this:
%%%
%%% \newcommand{\showDOI}[1]{\unskip}   % LaTeX syntax
%%%
%%% \def \showDOI #1{\unskip}           % plain TeX syntax
%%%
%%% ====================================================================

\ifx \showCODEN    \undefined \def \showCODEN     #1{\unskip}     \fi
\ifx \showDOI      \undefined \def \showDOI       #1{#1}\fi
\ifx \showISBNx    \undefined \def \showISBNx     #1{\unskip}     \fi
\ifx \showISBNxiii \undefined \def \showISBNxiii  #1{\unskip}     \fi
\ifx \showISSN     \undefined \def \showISSN      #1{\unskip}     \fi
\ifx \showLCCN     \undefined \def \showLCCN      #1{\unskip}     \fi
\ifx \shownote     \undefined \def \shownote      #1{#1}          \fi
\ifx \showarticletitle \undefined \def \showarticletitle #1{#1}   \fi
\ifx \showURL      \undefined \def \showURL       {\relax}        \fi
% The following commands are used for tagged output and should be
% invisible to TeX
\providecommand\bibfield[2]{#2}
\providecommand\bibinfo[2]{#2}
\providecommand\natexlab[1]{#1}
\providecommand\showeprint[2][]{arXiv:#2}

\bibitem[\protect\citeauthoryear{Arcuri and Briand}{Arcuri and Briand}{2011}]%
        {DBLP:conf/icse/ArcuriB11}
\bibfield{author}{\bibinfo{person}{Andrea Arcuri} {and}
  \bibinfo{person}{Lionel~C. Briand}.} \bibinfo{year}{2011}\natexlab{}.
\newblock \showarticletitle{A practical guide for using statistical tests to
  assess randomized algorithms in software engineering}. In
  \bibinfo{booktitle}{\emph{Proceedings of the 33rd International Conference on
  Software Engineering, {ICSE} 2011, Waikiki, Honolulu , HI, USA, May 21-28,
  2011}}. \bibinfo{pages}{1--10}.
\newblock


\bibitem[\protect\citeauthoryear{Auger, Bader, Brockhoff, and Zitzler}{Auger
  et~al\mbox{.}}{2009}]%
        {Auger2009}
\bibfield{author}{\bibinfo{person}{Anne Auger}, \bibinfo{person}{Johannes
  Bader}, \bibinfo{person}{Dimo Brockhoff}, {and} \bibinfo{person}{Eckart
  Zitzler}.} \bibinfo{year}{2009}\natexlab{}.
\newblock \showarticletitle{Articulating user preferences in many-objective
  problems by sampling the weighted hypervolume}. In
  \bibinfo{booktitle}{\emph{Proceedings of the 11th Annual conference on
  Genetic and evolutionary computation}}. \bibinfo{pages}{555--562}.
\newblock


\bibitem[\protect\citeauthoryear{Bao, Liu, Wang, and Fang}{Bao
  et~al\mbox{.}}{2019}]%
        {DBLP:conf/kbse/BaoLWF19}
\bibfield{author}{\bibinfo{person}{Liang Bao}, \bibinfo{person}{Xin Liu},
  \bibinfo{person}{Fangzheng Wang}, {and} \bibinfo{person}{Baoyin Fang}.}
  \bibinfo{year}{2019}\natexlab{}.
\newblock \showarticletitle{{ACTGAN:} Automatic Configuration Tuning for
  Software Systems with Generative Adversarial Networks}. In
  \bibinfo{booktitle}{\emph{34th {IEEE/ACM} International Conference on
  Automated Software Engineering, {ASE} 2019, San Diego, CA, USA, November
  11-15, 2019}}. \bibinfo{publisher}{{IEEE}}, \bibinfo{pages}{465--476}.
\newblock
\urldef\tempurl%
\url{https://doi.org/10.1109/ASE.2019.00051}
\showDOI{\tempurl}


\bibitem[\protect\citeauthoryear{Baresi, Pasquale, and Spoletini}{Baresi
  et~al\mbox{.}}{2010}]%
        {DBLP:conf/re/BaresiPS10}
\bibfield{author}{\bibinfo{person}{Luciano Baresi}, \bibinfo{person}{Liliana
  Pasquale}, {and} \bibinfo{person}{Paola Spoletini}.}
  \bibinfo{year}{2010}\natexlab{}.
\newblock \showarticletitle{Fuzzy Goals for Requirements-Driven Adaptation}. In
  \bibinfo{booktitle}{\emph{{RE} 2010, 18th {IEEE} International Requirements
  Engineering Conference, Sydney, New South Wales, Australia, September 27 -
  October 1, 2010}}. \bibinfo{publisher}{{IEEE} Computer Society},
  \bibinfo{pages}{125--134}.
\newblock
\urldef\tempurl%
\url{https://doi.org/10.1109/RE.2010.25}
\showDOI{\tempurl}


\bibitem[\protect\citeauthoryear{Bechikh, Kessentini, Said, and
  Gh{\'{e}}dira}{Bechikh et~al\mbox{.}}{2015}]%
        {DBLP:journals/ac/BechikhKSG15}
\bibfield{author}{\bibinfo{person}{Slim Bechikh}, \bibinfo{person}{Marouane
  Kessentini}, \bibinfo{person}{Lamjed~Ben Said}, {and} \bibinfo{person}{Khaled
  Gh{\'{e}}dira}.} \bibinfo{year}{2015}\natexlab{}.
\newblock \showarticletitle{Chapter Four - Preference Incorporation in
  Evolutionary Multiobjective Optimization: {A} Survey of the
  State-of-the-Art}.
\newblock \bibinfo{journal}{\emph{Adv. Comput.}}  \bibinfo{volume}{98}
  (\bibinfo{year}{2015}), \bibinfo{pages}{141--207}.
\newblock
\urldef\tempurl%
\url{https://doi.org/10.1016/bs.adcom.2015.03.001}
\showDOI{\tempurl}


\bibitem[\protect\citeauthoryear{Behzad, Luu, Huchette, Byna, Prabhat, Aydt,
  Koziol, and Snir}{Behzad et~al\mbox{.}}{2013}]%
        {DBLP:conf/sc/BehzadLHBPAKS13}
\bibfield{author}{\bibinfo{person}{Babak Behzad}, \bibinfo{person}{Huong
  Vu~Thanh Luu}, \bibinfo{person}{Joseph Huchette}, \bibinfo{person}{Surendra
  Byna}, \bibinfo{person}{Prabhat}, \bibinfo{person}{Ruth~A. Aydt},
  \bibinfo{person}{Quincey Koziol}, {and} \bibinfo{person}{Marc Snir}.}
  \bibinfo{year}{2013}\natexlab{}.
\newblock \showarticletitle{Taming parallel {I/O} complexity with auto-tuning}.
  In \bibinfo{booktitle}{\emph{International Conference for High Performance
  Computing, Networking, Storage and Analysis, SC'13, Denver, CO, {USA} -
  November 17 - 21, 2013}}, \bibfield{editor}{\bibinfo{person}{William Gropp}
  {and} \bibinfo{person}{Satoshi Matsuoka}} (Eds.). \bibinfo{publisher}{{ACM}},
  \bibinfo{pages}{68:1--68:12}.
\newblock
\urldef\tempurl%
\url{https://doi.org/10.1145/2503210.2503278}
\showDOI{\tempurl}


\bibitem[\protect\citeauthoryear{Bowers, Fredericks, and Cheng}{Bowers
  et~al\mbox{.}}{2018}]%
        {DBLP:conf/ssbse/BowersFC18}
\bibfield{author}{\bibinfo{person}{Kate~M. Bowers}, \bibinfo{person}{Erik~M.
  Fredericks}, {and} \bibinfo{person}{Betty H.~C. Cheng}.}
  \bibinfo{year}{2018}\natexlab{}.
\newblock \showarticletitle{Automated Optimization of Weighted Non-functional
  Objectives in Self-adaptive Systems}. In
  \bibinfo{booktitle}{\emph{Search-Based Software Engineering - 10th
  International Symposium, {SSBSE} 2018, Montpellier, France, September 8-9,
  2018, Proceedings}} \emph{(\bibinfo{series}{Lecture Notes in Computer
  Science}, Vol.~\bibinfo{volume}{11036})},
  \bibfield{editor}{\bibinfo{person}{Thelma~Elita Colanzi} {and}
  \bibinfo{person}{Phil McMinn}} (Eds.). \bibinfo{publisher}{Springer},
  \bibinfo{pages}{182--197}.
\newblock


\bibitem[\protect\citeauthoryear{Calinescu, Ceska, Gerasimou, Kwiatkowska, and
  Paoletti}{Calinescu et~al\mbox{.}}{2017}]%
        {Calinescu2017Designing}
\bibfield{author}{\bibinfo{person}{Radu Calinescu}, \bibinfo{person}{Milan
  Ceska}, \bibinfo{person}{Simos Gerasimou}, \bibinfo{person}{Marta
  Kwiatkowska}, {and} \bibinfo{person}{Nicola Paoletti}.}
  \bibinfo{year}{2017}\natexlab{}.
\newblock \showarticletitle{Designing Robust Software Systems through
  Parametric Markov Chain Synthesis}. In \bibinfo{booktitle}{\emph{IEEE
  International Conference on Software Architecture}}.
\newblock


\bibitem[\protect\citeauthoryear{Calinescu, Jr., Gerasimou, Kwiatkowska, and
  Paoletti}{Calinescu et~al\mbox{.}}{2018}]%
        {DBLP:journals/jss/CalinescuCGKP18}
\bibfield{author}{\bibinfo{person}{Radu Calinescu},
  \bibinfo{person}{Milan~Ceska Jr.}, \bibinfo{person}{Simos Gerasimou},
  \bibinfo{person}{Marta Kwiatkowska}, {and} \bibinfo{person}{Nicola
  Paoletti}.} \bibinfo{year}{2018}\natexlab{}.
\newblock \showarticletitle{Efficient synthesis of robust models for stochastic
  systems}.
\newblock \bibinfo{journal}{\emph{Journal of Systems and Software}}
  \bibinfo{volume}{143} (\bibinfo{year}{2018}), \bibinfo{pages}{140--158}.
\newblock


\bibitem[\protect\citeauthoryear{Chen, Xu, Chen, and Zhang}{Chen
  et~al\mbox{.}}{2021}]%
        {DBLP:conf/icse/0003XC021}
\bibfield{author}{\bibinfo{person}{Junjie Chen}, \bibinfo{person}{Ningxin Xu},
  \bibinfo{person}{Peiqi Chen}, {and} \bibinfo{person}{Hongyu Zhang}.}
  \bibinfo{year}{2021}\natexlab{}.
\newblock \showarticletitle{Efficient Compiler Autotuning via Bayesian
  Optimization}. In \bibinfo{booktitle}{\emph{43rd {IEEE/ACM} International
  Conference on Software Engineering, {ICSE} 2021, Madrid, Spain, 22-30 May
  2021}}. \bibinfo{publisher}{{IEEE}}, \bibinfo{pages}{1198--1209}.
\newblock
\urldef\tempurl%
\url{https://doi.org/10.1109/ICSE43902.2021.00110}
\showDOI{\tempurl}


\bibitem[\protect\citeauthoryear{Chen}{Chen}{2022}]%
        {DBLP:conf/wcre/Chen22}
\bibfield{author}{\bibinfo{person}{Tao Chen}.} \bibinfo{year}{2022}\natexlab{}.
\newblock \showarticletitle{Lifelong Dynamic Optimization for Self-Adaptive
  Systems: Fact or Fiction?}. In \bibinfo{booktitle}{\emph{{IEEE} International
  Conference on Software Analysis, Evolution and Reengineering, {SANER} 2022,
  Honolulu, HI, USA, March 15-18, 2022}}. \bibinfo{publisher}{{IEEE}},
  \bibinfo{pages}{78--89}.
\newblock
\urldef\tempurl%
\url{https://doi.org/10.1109/SANER53432.2022.00022}
\showDOI{\tempurl}


\bibitem[\protect\citeauthoryear{Chen and Bahsoon}{Chen and Bahsoon}{2015}]%
        {DBLP:journals/computer/ChenB15}
\bibfield{author}{\bibinfo{person}{Tao Chen} {and} \bibinfo{person}{Rami
  Bahsoon}.} \bibinfo{year}{2015}\natexlab{}.
\newblock \showarticletitle{Toward a Smarter Cloud: Self-Aware Autoscaling of
  Cloud Configurations and Resources}.
\newblock \bibinfo{journal}{\emph{Computer}} \bibinfo{volume}{48},
  \bibinfo{number}{9} (\bibinfo{year}{2015}), \bibinfo{pages}{93--96}.
\newblock
\urldef\tempurl%
\url{https://doi.org/10.1109/MC.2015.278}
\showDOI{\tempurl}


\bibitem[\protect\citeauthoryear{Chen and Bahsoon}{Chen and Bahsoon}{2017a}]%
        {DBLP:journals/tse/ChenB17}
\bibfield{author}{\bibinfo{person}{Tao Chen} {and} \bibinfo{person}{Rami
  Bahsoon}.} \bibinfo{year}{2017}\natexlab{a}.
\newblock \showarticletitle{Self-Adaptive and Online QoS Modeling for
  Cloud-Based Software Services}.
\newblock \bibinfo{journal}{\emph{{IEEE} Trans. Software Eng.}}
  \bibinfo{volume}{43}, \bibinfo{number}{5} (\bibinfo{year}{2017}),
  \bibinfo{pages}{453--475}.
\newblock
\urldef\tempurl%
\url{https://doi.org/10.1109/TSE.2016.2608826}
\showDOI{\tempurl}


\bibitem[\protect\citeauthoryear{Chen and Bahsoon}{Chen and Bahsoon}{2017b}]%
        {Chen2017Self}
\bibfield{author}{\bibinfo{person}{Tao Chen} {and} \bibinfo{person}{Rami
  Bahsoon}.} \bibinfo{year}{2017}\natexlab{b}.
\newblock \showarticletitle{Self-Adaptive Trade-off Decision Making for
  Autoscaling Cloud-Based Services}.
\newblock \bibinfo{journal}{\emph{IEEE Transactions on Services Computing}}
  \bibinfo{volume}{10}, \bibinfo{number}{4} (\bibinfo{year}{2017}),
  \bibinfo{pages}{618--632}.
\newblock


\bibitem[\protect\citeauthoryear{Chen, Bahsoon, and Yao}{Chen
  et~al\mbox{.}}{2018a}]%
        {DBLP:journals/csur/ChenBY18}
\bibfield{author}{\bibinfo{person}{Tao Chen}, \bibinfo{person}{Rami Bahsoon},
  {and} \bibinfo{person}{Xin Yao}.} \bibinfo{year}{2018}\natexlab{a}.
\newblock \showarticletitle{A Survey and Taxonomy of Self-Aware and
  Self-Adaptive Cloud Autoscaling Systems}.
\newblock \bibinfo{journal}{\emph{{ACM} Comput. Surv.}} \bibinfo{volume}{51},
  \bibinfo{number}{3} (\bibinfo{year}{2018}), \bibinfo{pages}{61:1--61:40}.
\newblock
\urldef\tempurl%
\url{https://doi.org/10.1145/3190507}
\showDOI{\tempurl}


\bibitem[\protect\citeauthoryear{Chen, Li, Bahsoon, and Yao}{Chen
  et~al\mbox{.}}{2018b}]%
        {Chen2018FEMOSAA}
\bibfield{author}{\bibinfo{person}{Tao Chen}, \bibinfo{person}{Ke Li},
  \bibinfo{person}{Rami Bahsoon}, {and} \bibinfo{person}{Xin Yao}.}
  \bibinfo{year}{2018}\natexlab{b}.
\newblock \showarticletitle{{FEMOSAA}: Feature Guided and Knee Driven
  Multi-Objective Optimization for Self-Adaptive Software}.
\newblock \bibinfo{journal}{\emph{ACM Transactions on Software Engineering and
  Methodology}} \bibinfo{volume}{27}, \bibinfo{number}{2}
  (\bibinfo{year}{2018}).
\newblock


\bibitem[\protect\citeauthoryear{Chen and Li}{Chen and Li}{2021}]%
        {DBLP:conf/sigsoft/0001Chen}
\bibfield{author}{\bibinfo{person}{Tao Chen} {and} \bibinfo{person}{Miqing
  Li}.} \bibinfo{year}{2021}\natexlab{}.
\newblock \showarticletitle{Multi-objectivizing software configuration tuning}.
  In \bibinfo{booktitle}{\emph{{ESEC/FSE} '21: 29th {ACM} Joint European
  Software Engineering Conference and Symposium on the Foundations of Software
  Engineering, Athens, Greece, August 23-28, 2021}},
  \bibfield{editor}{\bibinfo{person}{Diomidis Spinellis},
  \bibinfo{person}{Georgios Gousios}, \bibinfo{person}{Marsha Chechik}, {and}
  \bibinfo{person}{Massimiliano~Di Penta}} (Eds.). \bibinfo{publisher}{{ACM}},
  \bibinfo{pages}{453--465}.
\newblock
\urldef\tempurl%
\url{https://doi.org/10.1145/3468264.3468555}
\showDOI{\tempurl}


\bibitem[\protect\citeauthoryear{Chen and Li}{Chen and Li}{2022}]%
        {10.1145/3514233}
\bibfield{author}{\bibinfo{person}{Tao Chen} {and} \bibinfo{person}{Miqing
  Li}.} \bibinfo{year}{2022}\natexlab{}.
\newblock \showarticletitle{The Weights Can Be Harmful: Pareto Search versus
  Weighted Search in Multi-Objective Search-Based Software Engineering}.
\newblock \bibinfo{journal}{\emph{ACM Transactions on Software Engineering and
  Methodology}} (\bibinfo{date}{Jan} \bibinfo{year}{2022}).
\newblock
\showISSN{1049-331X}
\urldef\tempurl%
\url{https://doi.org/10.1145/3514233}
\showDOI{\tempurl}


\bibitem[\protect\citeauthoryear{Chen, Li, Li, and Deb}{Chen
  et~al\mbox{.}}{2020}]%
        {DBLP:journals/corr/abs-2001-08236}
\bibfield{author}{\bibinfo{person}{Tao Chen}, \bibinfo{person}{Miqing Li},
  \bibinfo{person}{Ke Li}, {and} \bibinfo{person}{Kalyanmoy Deb}.}
  \bibinfo{year}{2020}\natexlab{}.
\newblock \showarticletitle{Search-Based Software Engineering for Self-Adaptive
  Systems: Survey, Disappointments, Suggestions and Opportunities}.
\newblock \bibinfo{journal}{\emph{CoRR}}  \bibinfo{volume}{abs/2001.08236}
  (\bibinfo{year}{2020}).
\newblock


\bibitem[\protect\citeauthoryear{Chen, Li, and Yao}{Chen et~al\mbox{.}}{2019}]%
        {DBLP:journals/infsof/ChenLY19}
\bibfield{author}{\bibinfo{person}{Tao Chen}, \bibinfo{person}{Miqing Li},
  {and} \bibinfo{person}{Xin Yao}.} \bibinfo{year}{2019}\natexlab{}.
\newblock \showarticletitle{Standing on the shoulders of giants: Seeding
  search-based multi-objective optimization with prior knowledge for software
  service composition}.
\newblock \bibinfo{journal}{\emph{Inf. Softw. Technol.}}  \bibinfo{volume}{114}
  (\bibinfo{year}{2019}), \bibinfo{pages}{155--175}.
\newblock
\urldef\tempurl%
\url{https://doi.org/10.1016/j.infsof.2019.05.013}
\showDOI{\tempurl}


\bibitem[\protect\citeauthoryear{Dalpiaz, Dell'Anna, Aydemir, and
  {\c{C}}evikol}{Dalpiaz et~al\mbox{.}}{2019}]%
        {DBLP:conf/re/DalpiazDAC19}
\bibfield{author}{\bibinfo{person}{Fabiano Dalpiaz}, \bibinfo{person}{Davide
  Dell'Anna}, \bibinfo{person}{Fatma~Basak Aydemir}, {and}
  \bibinfo{person}{Sercan {\c{C}}evikol}.} \bibinfo{year}{2019}\natexlab{}.
\newblock \showarticletitle{Requirements Classification with Interpretable
  Machine Learning and Dependency Parsing}. In \bibinfo{booktitle}{\emph{27th
  {IEEE} International Requirements Engineering Conference, {RE} 2019, Jeju
  Island, Korea (South), September 23-27, 2019}},
  \bibfield{editor}{\bibinfo{person}{Daniela~E. Damian}, \bibinfo{person}{Anna
  Perini}, {and} \bibinfo{person}{Seok{-}Won Lee}} (Eds.).
  \bibinfo{publisher}{{IEEE}}, \bibinfo{pages}{142--152}.
\newblock
\urldef\tempurl%
\url{https://doi.org/10.1109/RE.2019.00025}
\showDOI{\tempurl}


\bibitem[\protect\citeauthoryear{Deb, Pratap, Agarwal, and Meyarivan}{Deb
  et~al\mbox{.}}{2002}]%
        {Deb2002}
\bibfield{author}{\bibinfo{person}{K. Deb}, \bibinfo{person}{A. Pratap},
  \bibinfo{person}{S. Agarwal}, {and} \bibinfo{person}{T. Meyarivan}.}
  \bibinfo{year}{2002}\natexlab{}.
\newblock \showarticletitle{A fast and elitist multiobjective genetic
  algorithm: NSGA-II}.
\newblock \bibinfo{journal}{\emph{IEEE Transactions on Evolutionary
  Computation}} \bibinfo{volume}{6}, \bibinfo{number}{2}
  (\bibinfo{year}{2002}), \bibinfo{pages}{182--197}.
\newblock


\bibitem[\protect\citeauthoryear{Do, Chekuri, and Bhowmik}{Do
  et~al\mbox{.}}{2019}]%
        {DBLP:conf/icsr/DoCB19}
\bibfield{author}{\bibinfo{person}{Quoc~Anh Do}, \bibinfo{person}{Surendra~Raju
  Chekuri}, {and} \bibinfo{person}{Tanmay Bhowmik}.}
  \bibinfo{year}{2019}\natexlab{}.
\newblock \showarticletitle{Automated Support to Capture Creative Requirements
  via Requirements Reuse}. In \bibinfo{booktitle}{\emph{Reuse in the Big Data
  Era - 18th International Conference on Software and Systems Reuse, {ICSR}
  2019, Cincinnati, OH, USA, June 26-28, 2019, Proceedings}}
  \emph{(\bibinfo{series}{Lecture Notes in Computer Science},
  Vol.~\bibinfo{volume}{11602})}, \bibfield{editor}{\bibinfo{person}{Xin Peng},
  \bibinfo{person}{Apostolos Ampatzoglou}, {and} \bibinfo{person}{Tanmay
  Bhowmik}} (Eds.). \bibinfo{publisher}{Springer}, \bibinfo{pages}{47--63}.
\newblock
\urldef\tempurl%
\url{https://doi.org/10.1007/978-3-030-22888-0\_4}
\showDOI{\tempurl}


\bibitem[\protect\citeauthoryear{Dom{\'\i}nguez-R{\'\i}os, Chicano, Alba, del
  {\'A}guila, and del Sagrado}{Dom{\'\i}nguez-R{\'\i}os et~al\mbox{.}}{2019}]%
        {Dominguez2019}
\bibfield{author}{\bibinfo{person}{Miguel~{\'A}ngel Dom{\'\i}nguez-R{\'\i}os},
  \bibinfo{person}{Francisco Chicano}, \bibinfo{person}{Enrique Alba},
  \bibinfo{person}{Isabel del {\'A}guila}, {and} \bibinfo{person}{Jos{\'e} del
  Sagrado}.} \bibinfo{year}{2019}\natexlab{}.
\newblock \showarticletitle{Efficient anytime algorithms to solve the
  bi-objective Next Release Problem}.
\newblock \bibinfo{journal}{\emph{Journal of Systems and Software}}
  \bibinfo{volume}{156} (\bibinfo{year}{2019}), \bibinfo{pages}{217--231}.
\newblock


\bibitem[\protect\citeauthoryear{Durillo and Nebro}{Durillo and Nebro}{2011}]%
        {DBLP:journals/aes/DurilloN11}
\bibfield{author}{\bibinfo{person}{Juan~Jos{\'{e}} Durillo} {and}
  \bibinfo{person}{Antonio~J. Nebro}.} \bibinfo{year}{2011}\natexlab{}.
\newblock \showarticletitle{jMetal: {A} Java framework for multi-objective
  optimization}.
\newblock \bibinfo{journal}{\emph{Adv. Eng. Softw.}} \bibinfo{volume}{42},
  \bibinfo{number}{10} (\bibinfo{year}{2011}), \bibinfo{pages}{760--771}.
\newblock
\urldef\tempurl%
\url{https://doi.org/10.1016/j.advengsoft.2011.05.014}
\showDOI{\tempurl}


\bibitem[\protect\citeauthoryear{Emmerich and Deutz}{Emmerich and
  Deutz}{2018}]%
        {Emmerich2018tutorial}
\bibfield{author}{\bibinfo{person}{Michael~TM Emmerich} {and}
  \bibinfo{person}{Andr{\'e}~H Deutz}.} \bibinfo{year}{2018}\natexlab{}.
\newblock \showarticletitle{A tutorial on multiobjective optimization:
  fundamentals and evolutionary methods}.
\newblock \bibinfo{journal}{\emph{Natural computing}} \bibinfo{volume}{17},
  \bibinfo{number}{3} (\bibinfo{year}{2018}), \bibinfo{pages}{585--609}.
\newblock


\bibitem[\protect\citeauthoryear{Esfahani, Kouroshfar, and Malek}{Esfahani
  et~al\mbox{.}}{2011}]%
        {DBLP:conf/sigsoft/EsfahaniKM11}
\bibfield{author}{\bibinfo{person}{Naeem Esfahani}, \bibinfo{person}{Ehsan
  Kouroshfar}, {and} \bibinfo{person}{Sam Malek}.}
  \bibinfo{year}{2011}\natexlab{}.
\newblock \showarticletitle{Taming uncertainty in self-adaptive software}. In
  \bibinfo{booktitle}{\emph{SIGSOFT/FSE'11 19th {ACM} {SIGSOFT} Symposium on
  the Foundations of Software Engineering {(FSE-19)} and ESEC'11: 13th European
  Software Engineering Conference (ESEC-13), Szeged, Hungary, September 5-9,
  2011}}, \bibfield{editor}{\bibinfo{person}{Tibor Gyim{\'{o}}thy} {and}
  \bibinfo{person}{Andreas Zeller}} (Eds.). \bibinfo{publisher}{{ACM}},
  \bibinfo{pages}{234--244}.
\newblock
\urldef\tempurl%
\url{https://doi.org/10.1145/2025113.2025147}
\showDOI{\tempurl}


\bibitem[\protect\citeauthoryear{Fekry, Carata, Pasquier, Rice, and
  Hopper}{Fekry et~al\mbox{.}}{2019}]%
        {DBLP:conf/icdcs/FekryCPRH19}
\bibfield{author}{\bibinfo{person}{Ayat Fekry}, \bibinfo{person}{Lucian
  Carata}, \bibinfo{person}{Thomas F.~J.{-}M. Pasquier},
  \bibinfo{person}{Andrew Rice}, {and} \bibinfo{person}{Andy Hopper}.}
  \bibinfo{year}{2019}\natexlab{}.
\newblock \showarticletitle{Towards Seamless Configuration Tuning of Big Data
  Analytics}. In \bibinfo{booktitle}{\emph{39th {IEEE} International Conference
  on Distributed Computing Systems, {ICDCS} 2019, Dallas, TX, USA, July 7-10,
  2019}}. \bibinfo{publisher}{{IEEE}}, \bibinfo{pages}{1912--1919}.
\newblock
\urldef\tempurl%
\url{https://doi.org/10.1109/ICDCS.2019.00189}
\showDOI{\tempurl}


\bibitem[\protect\citeauthoryear{Ferrari, Spagnolo, and Gnesi}{Ferrari
  et~al\mbox{.}}{2017}]%
        {DBLP:conf/re/FerrariSG17}
\bibfield{author}{\bibinfo{person}{Alessio Ferrari},
  \bibinfo{person}{Giorgio~Oronzo Spagnolo}, {and} \bibinfo{person}{Stefania
  Gnesi}.} \bibinfo{year}{2017}\natexlab{}.
\newblock \showarticletitle{{PURE:} {A} Dataset of Public Requirements
  Documents}. In \bibinfo{booktitle}{\emph{25th {IEEE} International
  Requirements Engineering Conference, {RE} 2017, Lisbon, Portugal, September
  4-8, 2017}}, \bibfield{editor}{\bibinfo{person}{Ana Moreira},
  \bibinfo{person}{Jo{\~{a}}o Ara{\'{u}}jo}, \bibinfo{person}{Jane Hayes},
  {and} \bibinfo{person}{Barbara Paech}} (Eds.). \bibinfo{publisher}{{IEEE}
  Computer Society}, \bibinfo{pages}{502--505}.
\newblock
\urldef\tempurl%
\url{https://doi.org/10.1109/RE.2017.29}
\showDOI{\tempurl}


\bibitem[\protect\citeauthoryear{Gao, Zhu, Zhang, Lin, and Yang}{Gao
  et~al\mbox{.}}{2021}]%
        {DBLP:conf/icse/GaoZ0LY21}
\bibfield{author}{\bibinfo{person}{Yanjie Gao}, \bibinfo{person}{Yonghao Zhu},
  \bibinfo{person}{Hongyu Zhang}, \bibinfo{person}{Haoxiang Lin}, {and}
  \bibinfo{person}{Mao Yang}.} \bibinfo{year}{2021}\natexlab{}.
\newblock \showarticletitle{Resource-Guided Configuration Space Reduction for
  Deep Learning Models}. In \bibinfo{booktitle}{\emph{43rd {IEEE/ACM}
  International Conference on Software Engineering, {ICSE} 2021, Madrid, Spain,
  22-30 May 2021}}. \bibinfo{publisher}{{IEEE}}, \bibinfo{pages}{175--187}.
\newblock
\urldef\tempurl%
\url{https://doi.org/10.1109/ICSE43902.2021.00028}
\showDOI{\tempurl}


\bibitem[\protect\citeauthoryear{Garlan, Cheng, Huang, Schmerl, and
  Steenkiste}{Garlan et~al\mbox{.}}{2004}]%
        {garlan2004rainbow}
\bibfield{author}{\bibinfo{person}{David Garlan}, \bibinfo{person}{S-W Cheng},
  \bibinfo{person}{A-C Huang}, \bibinfo{person}{Bradley Schmerl}, {and}
  \bibinfo{person}{Peter Steenkiste}.} \bibinfo{year}{2004}\natexlab{}.
\newblock \showarticletitle{Rainbow: Architecture-based self-adaptation with
  reusable infrastructure}.
\newblock \bibinfo{journal}{\emph{Computer}} \bibinfo{volume}{37},
  \bibinfo{number}{10} (\bibinfo{year}{2004}), \bibinfo{pages}{46--54}.
\newblock


\bibitem[\protect\citeauthoryear{Gerasimou, Calinescu, and
  Tamburrelli}{Gerasimou et~al\mbox{.}}{2018}]%
        {DBLP:journals/ase/GerasimouCT18}
\bibfield{author}{\bibinfo{person}{Simos Gerasimou}, \bibinfo{person}{Radu
  Calinescu}, {and} \bibinfo{person}{Giordano Tamburrelli}.}
  \bibinfo{year}{2018}\natexlab{}.
\newblock \showarticletitle{Synthesis of probabilistic models for
  quality-of-service software engineering}.
\newblock \bibinfo{journal}{\emph{Autom. Softw. Eng.}} \bibinfo{volume}{25},
  \bibinfo{number}{4} (\bibinfo{year}{2018}), \bibinfo{pages}{785--831}.
\newblock


\bibitem[\protect\citeauthoryear{Gerasimou, Tamburrelli, and
  Calinescu}{Gerasimou et~al\mbox{.}}{2016}]%
        {Gerasimou2016Search}
\bibfield{author}{\bibinfo{person}{Simos Gerasimou}, \bibinfo{person}{Giordano
  Tamburrelli}, {and} \bibinfo{person}{Radu Calinescu}.}
  \bibinfo{year}{2016}\natexlab{}.
\newblock \showarticletitle{Search-Based Synthesis of Probabilistic Models for
  Quality-of-Service Software Engineering (T)}. In
  \bibinfo{booktitle}{\emph{IEEE/ACM International Conference on Automated
  Software Engineering}}. \bibinfo{pages}{319--330}.
\newblock


\bibitem[\protect\citeauthoryear{Ghanbari, Simmons, Litoiu, and
  Iszlai}{Ghanbari et~al\mbox{.}}{2012}]%
        {DBLP:journals/fgcs/GhanbariSLI12}
\bibfield{author}{\bibinfo{person}{Hamoun Ghanbari}, \bibinfo{person}{Bradley
  Simmons}, \bibinfo{person}{Marin Litoiu}, {and} \bibinfo{person}{Gabriel
  Iszlai}.} \bibinfo{year}{2012}\natexlab{}.
\newblock \showarticletitle{Feedback-based optimization of a private cloud}.
\newblock \bibinfo{journal}{\emph{Future Gener. Comput. Syst.}}
  \bibinfo{volume}{28}, \bibinfo{number}{1} (\bibinfo{year}{2012}),
  \bibinfo{pages}{104--111}.
\newblock
\urldef\tempurl%
\url{https://doi.org/10.1016/j.future.2011.05.019}
\showDOI{\tempurl}


\bibitem[\protect\citeauthoryear{Gias, Casale, and Woodside}{Gias
  et~al\mbox{.}}{2019}]%
        {DBLP:conf/icdcs/GiasCW19}
\bibfield{author}{\bibinfo{person}{Alim~Ul Gias}, \bibinfo{person}{Giuliano
  Casale}, {and} \bibinfo{person}{Murray Woodside}.}
  \bibinfo{year}{2019}\natexlab{}.
\newblock \showarticletitle{{ATOM:} Model-Driven Autoscaling for
  Microservices}. In \bibinfo{booktitle}{\emph{39th {IEEE} International
  Conference on Distributed Computing Systems, {ICDCS} 2019, Dallas, TX, USA,
  July 7-10, 2019}}. \bibinfo{publisher}{{IEEE}}, \bibinfo{pages}{1994--2004}.
\newblock
\urldef\tempurl%
\url{https://doi.org/10.1109/ICDCS.2019.00197}
\showDOI{\tempurl}


\bibitem[\protect\citeauthoryear{Gong and Chen}{Gong and Chen}{2022}]%
        {DBLP:conf/msr/GongC22}
\bibfield{author}{\bibinfo{person}{Jingzhi Gong} {and} \bibinfo{person}{Tao
  Chen}.} \bibinfo{year}{2022}\natexlab{}.
\newblock \showarticletitle{Does Configuration Encoding Matter in Learning
  Software Performance? An Empirical Study on Encoding Schemes}. In
  \bibinfo{booktitle}{\emph{19th {IEEE/ACM} International Conference on Mining
  Software Repositories, {MSR} 2022, Pittsburgh, PA, USA, May 23-24, 2022}}.
  \bibinfo{publisher}{{ACM}}, \bibinfo{pages}{482--494}.
\newblock
\urldef\tempurl%
\url{https://doi.org/10.1145/3524842.3528431}
\showDOI{\tempurl}


\bibitem[\protect\citeauthoryear{Guo, Yi, and Qasem}{Guo et~al\mbox{.}}{2010}]%
        {guo2010evaluating}
\bibfield{author}{\bibinfo{person}{Jichi Guo}, \bibinfo{person}{Qing Yi}, {and}
  \bibinfo{person}{Apan Qasem}.} \bibinfo{year}{2010}\natexlab{}.
\newblock \showarticletitle{Evaluating the role of optimization-specific search
  heuristics in effective autotuning}.
\newblock \bibinfo{journal}{\emph{Technical report}} (\bibinfo{year}{2010}).
\newblock


\bibitem[\protect\citeauthoryear{Ha and Zhang}{Ha and Zhang}{2019}]%
        {DBLP:conf/icse/HaZ19}
\bibfield{author}{\bibinfo{person}{Huong Ha} {and} \bibinfo{person}{Hongyu
  Zhang}.} \bibinfo{year}{2019}\natexlab{}.
\newblock \showarticletitle{DeepPerf: performance prediction for configurable
  software with deep sparse neural network}. In
  \bibinfo{booktitle}{\emph{Proceedings of the 41st International Conference on
  Software Engineering, {ICSE} 2019, Montreal, QC, Canada, May 25-31, 2019}},
  \bibfield{editor}{\bibinfo{person}{Joanne~M. Atlee}, \bibinfo{person}{Tevfik
  Bultan}, {and} \bibinfo{person}{Jon Whittle}} (Eds.).
  \bibinfo{publisher}{{IEEE} / {ACM}}, \bibinfo{pages}{1095--1106}.
\newblock
\urldef\tempurl%
\url{https://doi.org/10.1109/ICSE.2019.00113}
\showDOI{\tempurl}


\bibitem[\protect\citeauthoryear{Han and Yu}{Han and Yu}{2016}]%
        {DBLP:conf/esem/HanY16}
\bibfield{author}{\bibinfo{person}{Xue Han} {and} \bibinfo{person}{Tingting
  Yu}.} \bibinfo{year}{2016}\natexlab{}.
\newblock \showarticletitle{An Empirical Study on Performance Bugs for Highly
  Configurable Software Systems}. In \bibinfo{booktitle}{\emph{Proceedings of
  the 10th {ACM/IEEE} International Symposium on Empirical Software Engineering
  and Measurement, {ESEM} 2016, Ciudad Real, Spain, September 8-9, 2016}}.
  \bibinfo{publisher}{{ACM}}, \bibinfo{pages}{23:1--23:10}.
\newblock
\urldef\tempurl%
\url{https://doi.org/10.1145/2961111.2962602}
\showDOI{\tempurl}


\bibitem[\protect\citeauthoryear{Harman, Mansouri, and Zhang}{Harman
  et~al\mbox{.}}{2012}]%
        {Harman2012}
\bibfield{author}{\bibinfo{person}{Mark Harman}, \bibinfo{person}{S~Afshin
  Mansouri}, {and} \bibinfo{person}{Yuanyuan Zhang}.}
  \bibinfo{year}{2012}\natexlab{}.
\newblock \showarticletitle{Search-based software engineering: Trends,
  techniques and applications}.
\newblock \bibinfo{journal}{\emph{ACM Computing Surveys (CSUR)}}
  \bibinfo{volume}{45}, \bibinfo{number}{1} (\bibinfo{year}{2012}),
  \bibinfo{pages}{11}.
\newblock


\bibitem[\protect\citeauthoryear{Hort, Kechagia, Sarro, and Harman}{Hort
  et~al\mbox{.}}{2021}]%
        {9397392}
\bibfield{author}{\bibinfo{person}{Max Hort}, \bibinfo{person}{Maria Kechagia},
  \bibinfo{person}{Federica Sarro}, {and} \bibinfo{person}{Mark Harman}.}
  \bibinfo{year}{2021}\natexlab{}.
\newblock \showarticletitle{A Survey of Performance Optimization for Mobile
  Applications}.
\newblock \bibinfo{journal}{\emph{IEEE Transactions on Software Engineering}}
  (\bibinfo{year}{2021}), \bibinfo{pages}{1--1}.
\newblock
\urldef\tempurl%
\url{https://doi.org/10.1109/TSE.2021.3071193}
\showDOI{\tempurl}


\bibitem[\protect\citeauthoryear{Iqbal, Krishna, Javidian, Ray, and
  Jamshidi}{Iqbal et~al\mbox{.}}{2022}]%
        {DBLP:conf/eurosys/IqbalKJRJ22}
\bibfield{author}{\bibinfo{person}{Md~Shahriar Iqbal}, \bibinfo{person}{Rahul
  Krishna}, \bibinfo{person}{Mohammad~Ali Javidian}, \bibinfo{person}{Baishakhi
  Ray}, {and} \bibinfo{person}{Pooyan Jamshidi}.}
  \bibinfo{year}{2022}\natexlab{}.
\newblock \showarticletitle{Unicorn: reasoning about configurable system
  performance through the lens of causality}. In
  \bibinfo{booktitle}{\emph{EuroSys '22: Seventeenth European Conference on
  Computer Systems, Rennes, France, April 5 - 8, 2022}},
  \bibfield{editor}{\bibinfo{person}{Y{\'{e}}rom{-}David Bromberg},
  \bibinfo{person}{Anne{-}Marie Kermarrec}, {and} \bibinfo{person}{Christos
  Kozyrakis}} (Eds.). \bibinfo{publisher}{{ACM}}, \bibinfo{pages}{199--217}.
\newblock
\urldef\tempurl%
\url{https://doi.org/10.1145/3492321.3519575}
\showDOI{\tempurl}


\bibitem[\protect\citeauthoryear{Iqbal, Su, Kotthoff, and Jamshidi}{Iqbal
  et~al\mbox{.}}{2020}]%
        {Iqbal2020}
\bibfield{author}{\bibinfo{person}{Md~Shahriar Iqbal}, \bibinfo{person}{Jianhai
  Su}, \bibinfo{person}{Lars Kotthoff}, {and} \bibinfo{person}{Pooyan
  Jamshidi}.} \bibinfo{year}{2020}\natexlab{}.
\newblock \showarticletitle{Flexibo: Cost-aware multi-objective optimization of
  deep neural networks}.
\newblock \bibinfo{journal}{\emph{arXiv preprint arXiv:2001.06588}}
  (\bibinfo{year}{2020}).
\newblock


\bibitem[\protect\citeauthoryear{Jamshidi and Casale}{Jamshidi and
  Casale}{2016}]%
        {DBLP:conf/mascots/JamshidiC16}
\bibfield{author}{\bibinfo{person}{Pooyan Jamshidi} {and}
  \bibinfo{person}{Giuliano Casale}.} \bibinfo{year}{2016}\natexlab{}.
\newblock \showarticletitle{An Uncertainty-Aware Approach to Optimal
  Configuration of Stream Processing Systems}. In
  \bibinfo{booktitle}{\emph{24th {IEEE} International Symposium on Modeling,
  Analysis and Simulation of Computer and Telecommunication Systems, {MASCOTS}
  2016, London, United Kingdom, September 19-21, 2016}}.
  \bibinfo{publisher}{{IEEE} Computer Society}, \bibinfo{pages}{39--48}.
\newblock


\bibitem[\protect\citeauthoryear{Jamshidi, Velez, K{\"{a}}stner, and
  Siegmund}{Jamshidi et~al\mbox{.}}{2018}]%
        {DBLP:conf/sigsoft/JamshidiVKS18}
\bibfield{author}{\bibinfo{person}{Pooyan Jamshidi}, \bibinfo{person}{Miguel
  Velez}, \bibinfo{person}{Christian K{\"{a}}stner}, {and}
  \bibinfo{person}{Norbert Siegmund}.} \bibinfo{year}{2018}\natexlab{}.
\newblock \showarticletitle{Learning to sample: exploiting similarities across
  environments to learn performance models for configurable systems}. In
  \bibinfo{booktitle}{\emph{Proceedings of the 2018 {ACM} Joint Meeting on
  European Software Engineering Conference and Symposium on the Foundations of
  Software Engineering, {ESEC/SIGSOFT} {FSE} 2018, Lake Buena Vista, FL, USA,
  November 04-09, 2018}}, \bibfield{editor}{\bibinfo{person}{Gary~T. Leavens},
  \bibinfo{person}{Alessandro Garcia}, {and} \bibinfo{person}{Corina~S.
  Pasareanu}} (Eds.). \bibinfo{publisher}{{ACM}}, \bibinfo{pages}{71--82}.
\newblock
\urldef\tempurl%
\url{https://doi.org/10.1145/3236024.3236074}
\showDOI{\tempurl}


\bibitem[\protect\citeauthoryear{Jamshidi, Velez, K{\"a}stner, Siegmund, and
  Kawthekar}{Jamshidi et~al\mbox{.}}{2017}]%
        {Jamshidi2017}
\bibfield{author}{\bibinfo{person}{Pooyan Jamshidi}, \bibinfo{person}{Miguel
  Velez}, \bibinfo{person}{Christian K{\"a}stner}, \bibinfo{person}{Norbert
  Siegmund}, {and} \bibinfo{person}{Prasad Kawthekar}.}
  \bibinfo{year}{2017}\natexlab{}.
\newblock \showarticletitle{Transfer learning for improving model predictions
  in highly configurable software}. In \bibinfo{booktitle}{\emph{2017 IEEE/ACM
  12th International Symposium on Software Engineering for Adaptive and
  Self-Managing Systems (SEAMS)}}. IEEE, \bibinfo{pages}{31--41}.
\newblock


\bibitem[\protect\citeauthoryear{Kampenes, Dyb{\aa}, Hannay, and
  Sj{\o}berg}{Kampenes et~al\mbox{.}}{2007}]%
        {DBLP:journals/infsof/KampenesDHS07}
\bibfield{author}{\bibinfo{person}{Vigdis~By Kampenes}, \bibinfo{person}{Tore
  Dyb{\aa}}, \bibinfo{person}{Jo~Erskine Hannay}, {and} \bibinfo{person}{Dag
  I.~K. Sj{\o}berg}.} \bibinfo{year}{2007}\natexlab{}.
\newblock \showarticletitle{A systematic review of effect size in software
  engineering experiments}.
\newblock \bibinfo{journal}{\emph{Information {\&} Software Technology}}
  \bibinfo{volume}{49}, \bibinfo{number}{11-12} (\bibinfo{year}{2007}),
  \bibinfo{pages}{1073--1086}.
\newblock


\bibitem[\protect\citeauthoryear{Kitchenham, Brereton, Budgen, Turner, Bailey,
  and Linkman}{Kitchenham et~al\mbox{.}}{2009}]%
        {DBLP:journals/infsof/KitchenhamBBTBL09}
\bibfield{author}{\bibinfo{person}{Barbara~A. Kitchenham},
  \bibinfo{person}{Pearl Brereton}, \bibinfo{person}{David Budgen},
  \bibinfo{person}{Mark Turner}, \bibinfo{person}{John Bailey}, {and}
  \bibinfo{person}{Stephen~G. Linkman}.} \bibinfo{year}{2009}\natexlab{}.
\newblock \showarticletitle{Systematic literature reviews in software
  engineering - {A} systematic literature review}.
\newblock \bibinfo{journal}{\emph{Inf. Softw. Technol.}} \bibinfo{volume}{51},
  \bibinfo{number}{1} (\bibinfo{year}{2009}), \bibinfo{pages}{7--15}.
\newblock
\urldef\tempurl%
\url{https://doi.org/10.1016/j.infsof.2008.09.009}
\showDOI{\tempurl}


\bibitem[\protect\citeauthoryear{Koziolek, Koziolek, and Reussner}{Koziolek
  et~al\mbox{.}}{2011}]%
        {DBLP:conf/qosa/KoziolekKR11}
\bibfield{author}{\bibinfo{person}{Anne Koziolek}, \bibinfo{person}{Heiko
  Koziolek}, {and} \bibinfo{person}{Ralf~H. Reussner}.}
  \bibinfo{year}{2011}\natexlab{}.
\newblock \showarticletitle{PerOpteryx: automated application of tactics in
  multi-objective software architecture optimization}. In
  \bibinfo{booktitle}{\emph{7th International Conference on the Quality of
  Software Architectures, QoSA 2011 and 2nd International Symposium on
  Architecting Critical Systems, {ISARCS} 2011. Boulder, CO, USA, June 20-24,
  2011, Proceedings}}, \bibfield{editor}{\bibinfo{person}{Ivica Crnkovic},
  \bibinfo{person}{Judith~A. Stafford}, \bibinfo{person}{Dorina~C. Petriu},
  \bibinfo{person}{Jens Happe}, {and} \bibinfo{person}{Paola Inverardi}}
  (Eds.). \bibinfo{publisher}{{ACM}}, \bibinfo{pages}{33--42}.
\newblock
\urldef\tempurl%
\url{https://doi.org/10.1145/2000259.2000267}
\showDOI{\tempurl}


\bibitem[\protect\citeauthoryear{Kumar, Chen, Bahsoon, and Buyya}{Kumar
  et~al\mbox{.}}{2020}]%
        {DBLP:conf/icse/Kumar0BB20}
\bibfield{author}{\bibinfo{person}{Satish Kumar}, \bibinfo{person}{Tao Chen},
  \bibinfo{person}{Rami Bahsoon}, {and} \bibinfo{person}{Rajkumar Buyya}.}
  \bibinfo{year}{2020}\natexlab{}.
\newblock \showarticletitle{{DATESSO:} self-adapting service composition with
  debt-aware two levels constraint reasoning}. In
  \bibinfo{booktitle}{\emph{{SEAMS} '20: {IEEE/ACM} 15th International
  Symposium on Software Engineering for Adaptive and Self-Managing Systems,
  Seoul, Republic of Korea, 29 June - 3 July, 2020}},
  \bibfield{editor}{\bibinfo{person}{Shinichi Honiden},
  \bibinfo{person}{Elisabetta~Di Nitto}, {and} \bibinfo{person}{Radu
  Calinescu}} (Eds.). \bibinfo{publisher}{{ACM}}, \bibinfo{pages}{96--107}.
\newblock
\urldef\tempurl%
\url{https://doi.org/10.1145/3387939.3391604}
\showDOI{\tempurl}


\bibitem[\protect\citeauthoryear{{Li}, {Liao}, {Deb}, {Min}, and {Yao}}{{Li}
  et~al\mbox{.}}{2020}]%
        {9066927}
\bibfield{author}{\bibinfo{person}{K. {Li}}, \bibinfo{person}{M. {Liao}},
  \bibinfo{person}{K. {Deb}}, \bibinfo{person}{G. {Min}}, {and}
  \bibinfo{person}{X. {Yao}}.} \bibinfo{year}{2020}\natexlab{}.
\newblock \showarticletitle{Does Preference Always Help? A Holistic Study on
  Preference-Based Evolutionary Multiobjective Optimization Using Reference
  Points}.
\newblock \bibinfo{journal}{\emph{IEEE Transactions on Evolutionary
  Computation}} \bibinfo{volume}{24}, \bibinfo{number}{6}
  (\bibinfo{year}{2020}), \bibinfo{pages}{1078--1096}.
\newblock
\urldef\tempurl%
\url{https://doi.org/10.1109/TEVC.2020.2987559}
\showDOI{\tempurl}


\bibitem[\protect\citeauthoryear{Li, Xiang, Chen, and Tan}{Li
  et~al\mbox{.}}{2020a}]%
        {DBLP:conf/kbse/LiXCT20}
\bibfield{author}{\bibinfo{person}{Ke Li}, \bibinfo{person}{Zilin Xiang},
  \bibinfo{person}{Tao Chen}, {and} \bibinfo{person}{Kay~Chen Tan}.}
  \bibinfo{year}{2020}\natexlab{a}.
\newblock \showarticletitle{BiLO-CPDP: Bi-Level Programming for Automated Model
  Discovery in Cross-Project Defect Prediction}. In
  \bibinfo{booktitle}{\emph{35th {IEEE/ACM} International Conference on
  Automated Software Engineering, {ASE} 2020, Melbourne, Australia, September
  21-25, 2020}}. \bibinfo{publisher}{{IEEE}}, \bibinfo{pages}{573--584}.
\newblock
\urldef\tempurl%
\url{https://doi.org/10.1145/3324884.3416617}
\showDOI{\tempurl}


\bibitem[\protect\citeauthoryear{Li, Xiang, Chen, Wang, and Tan}{Li
  et~al\mbox{.}}{2020b}]%
        {DBLP:conf/icse/LiX0WT20}
\bibfield{author}{\bibinfo{person}{Ke Li}, \bibinfo{person}{Zilin Xiang},
  \bibinfo{person}{Tao Chen}, \bibinfo{person}{Shuo Wang}, {and}
  \bibinfo{person}{Kay~Chen Tan}.} \bibinfo{year}{2020}\natexlab{b}.
\newblock \showarticletitle{Understanding the automated parameter optimization
  on transfer learning for cross-project defect prediction: an empirical
  study}. In \bibinfo{booktitle}{\emph{{ICSE} '20: 42nd International
  Conference on Software Engineering, Seoul, South Korea, 27 June - 19 July,
  2020}}, \bibfield{editor}{\bibinfo{person}{Gregg Rothermel} {and}
  \bibinfo{person}{Doo{-}Hwan Bae}} (Eds.). \bibinfo{publisher}{{ACM}},
  \bibinfo{pages}{566--577}.
\newblock
\urldef\tempurl%
\url{https://doi.org/10.1145/3377811.3380360}
\showDOI{\tempurl}


\bibitem[\protect\citeauthoryear{Li, Chen, and Yao}{Li et~al\mbox{.}}{2018}]%
        {li2018critical}
\bibfield{author}{\bibinfo{person}{Miqing Li}, \bibinfo{person}{Tao Chen},
  {and} \bibinfo{person}{Xin Yao}.} \bibinfo{year}{2018}\natexlab{}.
\newblock \showarticletitle{A Critical Review of "{A} Practical Guide to Select
  Quality Indicators for Assessing {Pareto}-Based Search Algorithms in
  Search-Based Software Engineering": Essay on Quality Indicator Selection for
  {SBSE}}. In \bibinfo{booktitle}{\emph{2018 IEEE/ACM 40th International
  Conference on Software Engineering: New Ideas and Emerging Technologies
  Results}}. \bibinfo{pages}{17--20}.
\newblock


\bibitem[\protect\citeauthoryear{{Li}, {Chen}, and {Yao}}{{Li}
  et~al\mbox{.}}{2022}]%
        {9252185}
\bibfield{author}{\bibinfo{person}{Miqing {Li}}, \bibinfo{person}{Tao {Chen}},
  {and} \bibinfo{person}{Xin {Yao}}.} \bibinfo{year}{2022}\natexlab{}.
\newblock \showarticletitle{How to Evaluate Solutions in Pareto-based
  Search-Based Software Engineering? A Critical Review and Methodological
  Guidance}.
\newblock \bibinfo{journal}{\emph{IEEE Transactions on Software Engineering}}
  \bibinfo{volume}{48}, \bibinfo{number}{5} (\bibinfo{year}{2022}),
  \bibinfo{pages}{1771--1799}.
\newblock
\urldef\tempurl%
\url{https://doi.org/10.1109/TSE.2020.3036108}
\showDOI{\tempurl}


\bibitem[\protect\citeauthoryear{Li, Yang, Li, and Liu}{Li
  et~al\mbox{.}}{2014a}]%
        {Li2014a}
\bibfield{author}{\bibinfo{person}{Miqing Li}, \bibinfo{person}{Shengxiang
  Yang}, \bibinfo{person}{Ke Li}, {and} \bibinfo{person}{Xiaohui Liu}.}
  \bibinfo{year}{2014}\natexlab{a}.
\newblock \showarticletitle{Evolutionary algorithms with segment-based search
  for multiobjective optimization problems}.
\newblock \bibinfo{journal}{\emph{IEEE Transactions on Cybernetics}}
  \bibinfo{volume}{44}, \bibinfo{number}{8} (\bibinfo{year}{2014}),
  \bibinfo{pages}{1295--1313}.
\newblock


\bibitem[\protect\citeauthoryear{Li and Yao}{Li and Yao}{2019}]%
        {li2019quality}
\bibfield{author}{\bibinfo{person}{Miqing Li} {and} \bibinfo{person}{Xin Yao}.}
  \bibinfo{year}{2019}\natexlab{}.
\newblock \showarticletitle{Quality Evaluation of Solution Sets in
  Multiobjective Optimisation: A Survey}.
\newblock \bibinfo{journal}{\emph{Comput. Surveys}} \bibinfo{volume}{52},
  \bibinfo{number}{2} (\bibinfo{year}{2019}).
\newblock


\bibitem[\protect\citeauthoryear{Li, Zeng, Meng, Tan, Zhang, Butt, and
  Fuller}{Li et~al\mbox{.}}{2014b}]%
        {DBLP:conf/hpdc/LiZMTZBF14}
\bibfield{author}{\bibinfo{person}{Min Li}, \bibinfo{person}{Liangzhao Zeng},
  \bibinfo{person}{Shicong Meng}, \bibinfo{person}{Jian Tan},
  \bibinfo{person}{Li Zhang}, \bibinfo{person}{Ali~Raza Butt}, {and}
  \bibinfo{person}{Nicholas~C. Fuller}.} \bibinfo{year}{2014}\natexlab{b}.
\newblock \showarticletitle{{MRONLINE:} MapReduce online performance tuning}.
  In \bibinfo{booktitle}{\emph{The 23rd International Symposium on
  High-Performance Parallel and Distributed Computing, HPDC'14, Vancouver, BC,
  Canada - June 23 - 27, 2014}}, \bibfield{editor}{\bibinfo{person}{Beth
  Plale}, \bibinfo{person}{Matei Ripeanu}, \bibinfo{person}{Franck Cappello},
  {and} \bibinfo{person}{Dongyan Xu}} (Eds.). \bibinfo{publisher}{{ACM}},
  \bibinfo{pages}{165--176}.
\newblock
\urldef\tempurl%
\url{https://doi.org/10.1145/2600212.2600229}
\showDOI{\tempurl}


\bibitem[\protect\citeauthoryear{Martens, Koziolek, Becker, and
  Reussner}{Martens et~al\mbox{.}}{2010}]%
        {DBLP:conf/wosp/MartensKBR10}
\bibfield{author}{\bibinfo{person}{Anne Martens}, \bibinfo{person}{Heiko
  Koziolek}, \bibinfo{person}{Steffen Becker}, {and} \bibinfo{person}{Ralf~H.
  Reussner}.} \bibinfo{year}{2010}\natexlab{}.
\newblock \showarticletitle{Automatically improve software architecture models
  for performance, reliability, and cost using evolutionary algorithms}. In
  \bibinfo{booktitle}{\emph{Proceedings of the first joint {WOSP/SIPEW}
  International Conference on Performance Engineering, San Jose, California,
  USA, January 28-30, 2010}}, \bibfield{editor}{\bibinfo{person}{Alan Adamson},
  \bibinfo{person}{Andre~B. Bondi}, \bibinfo{person}{Carlos Juiz}, {and}
  \bibinfo{person}{Mark~S. Squillante}} (Eds.). \bibinfo{publisher}{{ACM}},
  \bibinfo{pages}{105--116}.
\newblock
\urldef\tempurl%
\url{https://doi.org/10.1145/1712605.1712624}
\showDOI{\tempurl}


\bibitem[\protect\citeauthoryear{McHugh}{McHugh}{2012}]%
        {mchugh2012interrater}
\bibfield{author}{\bibinfo{person}{Mary~L McHugh}.}
  \bibinfo{year}{2012}\natexlab{}.
\newblock \showarticletitle{Interrater reliability: the kappa statistic}.
\newblock \bibinfo{journal}{\emph{Biochemia medica}} \bibinfo{volume}{22},
  \bibinfo{number}{3} (\bibinfo{year}{2012}), \bibinfo{pages}{276--282}.
\newblock


\bibitem[\protect\citeauthoryear{Menzies, Caglayan, Kocaguneli, Krall, Peters,
  and Turhan}{Menzies et~al\mbox{.}}{2012}]%
        {menzies2012promise}
\bibfield{author}{\bibinfo{person}{Tim Menzies}, \bibinfo{person}{Bora
  Caglayan}, \bibinfo{person}{Ekrem Kocaguneli}, \bibinfo{person}{Joe Krall},
  \bibinfo{person}{Fayola Peters}, {and} \bibinfo{person}{Burak Turhan}.}
  \bibinfo{year}{2012}\natexlab{}.
\newblock \bibinfo{title}{The promise repository of empirical software
  engineering data}.
\newblock
\newblock


\bibitem[\protect\citeauthoryear{Nair, Yu, Menzies, Siegmund, and Apel}{Nair
  et~al\mbox{.}}{2020}]%
        {nair2018finding}
\bibfield{author}{\bibinfo{person}{Vivek Nair}, \bibinfo{person}{Zhe Yu},
  \bibinfo{person}{Tim Menzies}, \bibinfo{person}{Norbert Siegmund}, {and}
  \bibinfo{person}{Sven Apel}.} \bibinfo{year}{2020}\natexlab{}.
\newblock \showarticletitle{Finding faster configurations using FLASH}.
\newblock \bibinfo{journal}{\emph{IEEE Transactions on Software Engineering}}
  \bibinfo{volume}{46}, \bibinfo{number}{7} (\bibinfo{year}{2020}).
\newblock


\bibitem[\protect\citeauthoryear{Odhnoff}{Odhnoff}{1965}]%
        {odhnoff1965techniques}
\bibfield{author}{\bibinfo{person}{Jan Odhnoff}.}
  \bibinfo{year}{1965}\natexlab{}.
\newblock \showarticletitle{On the techniques of optimizing and satisficing}.
\newblock \bibinfo{journal}{\emph{The Swedish Journal of Economics}}
  \bibinfo{volume}{67}, \bibinfo{number}{1} (\bibinfo{year}{1965}),
  \bibinfo{pages}{24--39}.
\newblock


\bibitem[\protect\citeauthoryear{Oh, Batory, Myers, and Siegmund}{Oh
  et~al\mbox{.}}{2017}]%
        {DBLP:conf/sigsoft/OhBMS17}
\bibfield{author}{\bibinfo{person}{Jeho Oh}, \bibinfo{person}{Don~S. Batory},
  \bibinfo{person}{Margaret Myers}, {and} \bibinfo{person}{Norbert Siegmund}.}
  \bibinfo{year}{2017}\natexlab{}.
\newblock \showarticletitle{Finding near-optimal configurations in product
  lines by random sampling}. In \bibinfo{booktitle}{\emph{Proceedings of the
  2017 11th Joint Meeting on Foundations of Software Engineering, {ESEC/FSE}
  2017, Paderborn, Germany, September 4-8, 2017}},
  \bibfield{editor}{\bibinfo{person}{Eric Bodden}, \bibinfo{person}{Wilhelm
  Sch{\"{a}}fer}, \bibinfo{person}{Arie van Deursen}, {and}
  \bibinfo{person}{Andrea Zisman}} (Eds.). \bibinfo{publisher}{{ACM}},
  \bibinfo{pages}{61--71}.
\newblock
\urldef\tempurl%
\url{https://doi.org/10.1145/3106237.3106273}
\showDOI{\tempurl}


\bibitem[\protect\citeauthoryear{Ramirez and Cheng}{Ramirez and Cheng}{2011}]%
        {DBLP:conf/models/RamirezC11}
\bibfield{author}{\bibinfo{person}{Andres~J. Ramirez} {and}
  \bibinfo{person}{Betty H.~C. Cheng}.} \bibinfo{year}{2011}\natexlab{}.
\newblock \showarticletitle{Automatic Derivation of Utility Functions for
  Monitoring Software Requirements}. In \bibinfo{booktitle}{\emph{Model Driven
  Engineering Languages and Systems, 14th International Conference, {MODELS}
  2011, Wellington, New Zealand, October 16-21, 2011. Proceedings}}
  \emph{(\bibinfo{series}{Lecture Notes in Computer Science},
  Vol.~\bibinfo{volume}{6981})}, \bibfield{editor}{\bibinfo{person}{Jon
  Whittle}, \bibinfo{person}{Tony Clark}, {and} \bibinfo{person}{Thomas
  K{\"{u}}hne}} (Eds.). \bibinfo{publisher}{Springer},
  \bibinfo{pages}{501--516}.
\newblock
\urldef\tempurl%
\url{https://doi.org/10.1007/978-3-642-24485-8\_37}
\showDOI{\tempurl}


\bibitem[\protect\citeauthoryear{Ramirez, Knoester, Cheng, and
  McKinley}{Ramirez et~al\mbox{.}}{2009}]%
        {DBLP:conf/icac/RamirezKCM09}
\bibfield{author}{\bibinfo{person}{Andres~J. Ramirez},
  \bibinfo{person}{David~B. Knoester}, \bibinfo{person}{Betty H.~C. Cheng},
  {and} \bibinfo{person}{Philip~K. McKinley}.} \bibinfo{year}{2009}\natexlab{}.
\newblock \showarticletitle{Applying genetic algorithms to decision making in
  autonomic computing systems}. In \bibinfo{booktitle}{\emph{Proceedings of the
  6th International Conference on Autonomic Computing, {ICAC} 2009, June 15-19,
  2009, Barcelona, Spain}}, \bibfield{editor}{\bibinfo{person}{Simon~A.
  Dobson}, \bibinfo{person}{John Strassner}, \bibinfo{person}{Manish Parashar},
  {and} \bibinfo{person}{Onn Shehory}} (Eds.). \bibinfo{publisher}{{ACM}},
  \bibinfo{pages}{97--106}.
\newblock
\urldef\tempurl%
\url{https://doi.org/10.1145/1555228.1555258}
\showDOI{\tempurl}


\bibitem[\protect\citeauthoryear{Sayagh, Kerzazi, Adams, and Petrillo}{Sayagh
  et~al\mbox{.}}{2020}]%
        {DBLP:journals/tse/SayaghKAP20}
\bibfield{author}{\bibinfo{person}{Mohammed Sayagh},
  \bibinfo{person}{Noureddine Kerzazi}, \bibinfo{person}{Bram Adams}, {and}
  \bibinfo{person}{F{\'{a}}bio Petrillo}.} \bibinfo{year}{2020}\natexlab{}.
\newblock \showarticletitle{Software Configuration Engineering in Practice
  Interviews, Survey, and Systematic Literature Review}.
\newblock \bibinfo{journal}{\emph{{IEEE} Trans. Software Eng.}}
  \bibinfo{volume}{46}, \bibinfo{number}{6} (\bibinfo{year}{2020}),
  \bibinfo{pages}{646--673}.
\newblock
\urldef\tempurl%
\url{https://doi.org/10.1109/TSE.2018.2867847}
\showDOI{\tempurl}


\bibitem[\protect\citeauthoryear{Shahbazian, Karthik, Brun, and
  Medvidovic}{Shahbazian et~al\mbox{.}}{2020}]%
        {DBLP:conf/sigsoft/ShahbazianKBM20}
\bibfield{author}{\bibinfo{person}{Arman Shahbazian}, \bibinfo{person}{Suhrid
  Karthik}, \bibinfo{person}{Yuriy Brun}, {and} \bibinfo{person}{Nenad
  Medvidovic}.} \bibinfo{year}{2020}\natexlab{}.
\newblock \showarticletitle{eQual: informing early design decisions}. In
  \bibinfo{booktitle}{\emph{{ESEC/FSE} '20: 28th {ACM} Joint European Software
  Engineering Conference and Symposium on the Foundations of Software
  Engineering, Virtual Event, USA, November 8-13, 2020}},
  \bibfield{editor}{\bibinfo{person}{Prem Devanbu}, \bibinfo{person}{Myra~B.
  Cohen}, {and} \bibinfo{person}{Thomas Zimmermann}} (Eds.).
  \bibinfo{publisher}{{ACM}}, \bibinfo{pages}{1039--1051}.
\newblock
\urldef\tempurl%
\url{https://doi.org/10.1145/3368089.3409749}
\showDOI{\tempurl}


\bibitem[\protect\citeauthoryear{Shaukat, Naseem, and Zubair}{Shaukat
  et~al\mbox{.}}{2018}]%
        {DBLP:conf/cse/ShaukatNZ18}
\bibfield{author}{\bibinfo{person}{Zain~Shaukat Shaukat},
  \bibinfo{person}{Rashid Naseem}, {and} \bibinfo{person}{Muhammad Zubair}.}
  \bibinfo{year}{2018}\natexlab{}.
\newblock \showarticletitle{A Dataset for Software Requirements Risk
  Prediction}. In \bibinfo{booktitle}{\emph{2018 {IEEE} International
  Conference on Computational Science and Engineering, {CSE} 2018, Bucharest,
  Romania, October 29-31, 2018}}, \bibfield{editor}{\bibinfo{person}{Florin
  Pop}, \bibinfo{person}{Catalin Negru}, \bibinfo{person}{Horacio
  Gonz{\'{a}}lez{-}V{\'{e}}lez}, {and} \bibinfo{person}{Jacek Rak}} (Eds.).
  \bibinfo{publisher}{{IEEE} Computer Society}, \bibinfo{pages}{112--118}.
\newblock
\urldef\tempurl%
\url{https://doi.org/10.1109/CSE.2018.00022}
\showDOI{\tempurl}


\bibitem[\protect\citeauthoryear{Singh, Bezemer, Shang, and Hassan}{Singh
  et~al\mbox{.}}{2016}]%
        {DBLP:conf/wosp/SinghBSH16}
\bibfield{author}{\bibinfo{person}{Ravjot Singh}, \bibinfo{person}{Cor{-}Paul
  Bezemer}, \bibinfo{person}{Weiyi Shang}, {and} \bibinfo{person}{Ahmed~E.
  Hassan}.} \bibinfo{year}{2016}\natexlab{}.
\newblock \showarticletitle{Optimizing the Performance-Related Configurations
  of Object-Relational Mapping Frameworks Using a Multi-Objective Genetic
  Algorithm}. In \bibinfo{booktitle}{\emph{Proceedings of the 7th {ACM/SPEC}
  International Conference on Performance Engineering, {ICPE} 2016, Delft, The
  Netherlands, March 12-16, 2016}}, \bibfield{editor}{\bibinfo{person}{Alberto
  Avritzer}, \bibinfo{person}{Alexandru Iosup}, \bibinfo{person}{Xiaoyun Zhu},
  {and} \bibinfo{person}{Steffen Becker}} (Eds.). \bibinfo{publisher}{{ACM}},
  \bibinfo{pages}{309--320}.
\newblock
\urldef\tempurl%
\url{https://doi.org/10.1145/2851553.2851576}
\showDOI{\tempurl}


\bibitem[\protect\citeauthoryear{Sinha, Cashman, and Cohen}{Sinha
  et~al\mbox{.}}{2020}]%
        {DBLP:conf/ssbse/SinhaCC20}
\bibfield{author}{\bibinfo{person}{Urjoshi Sinha}, \bibinfo{person}{Mikaela
  Cashman}, {and} \bibinfo{person}{Myra~B. Cohen}.}
  \bibinfo{year}{2020}\natexlab{}.
\newblock \showarticletitle{Using a Genetic Algorithm to Optimize
  Configurations in a Data-Driven Application}. In
  \bibinfo{booktitle}{\emph{Search-Based Software Engineering - 12th
  International Symposium, {SSBSE} 2020, Bari, Italy, October 7-8, 2020,
  Proceedings}} \emph{(\bibinfo{series}{Lecture Notes in Computer Science},
  Vol.~\bibinfo{volume}{12420})}, \bibfield{editor}{\bibinfo{person}{Aldeida
  Aleti} {and} \bibinfo{person}{Annibale Panichella}} (Eds.).
  \bibinfo{publisher}{Springer}, \bibinfo{pages}{137--152}.
\newblock
\urldef\tempurl%
\url{https://doi.org/10.1007/978-3-030-59762-7\_10}
\showDOI{\tempurl}


\bibitem[\protect\citeauthoryear{Vargha and Delaney}{Vargha and
  Delaney}{2000}]%
        {Vargha2000ACA}
\bibfield{author}{\bibinfo{person}{Andr{\'a}s Vargha} {and}
  \bibinfo{person}{Harold~D. Delaney}.} \bibinfo{year}{2000}\natexlab{}.
\newblock \showarticletitle{A Critique and Improvement of the CL Common
  Language Effect Size Statistics of McGraw and Wong}.
\newblock


\bibitem[\protect\citeauthoryear{Veerapen, Ochoa, Harman, and Burke}{Veerapen
  et~al\mbox{.}}{2015}]%
        {Veerapen2015}
\bibfield{author}{\bibinfo{person}{Nadarajen Veerapen},
  \bibinfo{person}{Gabriela Ochoa}, \bibinfo{person}{Mark Harman}, {and}
  \bibinfo{person}{Edmund~K Burke}.} \bibinfo{year}{2015}\natexlab{}.
\newblock \showarticletitle{An integer linear programming approach to the
  single and bi-objective next release problem}.
\newblock \bibinfo{journal}{\emph{Information and Software Technology}}
  \bibinfo{volume}{65} (\bibinfo{year}{2015}), \bibinfo{pages}{1--13}.
\newblock


\bibitem[\protect\citeauthoryear{Wang, Olhofer, and Jin}{Wang
  et~al\mbox{.}}{2017}]%
        {wang2017mini}
\bibfield{author}{\bibinfo{person}{Handing Wang}, \bibinfo{person}{Markus
  Olhofer}, {and} \bibinfo{person}{Yaochu Jin}.}
  \bibinfo{year}{2017}\natexlab{}.
\newblock \showarticletitle{A mini-review on preference modeling and
  articulation in multi-objective optimization: current status and challenges}.
\newblock \bibinfo{journal}{\emph{Complex \& Intelligent Systems}}
  \bibinfo{volume}{3}, \bibinfo{number}{4} (\bibinfo{year}{2017}),
  \bibinfo{pages}{233--245}.
\newblock


\bibitem[\protect\citeauthoryear{Whittle, Sawyer, Bencomo, Cheng, and
  Bruel}{Whittle et~al\mbox{.}}{2009}]%
        {DBLP:conf/re/WhittleSBCB09}
\bibfield{author}{\bibinfo{person}{Jon Whittle}, \bibinfo{person}{Peter
  Sawyer}, \bibinfo{person}{Nelly Bencomo}, \bibinfo{person}{Betty H.~C.
  Cheng}, {and} \bibinfo{person}{Jean{-}Michel Bruel}.}
  \bibinfo{year}{2009}\natexlab{}.
\newblock \showarticletitle{{RELAX:} Incorporating Uncertainty into the
  Specification of Self-Adaptive Systems}. In \bibinfo{booktitle}{\emph{{RE}
  2009, 17th {IEEE} International Requirements Engineering Conference, Atlanta,
  Georgia, USA, August 31 - September 4, 2009}}. \bibinfo{publisher}{{IEEE}
  Computer Society}, \bibinfo{pages}{79--88}.
\newblock
\urldef\tempurl%
\url{https://doi.org/10.1109/RE.2009.36}
\showDOI{\tempurl}


\bibitem[\protect\citeauthoryear{Wilcoxon}{Wilcoxon}{1945}]%
        {Wilcoxon1945IndividualCB}
\bibfield{author}{\bibinfo{person}{Frank Wilcoxon}.}
  \bibinfo{year}{1945}\natexlab{}.
\newblock \showarticletitle{Individual Comparisons by Ranking Methods}.
\newblock


\bibitem[\protect\citeauthoryear{Xi, Liu, Raghavachari, Xia, and Zhang}{Xi
  et~al\mbox{.}}{2004}]%
        {DBLP:conf/www/XiLRXZ04}
\bibfield{author}{\bibinfo{person}{Bowei Xi}, \bibinfo{person}{Zhen Liu},
  \bibinfo{person}{Mukund Raghavachari}, \bibinfo{person}{Cathy~H. Xia}, {and}
  \bibinfo{person}{Li Zhang}.} \bibinfo{year}{2004}\natexlab{}.
\newblock \showarticletitle{A smart hill-climbing algorithm for application
  server configuration}. In \bibinfo{booktitle}{\emph{Proceedings of the 13th
  international conference on World Wide Web, {WWW} 2004, New York, NY, USA,
  May 17-20, 2004}}, \bibfield{editor}{\bibinfo{person}{Stuart~I. Feldman},
  \bibinfo{person}{Mike Uretsky}, \bibinfo{person}{Marc Najork}, {and}
  \bibinfo{person}{Craig~E. Wills}} (Eds.). \bibinfo{publisher}{{ACM}},
  \bibinfo{pages}{287--296}.
\newblock
\urldef\tempurl%
\url{https://doi.org/10.1145/988672.988711}
\showDOI{\tempurl}


\bibitem[\protect\citeauthoryear{Xu, Jin, Fan, Zhou, Pasupathy, and
  Talwadker}{Xu et~al\mbox{.}}{2015}]%
        {DBLP:conf/sigsoft/XuJFZPT15}
\bibfield{author}{\bibinfo{person}{Tianyin Xu}, \bibinfo{person}{Long Jin},
  \bibinfo{person}{Xuepeng Fan}, \bibinfo{person}{Yuanyuan Zhou},
  \bibinfo{person}{Shankar Pasupathy}, {and} \bibinfo{person}{Rukma
  Talwadker}.} \bibinfo{year}{2015}\natexlab{}.
\newblock \showarticletitle{Hey, you have given me too many knobs!:
  understanding and dealing with over-designed configuration in system
  software}. In \bibinfo{booktitle}{\emph{Proceedings of the 2015 10th Joint
  Meeting on Foundations of Software Engineering, {ESEC/FSE} 2015, Bergamo,
  Italy, August 30 - September 4, 2015}},
  \bibfield{editor}{\bibinfo{person}{Elisabetta~Di Nitto},
  \bibinfo{person}{Mark Harman}, {and} \bibinfo{person}{Patrick Heymans}}
  (Eds.). \bibinfo{publisher}{{ACM}}, \bibinfo{pages}{307--319}.
\newblock
\urldef\tempurl%
\url{https://doi.org/10.1145/2786805.2786852}
\showDOI{\tempurl}


\bibitem[\protect\citeauthoryear{Ye and Kalyanaraman}{Ye and
  Kalyanaraman}{2003}]%
        {DBLP:conf/sigmetrics/YeK03}
\bibfield{author}{\bibinfo{person}{Tao Ye} {and} \bibinfo{person}{Shivkumar
  Kalyanaraman}.} \bibinfo{year}{2003}\natexlab{}.
\newblock \showarticletitle{A recursive random search algorithm for large-scale
  network parameter configuration}. In \bibinfo{booktitle}{\emph{Proceedings of
  the International Conference on Measurements and Modeling of Computer
  Systems, {SIGMETRICS} 2003, June 9-14, 2003, San Diego, CA, {USA}}},
  \bibfield{editor}{\bibinfo{person}{Bill Cheng}, \bibinfo{person}{Satish~K.
  Tripathi}, \bibinfo{person}{Jennifer Rexford}, {and}
  \bibinfo{person}{William~H. Sanders}} (Eds.). \bibinfo{publisher}{{ACM}},
  \bibinfo{pages}{196--205}.
\newblock
\urldef\tempurl%
\url{https://doi.org/10.1145/781027.781052}
\showDOI{\tempurl}


\bibitem[\protect\citeauthoryear{Yu, Jin, and Olhofer}{Yu
  et~al\mbox{.}}{2019}]%
        {DBLP:conf/cec/YuJO19}
\bibfield{author}{\bibinfo{person}{Guo Yu}, \bibinfo{person}{Yaochu Jin}, {and}
  \bibinfo{person}{Markus Olhofer}.} \bibinfo{year}{2019}\natexlab{}.
\newblock \showarticletitle{References or Preferences - Rethinking
  Many-objective Evolutionary Optimization}. In
  \bibinfo{booktitle}{\emph{{IEEE} Congress on Evolutionary Computation, {CEC}
  2019, Wellington, New Zealand, June 10-13, 2019}}.
  \bibinfo{publisher}{{IEEE}}, \bibinfo{pages}{2410--2417}.
\newblock
\urldef\tempurl%
\url{https://doi.org/10.1109/CEC.2019.8790106}
\showDOI{\tempurl}


\bibitem[\protect\citeauthoryear{Yu, Zheng, Shen, and Li}{Yu
  et~al\mbox{.}}{2016}]%
        {Yu2016}
\bibfield{author}{\bibinfo{person}{Guo Yu}, \bibinfo{person}{Jinhua Zheng},
  \bibinfo{person}{Ruimin Shen}, {and} \bibinfo{person}{Miqing Li}.}
  \bibinfo{year}{2016}\natexlab{}.
\newblock \showarticletitle{Decomposing the user-preference in multiobjective
  optimization}.
\newblock \bibinfo{journal}{\emph{Soft Computing}} \bibinfo{volume}{20},
  \bibinfo{number}{10} (\bibinfo{year}{2016}), \bibinfo{pages}{4005--4021}.
\newblock


\bibitem[\protect\citeauthoryear{Zhang and Li}{Zhang and Li}{2007}]%
        {DBLP:journals/tec/ZhangL07}
\bibfield{author}{\bibinfo{person}{Qingfu Zhang} {and} \bibinfo{person}{Hui
  Li}.} \bibinfo{year}{2007}\natexlab{}.
\newblock \showarticletitle{{MOEA/D:} {A} Multiobjective Evolutionary Algorithm
  Based on Decomposition}.
\newblock \bibinfo{journal}{\emph{{IEEE} Trans. Evol. Comput.}}
  \bibinfo{volume}{11}, \bibinfo{number}{6} (\bibinfo{year}{2007}),
  \bibinfo{pages}{712--731}.
\newblock


\bibitem[\protect\citeauthoryear{Zhu, Liu, Guo, Bao, Ma, Liu, Song, and
  Yang}{Zhu et~al\mbox{.}}{2017}]%
        {DBLP:conf/cloud/ZhuLGBMLSY17}
\bibfield{author}{\bibinfo{person}{Yuqing Zhu}, \bibinfo{person}{Jianxun Liu},
  \bibinfo{person}{Mengying Guo}, \bibinfo{person}{Yungang Bao},
  \bibinfo{person}{Wenlong Ma}, \bibinfo{person}{Zhuoyue Liu},
  \bibinfo{person}{Kunpeng Song}, {and} \bibinfo{person}{Yingchun Yang}.}
  \bibinfo{year}{2017}\natexlab{}.
\newblock \showarticletitle{BestConfig: tapping the performance potential of
  systems via automatic configuration tuning}. In
  \bibinfo{booktitle}{\emph{Proceedings of the 2017 Symposium on Cloud
  Computing, SoCC 2017, Santa Clara, CA, USA, September 24-27, 2017}}.
  \bibinfo{publisher}{{ACM}}, \bibinfo{pages}{338--350}.
\newblock
\urldef\tempurl%
\url{https://doi.org/10.1145/3127479.3128605}
\showDOI{\tempurl}


\bibitem[\protect\citeauthoryear{Zitzler, Brockhoff, and Thiele}{Zitzler
  et~al\mbox{.}}{2007}]%
        {Zitzler2007a}
\bibfield{author}{\bibinfo{person}{E. Zitzler}, \bibinfo{person}{D. Brockhoff},
  {and} \bibinfo{person}{L. Thiele}.} \bibinfo{year}{2007}\natexlab{}.
\newblock \showarticletitle{The hypervolume indicator revisited: On the design
  of Pareto-compliant indicators via weighted integration}. In
  \bibinfo{booktitle}{\emph{International Conference on Evolutionary
  Multi-Criterion Optimization}}. Springer, \bibinfo{pages}{862--876}.
\newblock


\bibitem[\protect\citeauthoryear{Zitzler and K{\"{u}}nzli}{Zitzler and
  K{\"{u}}nzli}{2004}]%
        {DBLP:conf/ppsn/ZitzlerK04}
\bibfield{author}{\bibinfo{person}{Eckart Zitzler} {and} \bibinfo{person}{Simon
  K{\"{u}}nzli}.} \bibinfo{year}{2004}\natexlab{}.
\newblock \showarticletitle{Indicator-Based Selection in Multiobjective
  Search}. In \bibinfo{booktitle}{\emph{Parallel Problem Solving from Nature -
  {PPSN} VIII, 8th International Conference, Birmingham, UK, September 18-22,
  2004, Proceedings}} \emph{(\bibinfo{series}{Lecture Notes in Computer
  Science}, Vol.~\bibinfo{volume}{3242})},
  \bibfield{editor}{\bibinfo{person}{Xin Yao}, \bibinfo{person}{Edmund~K.
  Burke}, \bibinfo{person}{Jos{\'{e}}~Antonio Lozano}, \bibinfo{person}{Jim
  Smith}, \bibinfo{person}{Juan Juli{\'{a}}n~Merelo Guerv{\'{o}}s},
  \bibinfo{person}{John~A. Bullinaria}, \bibinfo{person}{Jonathan~E. Rowe},
  \bibinfo{person}{Peter Ti{\~{n}}o}, \bibinfo{person}{Ata Kab{\'{a}}n}, {and}
  \bibinfo{person}{Hans{-}Paul Schwefel}} (Eds.).
  \bibinfo{publisher}{Springer}, \bibinfo{pages}{832--842}.
\newblock


\bibitem[\protect\citeauthoryear{Zitzler and Thiele}{Zitzler and
  Thiele}{1998}]%
        {Zitzler1998}
\bibfield{author}{\bibinfo{person}{E. Zitzler} {and} \bibinfo{person}{L.
  Thiele}.} \bibinfo{year}{1998}\natexlab{}.
\newblock \showarticletitle{Multiobjective optimization using evolutionary
  algorithms - A comparative case study}.
\newblock In \bibinfo{booktitle}{\emph{Proceedings of the International
  Conference on Parallel Problem Solving from Nature (PPSN)}}.
  \bibinfo{pages}{292--301}.
\newblock


\bibitem[\protect\citeauthoryear{Zitzler, Thiele, Laumanns, Fonseca, and
  Da~Fonseca}{Zitzler et~al\mbox{.}}{2003}]%
        {Zitzler2003}
\bibfield{author}{\bibinfo{person}{E. Zitzler}, \bibinfo{person}{L. Thiele},
  \bibinfo{person}{M. Laumanns}, \bibinfo{person}{C.~M. Fonseca}, {and}
  \bibinfo{person}{V.~G. Da~Fonseca}.} \bibinfo{year}{2003}\natexlab{}.
\newblock \showarticletitle{{Performance assessment of multiobjective
  optimizers: An analysis and review}}.
\newblock \bibinfo{journal}{\emph{IEEE Transactions on Evolutionary
  Computation}} \bibinfo{volume}{7}, \bibinfo{number}{2}
  (\bibinfo{year}{2003}), \bibinfo{pages}{117--132}.
\newblock


\end{thebibliography}
